\documentclass{aa}  

\usepackage{graphicx}
\usepackage{placeins}
\usepackage{txfonts}
\usepackage{ulem}
\usepackage{pdflscape}
\usepackage{amsmath}
\usepackage{hyperref}
\hypersetup{colorlinks,allcolors=blue}

\makeatletter
\renewcommand*\aa@pageof{, page \thepage{} of \pageref*{LastPage}}
\makeatother

\newcommand{\MJup}{M$_{\mathrm{Jup}}$\xspace}

\newcommand{\MSun}{M$_{\odot}$\xspace}

\newcommand{\teff}{$T_{\rm eff}$\xspace}

\newcommand{\kms}{km\, s$^{-1}$\xspace}

\begin{document}

   \title{Stellar companions and Jupiter-like planets in young associations }


   \author{R. Gratton\inst{1}, M. Bonavita\inst{1,2}, D. Mesa\inst{1}, S. Desidera\inst{1}, A. Zurlo\inst{3,4,5}, S. Marino\inst{6}, V. D'Orazi\inst{1,7}, E. Rigliaco\inst{1}, V. Nascimbeni\inst{1}, D. Barbato\inst{1}, G. Columba\inst{1,8}, V.  Squicciarini\inst{1,9} 
   }
  \institute{
   \inst{1}INAF-Osservatorio Astronomico di Padova, Vicolo dell'Osservatorio 5, Padova, Italy, 35122-I \\
   \inst{2}Institute for Astronomy, University of Edinburgh Royal Observatory, Blackford Hill, EH9 3HJ, Edinburgh, UK \\
   \inst{3}Instituto de Estudios Astrof\'isicos, Facultad de Ingenier\'ia y Ciencias, Universidad Diego Portales, Av. Ej\'ercito 441, Santiago, Chile\\
   \inst{4}Escuela de Ingenier\'ia Industrial, Facultad de Ingenier\'ia y Ciencias, Universidad Diego Portales, Av. Ej\'ercito 441, Santiago, Chile\\
   \inst{5}Millennium Nucleus on Young Exoplanets and their Moons (YEMS)\\
   \inst{6}Department of Physics and Astronomy, University of Exeter, Stocker Road, Exeter, EX4 4QL, UK\\
   \inst{7}Dipartimento di Fisica, Universit\`{a} di Roma Tor Vergata, via della Ricerca Scientifica 1, 00133, Roma, Italy\\
   \inst{8}Dipartimento di Fisica e Astronomia G. Galilei, Universit\`{a} di Padova, Via Francesco Marzolo, 8, 35121 Padova, Italy\\
   \inst{9}LESIA, Observatoire de Paris, Universit\'e PSL, CNRS, Sorbonne Universit\'e, Universit\'e Paris Cit\'e, 5 place Jules Janssen, 92195 Meudon, France
} 

   \date{Received; accepted }

 
  \abstract
   {The formation mechanisms of stellar, brown dwarf, and planetary companions, their dependencies on the environment and their interactions with each other are still not well established. Recently, combining high-contrast imaging and space astrometry we found that Jupiter-like (JL) planets are frequent in the $\beta$ Pic moving group (BPMG) around those stars where their orbit can be stable, prompting further analysis and discussion.}
   {We broaden our previous analysis to other young nearby associations to determine the frequency, mass and separation of companions in general and JL in particular and their dependencies on the mass and age of the associations.}
   {We collected available data about companions to the stars in the BPMG and seven additional young associations, including those revealed by visual observations, eclipses, spectroscopy and astrometry. }
   {
   We determined search completeness and found that it is very high for stellar companions, while completeness corrections are still large for JL companions. Once these corrections are included, we found a high frequency of companions, both stellar ($>0.52\pm 0.03$) and JL ($0.57\pm 0.11$). The two populations are clearly separated by a gap that corresponds to the well-known brown dwarf desert. Within the population of massive companions, we found clear trends in frequency, separation, and mass ratios with stellar mass. Planetary companions pile up in the region just outside the ice line and we found them to be frequent once completeness was considered. The frequency of JL planets decreases with the overall mass and possibly the age of the association.}
   {We tentatively identify the two populations as due to disk fragmentation and core accretion, respectively. The distributions of stellar companions with a semi-major axis $<1000$ au is indeed well reproduced by a simple model of formation by disk fragmentation. The observed trends with stellar mass can be explained by a shorter but much more intense phase of accretion onto the disk of massive stars and by a more steady and prolonged accretion on solar-type stars. Possible explanations for the trends in the population of JL planets with association mass and age are briefly discussed.}


\titlerunning{Companions in young associations }
\authorrunning{R. Gratton et al.}

   \maketitle
%

\section{Introduction}

A large fraction of the stars have companions over a wide range of masses: stars (mass $M>0.075$ \MSun), brown dwarfs (BDs: $0.013<M<0.075$ \MSun), or planets ($M<0.013$ \MSun). The mechanisms for the formation of multiple systems, how they depend on the environment, and how they interact with each other are still not well established. The favoured scenarios for the formation of stellar companions are turbulent fragmentation of clouds for a separation $>500$ au \citep{Offner2010, Offner2016} and disk fragmentation for a separation $<500$ au \citep{Kratter2010}. However, the details of which are far from being well understood (see e.g. \citealt{Tohline2002}). Disk fragmentation is expected to be more efficient around massive stars because of the larger value of the accretion rate from the natal cloud and hence the larger expected disk-to-star mass ratio during early phases of formation, when binaries are likely to form \citep{Machida2010, Kratter2016, Elbakyan2023}. In disk fragmentation, mass accretion on the secondary may be favoured with respect to accretion on the primary \citep{Clarke2012}; if the disk survives long enough, this would lead to a preference for equal mass binaries \citep{Kratter2010}. Planets and BDs may either form by disk fragmentation \citep{Boss1997} or more likely by core accretion \citep{Pollack1996, Mordasini2012, Bitsch2015}. While disk fragmentation is a fast process (at most a few $10^4$ yr), core accretion is a slow process requiring a few million of years. We then expect that binary formation by disk fragmentation occurs much earlier and sets the stage for the later formation of planets by the core accretion process. Early formation of stellar companions may be an obstacle for the later formation of planets if they reside at a distance from the central star similar to that of planet formation because they may destroy the proto-planetary disk or make planetary orbits unstable (see e.g. \citealt{Holman1999}).

Related to these issues are basic questions in the study of exoplanets, such as whether the Solar System is an anomaly in the exoplanetary context and where and how it formed. The answers to these questions are uncertain \citep{Martin2015, Parker2020, Gaudi2022, Raymond2020, Raymond2022} and observations of extrasolar planetary systems may help. Most of the companion mass of our Solar System is represented by the giant planets lying just beyond the ice-line, which is the region of the system where ice can survive the Sun’s heat. We can call planets found in this region (semi-major axis of the orbits between 3 and 15 au: \citealt{Hayashi1981, Podolak2004, Martin2012}) in extrasolar systems Jupiter-like  (JL) planets if they have a mass equal or larger than that of Jupiter. Apart from transits, an external observer would likely discover JL planets long before detecting smaller inner rocky planets such as the Earth. Furthermore, a correlation is found between the presence of JL planets and inner smaller planets \citep{Bryan2016, Rosenthal2022}. Therefore, a basic step to answer the above questions is to establish how common JL planets are around Sun-like stars, as it is already known that giant planets are rare amongst stars of a lower mass (see e.g. \citealt{Laughlin2004, Alibert2011, Burn2021, Johnson2010}). In this respect, it is remarkable that the peak of the distribution of stellar companions is only slightly further out than the ice-line in solar-type stars (see e.g. \citealt{Raghavan2010, Gratton2023}).

Over the last twenty years, surveys based on variations of radial velocities (RVs) have been used to detect planets outside the Solar System \citep{Cumming2008,Mayor2011,Wittenmyer2016,Fernandes2019,Zhu2022,Wolthoff2022}. These surveys have found that only about 6\% of Sun-like stars host JL planets with a mass $>1$ \MJup \footnote{We adopt here this limit for the JL planets because detection of smaller mass JL planets is very difficult, see e.g. the discussion in \citet{Lagrange2023}.} and only 2\% host a massive JL planet ($M>3$ \MJup). However, planet parameters become highly uncertain when the periods are longer than the time coverage of the RV series \citep{Lagrange2023}, so these results should be taken with care. A similar, though very uncertain, result has been obtained using a single microlensing event involving a massive JL planet \citep{Gould2010}, while a large incidence of wide-orbit planets over a wider range of masses has been found by \citet{Poleski2021}. This would indicate that our Solar System is relatively unusual. However, JL planets are difficult to discover through RV, because of the long periods and low amplitudes of RV variations. Typically, these surveys target stars with ages of several billion years, whose sites of formation are essentially unknown. Therefore, they are of limited use to understand where and how our Solar System formed.

A new observation may provide important hints. Using direct imaging, very recently three independent papers \citep{Mesa2023, DeRosa2023, Franson:2023arXiv} found a third star (AF Lep) hosting a JL planet in the $\beta$ Pic moving group (BPMG), in addition to $\beta$ Pic itself and 51 Eri. The BPMG is a sparse group of 20-million-year-old stars, likely born in a single episode of star formation. The BPMG includes only about 30 Sun-like stars ($M>0.8$ \MSun). Due to its young age and close proximity to the Sun, the BPMG is the most favourable star ensemble for planet detection by means of direct high contrast imaging (HCI). Even so, the discovery of JL planets with this technique is still difficult, and only JL planets with a mass greater than three times that of Jupiter (i.e., roughly a third of all such planets) can be detected, and that can only occur under favourable conditions. The HCI discovery of three stars hosting JL planets in this small group of stars was therefore rather unexpected. 

In a previous paper \citep{Gratton2023b} we provided a reference framework for these surprising discoveries. We found that there are 17 stars with sufficient data and that can potentially host JL planets with stable orbits in the BPMG. We then estimated various selection effects in HCI as well as information from astrometry and found that seven out of 17 stars in the BPMG are likely hosting massive JL planets ($M>3$ \MJup). If it is then considered that only a third of JL planets are sufficiently massive to be detected by any of the two methods, the conclusion being that nearly all Sun-like stars in the BPMG are likely to host a JL planet. This tells a very different story about how planets form than the result obtained with RV surveys: the frequency of JL planets is overall low, but it is large in the BPMG.

There are several possible explanations for the lack of convergence between RV and imaging methods. The presence of giant planets is known to depend on the metal content and mass of their host stars \citep{Kennedy2008, Mordasini2012, Johnson2010}; however, as discussed in \citet{Gratton2023b} this is insufficient to explain the large discrepancy observed across methods. Stars may lose planets due to interactions with other objects that are passing by, whose perturbations may cause long-term instabilities in the systems, especially when several planets are present. A final possibility is that JL planets preferably form in low-density environments, such as the BPMG, while most old stars surveyed through RV formed in larger and higher density environment.

An open question is why it is more likely for JL planets to form in low-density environments. The most widely accepted mechanism for planet formation is the core accretion scenario \citep{Pollack1996, Mordasini2012, Bitsch2015} a process requiring a few million years. Perhaps proto-planetary disks can survive that long only in low-density environments, because in denser ones perturbations or photoionisation by nearby massive stars likely destroy the disks, thus preventing the formation of JL planets \citep{Adams2001, Adams2004, Winter2018, Parker2020, Parker2021, Winter2022, Wright2022}. We would then predict that the Solar System was likely formed in a low-density environment, that is to say not the commonly accepted scenario (see e.g. \citealt{Pfalzner2013, Pfalzner2020}). 

In this paper we expand this analysis to the remaining young nearby associations. This not only allows us to set this result on a more robust basis, but also to have a first determination of the dependencies on the mass and age of the associations. Young associations are very useful in this context because they had limited, if any, dynamical evolution affecting the properties of their multiple systems (e.g. mass segregation). Star formation is essentially complete in these associations. Stars are young enough (age $<200$~Myr) and associations are so loose (density $<0.1$~star/pc$^3$) that the long-term evolution of multiple systems related to the environment \citep{Heggie1975, Binney1987, Kaczmarek2011} is likely not strongly influencing their properties even for separation as large as a few thousands au. We notice that since the density of stars in these associations is lower than that of the solar neighbourhood, the formalism of \citet{Binney1987} indicates that encounters with field stars are  more probable than those with other members of the association. This is due to the much larger spread in velocities. Possibly, encounters are more probable in the birthplace of individual stars, but this can be considered as a component of the companion formation mechanisms. On the other hand, to avoid incompleteness that is  too large, the analysis must consider only stars close to the Sun. We adopted an upper limit of 100 pc to their distance.

As was done for the case of the BPMG, we combined the search for JL planets with a rather complete census of stellar binary companions. This helps to clarify the relative role of disk fragmentation and core accretion in the formation of companions and to correctly establish what the fraction of stars that host JL planets is over the mother population of the stars that may potentially have them. This paper is organised as follows: in Section 2 we present the properties of the associations, the data about the multiplicity of their members, the basic parameters of the stars and of their companions, and discuss the completeness of our search. In Section 3 we discuss the main properties of the population of the stellar companions. In Section 4 we present the statistics about the JL companions. The trends we observed are discussed in Section 5 and conclusions are drawn in Section 6. The Appendices present a discussion on the jitter in Gaia RVs and extensive tables with the relevant data for all stars considered to be a member of the associations and their companions.

\section{Data}

\begin{table*}
\caption{Main parameters for young nearby associations. }
\scriptsize
\begin{tabular}{lccccccccl}
\hline
Association          & Age & Distance & N. prim. & Mass & Size X & Size Y & Size Z &Gaia lim & Members source \\
& Myr & pc & $M>0.8$ \MSun & \MSun &pc &pc & pc & \MSun & \\
\hline
AB Dor  &  $137\pm 17$ & $50.7\pm 21.1$ & 55 & 191 & $33.9\pm 4.6$ & $33.1\pm 4.5$ &$16.5\pm 2.2$&0.038 &\citet{Zuckerman2004} \\
&&        &      &      &    & & & & \citet{Torres2008} \\
&&        &      &      &    & & & & \citet{Gagne2018} \\ 
&&        &      &      &    & & & & \citet{Faherty2018}\\
&&        &      &      &    & & & & \citet{Gagne2018c}\\
\hline
Argus  & $48\pm 10$ & $82.6\pm 34.8$ & 35 & 118 & $25.2\pm 34.2$ &$50.8\pm 8.5$ &$20.4\pm 3.4$& 0.030 &\citet{Zuckerman2019} \\
&&       &      &      &   & & & & \citet{DeSilva2013} \\
\hline
$\beta$ Pic & $21\pm 4$ & $46.6\pm 16.1$ & 27 & 94 & $36.4\pm 7.0$ &$16.7\pm 3.2$ &$9.2\pm 1.8$& 0.013 &\citet{Shkolnik2017} \\
&&      &      &      &   & & & &\citet{Gratton2023b} \\
\hline
Carina &  $28\pm 11$ & $79.6\pm 7.7$ & 17 & 51 & $11.0\pm 2.7$ & $18.8\pm 4.5$&$11.7\pm 2.8$& 0.020 & \citet{Torres2008} \\
&&       &      &      &   & & & &  \citet{Booth2021} \\
&&       &      &      &   & & & &  \citet{Kerr2022}\\
\hline
Carina Near&  $200\pm 20$ &  $34.3\pm 8.9$ & 12 & 41 & $9.3\pm 2.6$ & $23.2\pm 6.4$&$19.5\pm 5.4$& 0.045 &\citet{Zuckerman2006} \\
&&        &      &      &    & & & & \citet{Faherty2018}\\
&&      &      &      &  & & & &\citet{Stahl2022}\\
\hline
Columba & $36\pm 8$ &  $70.2\pm 23.7$ &  52 & 155 & $21.5\pm 3.0$ &$32.8\pm 4.6$ &$20.7\pm 2.9$& 0.030 & \citet{Torres2008} \\       
&&        &      &      &  & & &  & \citet{Moor2013} \\
&&        &      &      &  & & &  & \citet{Elliott2016} \\
&&        &      &      &  & & &  & \citet{Kerr2022} \\
\hline
Tuc-Hor & $37\pm 11$ & $48.8\pm 9.1$ & 41 & 158 & $23.9\pm 3.7$ &$14.7\pm 2.3$ &$3.8\pm 0.6$& 0.023 &\citet{Torres2008} \\
&&        &      &      &    & & & & \citet{Gagne2018e}\\
&&       &      &      & &  &  & & \citet{Zuniga2021} \\
&&       &      &      & &  &  & & \citet{Popinchalk2023} \\
\hline
Volans/Crius 221  & $89\pm 6$ & $82.1\pm 23.5$ & 41 & 123 & $18.5\pm 2.9$& $28.7\pm 4.4$ &$12.5\pm 1.9$& 0.040 &\citet{Gagne2018d} \\
&&        &      &    &  &  &  &  & \citet{Moranta2022} \\
\hline
\normalsize
\end{tabular}

In this table N. prim is the number of primaries with a mass $>0.8$ \MSun. The size X, Y, Z are the quadratic sum of the standard deviation of the X, Y, and Z Cartesian coordinates on the galactic frame (X is towards the centre of the Galaxy), respectively. Gaia lim is the mass corresponding to a magnitude of $G=19$.
\label{tab:associations}
\end{table*}

\subsection{Properties of young associations}

\subsubsection{Choice of associations}

We considered known associations and moving groups within 100 pc from the Sun and with ages in the range $10-200$ Myr: AB Doradus \citep{Zuckerman2004}, Argus \citep{DeSilva2013}, $\beta$ Pic moving group (BPMG) \citep{Shkolnik2017}, Carina \citep{Torres2008}, Carina Near \citep{Zuckerman2006}, Columba \citep{Torres2008}, Tucana-Horologium (Tuc-Hor, \citealt{Torres2008}), and Volans-Carina \citep{Gagne2018}. 

The case of Volans-Carina should be discussed further. In fact, the membership criteria considered by \citep{Gagne2018} are very strict and only leave the core members. In a recent study, \citet{Moranta2022} found that a group they considered in their analysis (Crius 221) may be considered as the corona of this association. While membership of some particular star to this group may be questioned, hereinafter we will assume that this is indeed the case, and extended the list of the Volans-Carina association including members of this group. Hereinafter, we will call this extended association Volans/Crius 221.

The main parameters for these associations are given in Table \ref{tab:associations}. Ages for these associations are an average from a number of literature determinations as given in Section 2.1.4.
Distances are the median of distances of the individual members with mass $>0.8$ \MSun obtained from the parallaxes listed in the Data Release 3 (DR3) of results from European Space Agency astrometric Gaia satellite \citep{Gaia_DR3}. The Gaia limit is the mass of a BD with $G=19$ at the average distance and age of each association. This was obtained using the isochrones by \citet{Baraffe2015} for the age and median distance of the associations. It provides a first estimate of the completeness in the search for wide stellar and BD companions using Gaia.

\subsubsection{Membership to the associations}


We considered members of the various associations from the sources listed in Table \ref{tab:associations}. Only systems where the primaries have a mass above 0.8 \MSun were kept. All stars were checked for membership to the respective clusters using the Banyan $\Sigma$ code \citep{Gagne2018}\footnote{\url{https://www.exoplanetes.umontreal.ca/banyan/banyansigma.php}}, but in the case of Volans/Crius 221 for the reason mentioned above we kept also those objects that have low membership probability. For the case of Argus, whose membership is not well defined, we checked that available information supports a young age for the proposed members. We dropped a few proposed objects because there is evidence they are much older than assumed for the association. In total we considered  296 stars in 280 stellar systems with the primary above the mass limit mentioned above as members of these associations. Some very wide binaries have entries for the individual objects in our list if the mass of each of the component is $>0.8$ \MSun. The presence of the wide (massive) companion has been considered when estimating multiplicity.

\subsubsection{Basic parameters for programme stars}


In this paper we considered the Gaia $G$ \citep{Gaia_DR3} and 2MASS $K$ magnitudes \citep{2mass}. Distances were taken from Gaia DR3, whenever possible; else from Gaia Data Release 2 (DR2) or, as a last resort, from Hipparcos \citep{1997A&A...323L..49P}. Interstellar absorption was neglected, considering the close distance of target stars. We checked that this is appropriate using the 3-d map by \citet{Green2019} for all stars for which they give results.

Masses for the stars are derived from the absolute Gaia $G$ magnitudes using the calibration by Pecaut \& Mamajek \citep{Pecaut2013} if $M_G<3$, else from the isochrone by \citet{Baraffe2015} of appropriate age. In case the star is an unresolved binary in the Gaia Catalogue, we corrected the Gaia $G$ magnitudes for the contribution by the secondary before extracting the masses. This was done on an iterative way, but convergence was always very fast.

\begin{figure}[htb]
\centering
\includegraphics[width=\linewidth]{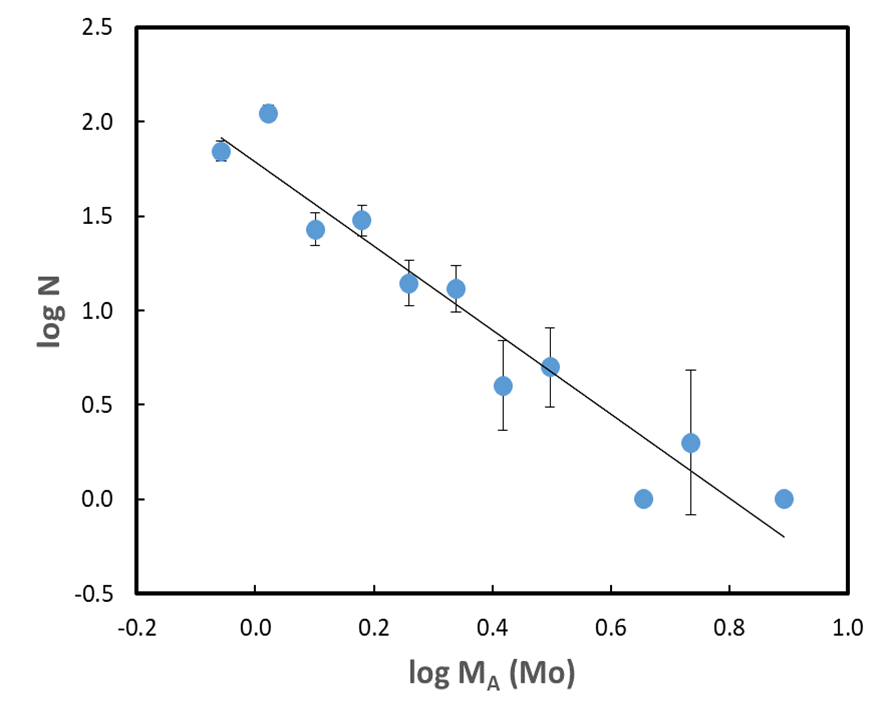}
\caption{Mass function for the primaries in the young associations. The slope of the best fitting line (solid line) is $-$2.24.
}
\label{fig:mass_function}
\end{figure}

The median mass of the primaries is 1.04 \MSun. The masses of the primary stars in the range 0.8-5 \MSun distribute close to a Salpeter mass function \citep{Salpeter1955}, with a slope of $-$2.24 (see Figure~\ref{fig:mass_function}). 

\subsubsection{Age and size of the associations}

\begin{figure}[htb]
\centering
\includegraphics[width=8.8truecm]{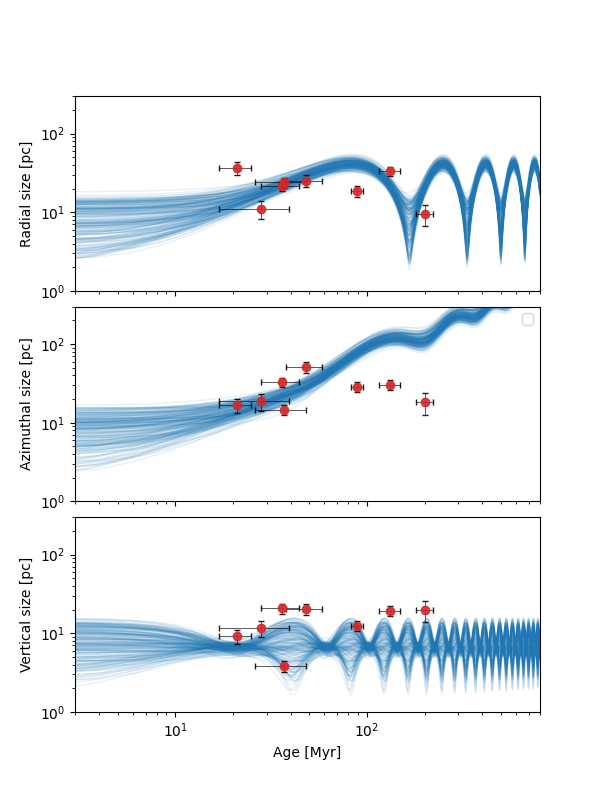}
\caption{Run of the size of the young associations considered in this paper with age (red dots).  $X$ is the direction of the galactic centre (upper panel); $Y$ is the direction perpendicular to $X$ on the galactic plane (middle panel); $Z$ is the vertical direction (lower panel). For comparison, expectations for a simple model of expansion of an association in the galactic potential is also shown; results for 200 random extractions for the initial sizes along $X$, $Y$, and $Z$ axes with uniform distribution between 2 and 15 pc are shown. In all cases we assumed an initial isotropic velocity dispersion of 0.5 \kms.
}
\label{fig:ass_size}
\end{figure}

\begin{table*}
\caption{Ages for associations and moving groups (in million years)}
\begin{tabular}{lcccccc}
\hline
Reference & BPMG & Columba &	Carina	& Tuc-Hor	& Argus	& AB Doradus\\
\hline
\multicolumn{6}{c}{Isochrones}\\
\hline
\citet{2015MNRAS.454..593B} & 24 & 42 & 45 & 45 & 58 & 149 \\
\citet{Booth2021}           &    &    & 13 &    &    &     \\
\hline
\multicolumn{6}{c}{Lithium}\\
\hline
\citet{Barrado2004}	        &    &    &    &    & 50 &     \\	
\citet{Mentuch2008}         & 21 &    &    & 27 &    &$>$45 \\
\citet{Binks2014}           & 21 &    &    &    &    &      \\
\citet{Malo2014}            & 26 &    &    &    &    &      \\
\citet{Shkolnik2017}        & 22 &    &    &    &    &      \\
\citet{Schneider2019}       & 22 &	  & 22 &    &    &      \\
\citet{Wood2023}            &    &    &	40 &    &    &      \\			
\hline
\multicolumn{6}{c}{Kinematics}\\
\hline
\citet{McCarthy2014}        &    &    &    &    &    & 125 \\
\citet{Miret-Roig2020}      & 13 &$>$40&$>$28&$>$28&37&    \\	
\citet{Miret-Roig2020}      & 19 &    &    &    &    &     \\
\citet{Booth2021}           &    &    & 20 &    &    &     \\
\citet{Kerr2022}            &    & 26 & 26 & 46 &    &     \\
\citet{Couture2023}         & 20 &    &    &    &    &     \\
\hline								
Mean                        & 21 & 36 & 28 & 37 & 48 & 137 \\
Standard deviation          &  4 &  8 & 11 & 11 & 10 &  17 \\
\hline
\end{tabular}
\label{t:age1}
\end{table*}

In our discussion, we wish to link the observed frequency of JL planets to the properties of the environment where they formed. Even selecting young associations and moving groups, they still are tens to hundreds million years old. Their environment has substantially changed within this interval of time and some aspects cannot be recognised any more. We focus here on two of the simplest parameters, that are the age and mass of the star forming region that originated the observed association. 

We report in Tables~\ref{t:age1} a number of literature determination of ages for six of the moving groups considered in this paper. The values we adopted are the straight averages. In this paper, we did not consider the ages by \citet{Ujjwal2020} because they are much lower than and uncorrelated with other estimates. For Carina Near we adopted the age of 200 Myr given by \citet{Zuckerman2006}, while for Volans/Crius 221 that of $89\pm 6$ Myr given by \citet{Gagne2018d}.

The mass of associations was obtained by summing up the mass of all components found and then correcting this value for the mass of the stars with masses lower than 0.8 \MSun . We assumed that the same correction found for the case of the BPMG applies to the other clusters too. A total mass of the BPMG of 94 \MSun was derived by \citet{Gratton2023b}. Masses estimated in this way are given in Table \ref{tab:associations}. Summing all eight associations, the total mass is 932 \MSun.

The approach we followed - considering known members - likely underestimates the actual values. Furthermore, it might introduce biases, because members may be easier to recognise in the youngest associations. One possible reason is that the older associations are more dispersed. To have a first estimate of the relevance of this effect we determined the actual sizes of the associations and considered if they change systematically with age. We considered separately the size along the three different galactic coordinates because it is expected that they obey different runs with time. In fact the expansion of the associations occurs within the galactic potential (see e.g. \citealt{Binney1987, Asiain1999}). If we neglect perturbations due to other stars, molecular clouds and spirals (disk heating), an approximation that is not too bad in view of the very long relaxation time and young ages of these associations, we might expect that after an epicyclic period stars return to a compact configuration in the galactic radial direction $X$, a focusing effect noticed many years ago by \citet{Yuan1977}. Similarly, a compact configuration in the vertical direction $Z$ is obtained twice each vertical period. This argument is similar to that considered by \citet{Dinnbier2022}. The run is more complex in the galactic azimuthal direction $Y$, where the focusing effect is lower and stars spread over time along an arc that grows almost linearly with time. We use the publicly available code \texttt{galpy} \citep{Bovy2015}\footnote{\url{https://docs.galpy.org/en/v1.9.0/}} to model the motion of the stars of an association in the Milky Way galactic potential, neglecting self-gravity. We considered the evolution of the position dispersion with time for samples of 100 stars extracted with Gaussian distributions in the original positions and velocities. We run this exercise 200 times, each one with a random extractions of the initial spread along $X$, $Y$, and $Z$ axes with uniform distribution between 2 and 15 pc, comparable to the original size of the BPMG as derived by \citet{Miret-Roig2020}. In all cases we assumed an initial isotropic velocity dispersion of 0.5 \kms that is at the lower edge of the typical velocity distribution in nearby low-mass star forming regions (0.5-1.0 \kms \citealt{Hennebelle2012, Heyer2015}) and it is similar to the velocity dispersion in the BPMG \citep{Miret-Roig2020}. The galactic potential has epicyclic and vertical periods of 170 and 80 Myr, respectively, at the Sun position; this corresponds to the galactic constants of \citet{Kerr1986}. Figure~\ref{fig:ass_size} compares the results of this model with the observed association sizes on the three galactic coordinates. We found that indeed the size of the associations should remain limited over time in $X$ and $Z$ direction, as observed. However, we expect that the associations should become very elongated in the $Y$ direction after 50 Myr, only marginally modulated on the epicyclic period. This is not observed: the oldest associations have dispersion in $Y$ similar to the youngest ones. We notice that the larger the initial velocity dispersion (that is $\sigma(V)$ in the usual notation used in galactic dynamics), the highest would be the dispersion along the $Y$ direction, and that the value we assumed is actually at the lower edge of the expected distribution. To eliminate this discrepancy, we should assume values of $\sigma(V)=0.1-0.2$ \kms, while higher values would be needed for the other directions to match observations. While this is perhaps not impossible (for instance, \citealt{Miret-Roig2020} found $\sigma(V)=0.43$ \kms for the BPMG), such a systematic effect would require an explanation. 

Rather, we consider the possibility that while there are not significant selection effects for the associations with ages $<50$ Myr, the case can be very different for the oldest ones. In fact, the small sizes along $Y$ of the older associations may be an artefact of selection effects: only those stars that are in the region of the association that is closest to the Sun (75 pc for AB Dor, 50 pc for Carina Near) or in one of the two hemispheres (southern one for Carina Near and Volans/Crius 221) are considered as members in the papers from which we extracted membership (and in Banyan $\Sigma$). At their age and using the same initial expansion velocity considered for the other associations, we expect them to be more extended than observed along the $Y$ direction, by factors of 2.7, 3,7, and 5.6 for Volans/Crius 221, AB Dor and Carina Near, respectively.  The missing stars should be searched in the directions around the galactic coordinates ($l=90, b=0$ for Carina Near and Volans/Crius 221) and ($l=270, b=0$) at distances of the order of 100 pc or more. They should not be easily separable from field objects because they should have small proper motions and RVs with low errors are required for their selection. We then consider the possibility that these associations were much more massive at birth and that their total mass should be corrected upwards by a factor of at least two.



\subsection{Companion detections}

We carried out an extensive search for companions to the programme stars. We considered visual, spectroscopic, eclipsing, and astrometric binaries, as well as substellar companions. 

\subsubsection{Visual binaries}


We considered visual binaries companions from several sources. First we considered if there are companions with similar distances and proper motion within 600 arcsec listed as separate entries in the Gaia DR3 catalogue \citep{Gaia_DR3}. The limiting contrast of Gaia is discussed in \citet{Brandeker2019}.  Given the very low surface densities of these associations (typical values are 0.01-0.02 stars per square degree), the probability that physically unbound objects of the same association are projected within this radius from a star is of about $5\times 10^{-4}$ and can then be neglected. We then checked if the stars are in the Washington Double Star (WDS) visual binary catalogue \citep{Mason2001}, in the Multiple Star Catalogue by \citet{Tokovinin2018} (we found 44 entries), or were detected in HCI \citep{Biller2007, Brandt2014, DeRosa2014, Elliott2015, Galicher2016, Bonavita2016, Janson2017, Stone2018, Asensio-Torres2018, Nielsen2019, Vigan2021, Zhou2022, Dahlqvist2022, Bonavita2022, Mesa2022}. Gaia data are available for all stars in the sample, and HCI is available for 194 stars. Orbital parameters for HIP3589  were from \citet{Cvetkovic2011}.

\subsubsection{Spectroscopic binaries}


We checked if the programme stars have entries in the spectroscopic binary catalogue \citep{Pourbaix2004}, in the catalogue obtained from GALactic Archaeology with HERMES\footnote{HERMES is the High Efficiency and Resolution Multi-Element Spectrograph at the Anglo Australian Telescope.} (GALAH) survey data \citep{Traven2020}, and from Gaia RVs ($gaiadr3.nss\_two\_body\_orbit$). We also considered data from the catalogues of high precision RV by \citet{Butler2017, Tal-Or2019, Trifonov2020, Grandjean2020, Grandjean2021, Grandjean2023}. Overall, high precision RV are available for 79 stars; lower precision RV are available for 181 additional stars; there is no RV sequence for 37 stars. Whenever data were available, they were used to constrain the orbital parameter of the companion. For the double line spectroscopic binary HIP89925 (108 Her) data were from \citet{Fekel2009} and for HIP100751 ($\alpha$ Pav) from \citet{Luyten1936}.

A few planets were discovered around stars claimed to be members of young associations using high precision RV:  $\beta$ Pic c \citep{Lacour2021} in the BPMG; 
HIP50786 \citep{Korzennik2000, Wittenmyer2019} and HIP72339 \citep{Udry2000} in the AB Dor moving group. Most of them are giant planets with $a>1$ au, but HIP72339b is a hot-Jupiter. However, HIP50786 and HIP72339 are likely old stars, so we will not consider them any more in this paper.

In general, when using RVs of young stars the large value of the stellar jitter related to the high activity level must be considered (see e.g. discussion in \citealt{Lagrange2023}). The RV jitter from high precision measurements on timescales of decades decreases from over 500 m s$^ {-1}$ for 5 Myr-old stars to 2.3 m s$^ {-1}$ for stars with ages of around 5 Gyr \citep{Brems2019, Tran2021}. Special techniques may limit its impact, but this requires extensive observation and thorough discussion, possible only for individual cases {such as that of  $\beta$ Pic \citep{Lacour2021}.  We also notice that the Gaia robust RV amplitude is a less good indicator for young stars because of the higher activity and faster rotation. In particular, it is unreliable for fast rotating K-stars (such as TYC 7100-2112-1 in Columba), likely because of severe blending in the spectra. These stars can be identified as short period ($\sim$1 day) quasi-periodic variable thanks to the TESS light curves. We considered the higher jitter in Gaia RVs for young stars as described in Appendix A.}.

\subsubsection{Eclipsing binaries}

Obviously, to detect a companion by eclipses/transits the Sun should be close to the orbital plane. Hence only a fraction of the companions can be detected this way. We looked for entries corresponding to the stars in the eclipsing binary (EB) catalogue by \citet{Avvakumova2013} and in the Kepler EB catalogue \citet{Slawson2011}. In addition, all programme stars but 15 have been observed by the NASA Transiting Exoplanet Survey Satellite (TESS) satellite, though only 260 of them have been observed with a 2-minute cadence. The 21 objects observed but missing the 2-minute cadence data are generally faint, and so of low-mass. We inspected the TESS catalogues of EB \citep{Prsa2022, IJspeert2021} and found two EBs listed: HIP 57013 in the Argus moving group and HIP28474 (RT Pic) in the Columba moving group. However, the second star is likely not an EBs, as discussed in the Appendix\footnote{The TESS light curve shows two periods with an amplitude larger than 0.01 mag and a length of about six days, but no obvious transit.}. A single EB detected would be consistent with the total number of seven detected stellar companions with $a<0.165$ au and $M_B>0.075$ \MSun, that roughly corresponds to a period of 20 days where the search for EB should be complete over at least 89\% of the targets. 

We also inspected the TESS Target of Interest catalogue where we found a hot Neptune transiting DS Tuc A in the Tuc-Hor moving group \citep{Newton2019, Benatti2019} and a mini-Neptune (3 $R_{Earth}$) transiting HIP 94235 in the AB Dor moving group \citep{Zhou2022}. Since mass is not available for this second planet, it will not be considered in the Tables in the Appendix.


\subsubsection{Astrometric binaries}


We inspected the Gaia $gaiadr3.nss\_two\_body\_orbit$ catalogue \citep{2023A&A...674A..34G} and found entries for 12 objects: HIP12838 and HIP37855 in AB Dor; TYC 9412-1370-1 in Argus; HIP76629 in BPMG; HIP26144 in Carina; HIP35564 and HIP37918 in Carina-Near; HIP2729, HIP16853, and HIP108422 in Tuc-Hor; and TYC 8933-327-1 in Volans/Crius 221. An orbit solution is available for these stars. We found five entries in the $gaiadr3.nss\_acceleration\_astro$ catalogue \citep{2023A&A...674A..34G}: HIP55746 in AB Dor, CD-58 2194 in Argus, TYC 8602-718-1 in Carina, HIP37635 in Carina Near, and HIP18714 in Tuc-Hor. We searched for entries corresponding to the programme stars in the catalogue of astrometric orbit determination with Markov chain Monte Carlo and genetic algorithms by \citet{Holl2023} and found two entries: HIP16853 and HIP102626 (BO Mic), both in Tuc-Hor.

We considered the Proper Motion Acceleration (PMa) from \citet{Kervella2022}. The PMa is the difference between the proper motion in Gaia DR3 (baseline of 34 months) and that determined using the position at Hipparcos (1991.25) and Gaia DR3 (2016.0) epochs. This quantity is available for 178 stars. The PMa is sensitive to binaries with separation between 1 and 100 au. We considered as an indication for the presence of companions any value of the PMa with a signal-to-noise ratio $S/N>3$. 

We also considered the renormalised unit weight error (RUWE) as an indication of binarity. This parameter is an indication of the goodness of the 5-parameters solution found by Gaia \citep{Lindegren2018}. \citet{Belokurov2020} showed that a value $>1.4$ of this parameter is an indication of binarity, at least for stars that are not too bright ($G>4$) and saturated in the Gaia scans. This method is sensitive to systems with periods from a few months to a decade \citep{Penoyre2021}. The RUWE parameter is available for all but seven stars. Orbital parameters for HIP21965 were retrieved from \citet{Goldin2007}. All these data were considered in our analysis.

\subsubsection{Substellar companions}



We searched for substellar companions in the NASA Exoplanet Archive\footnote{ \url{https://exoplanetarchive.ipac.caltech.edu/applications/Inventory/search.html}}. For all objects with entries in this catalogue, we looked for the individual references to derive parameters for the substellar companions:
HIP30034: \citet{Vigan2021};
HIP77797: \citet{Huelamo2010, Nielsen2013};
HIP99770: \citet{Currie2023};
HIP114189 (HR8799): \citet{Zurlo2022};
HIP116805 ($\kappa$ And): \citet{Carson2013, Uyama2020};
TYC 8047-232-1: \citet{Chauvin2010}. 

We also checked that none of the free-floating substellar objects listed in \citet{Aller2016, Artigau2015, Best2020, Bowler2012, Bowler2017, Delorme2013, Desrochers2018, Gagne2015, Gizis2002, Gizis2015, Liu2013, Liu2016, Naud2014, Patience2012, Schneider2014, Schneider2017, Schneider2019, Schneider2023, Shkolnik2017} and \citet{Zhang2021}  is projected within 600 arcsec of any of the target stars.

\subsection{Companion masses and semi-major axis}

\begin{figure*}[htb]
    \centering
    \includegraphics[width=6.5cm]{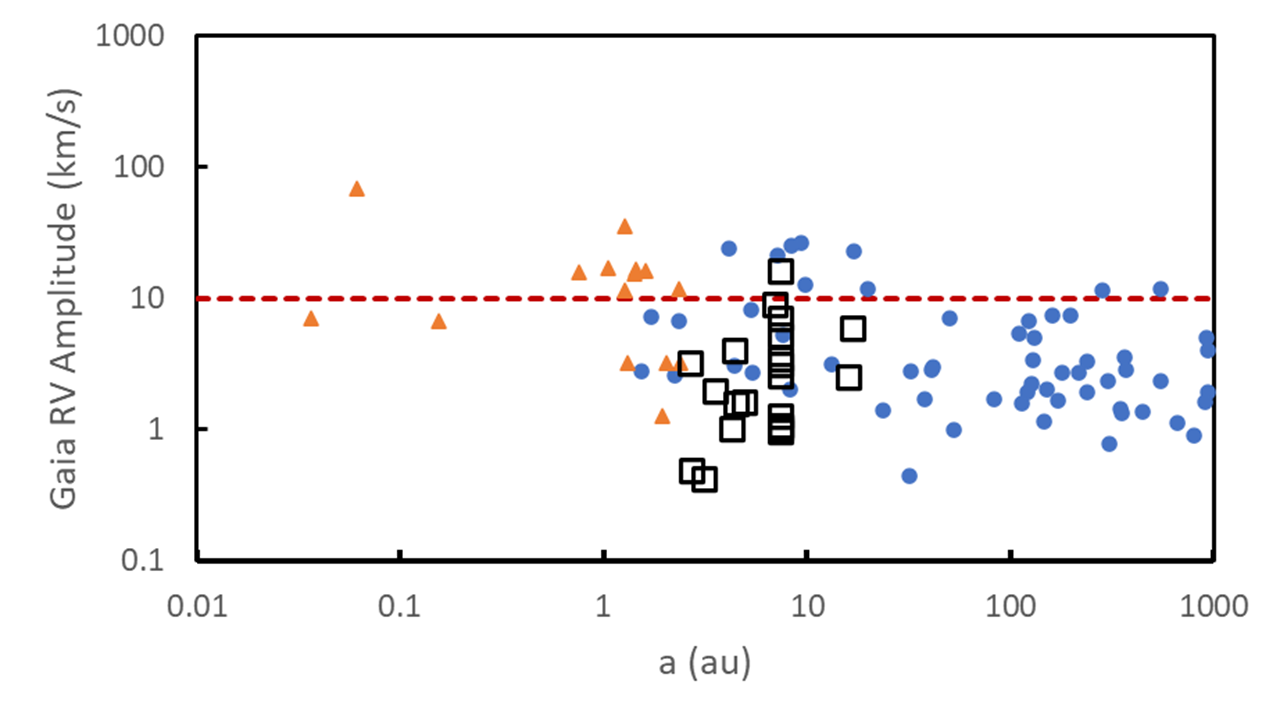}
    \includegraphics[width=6.5cm]{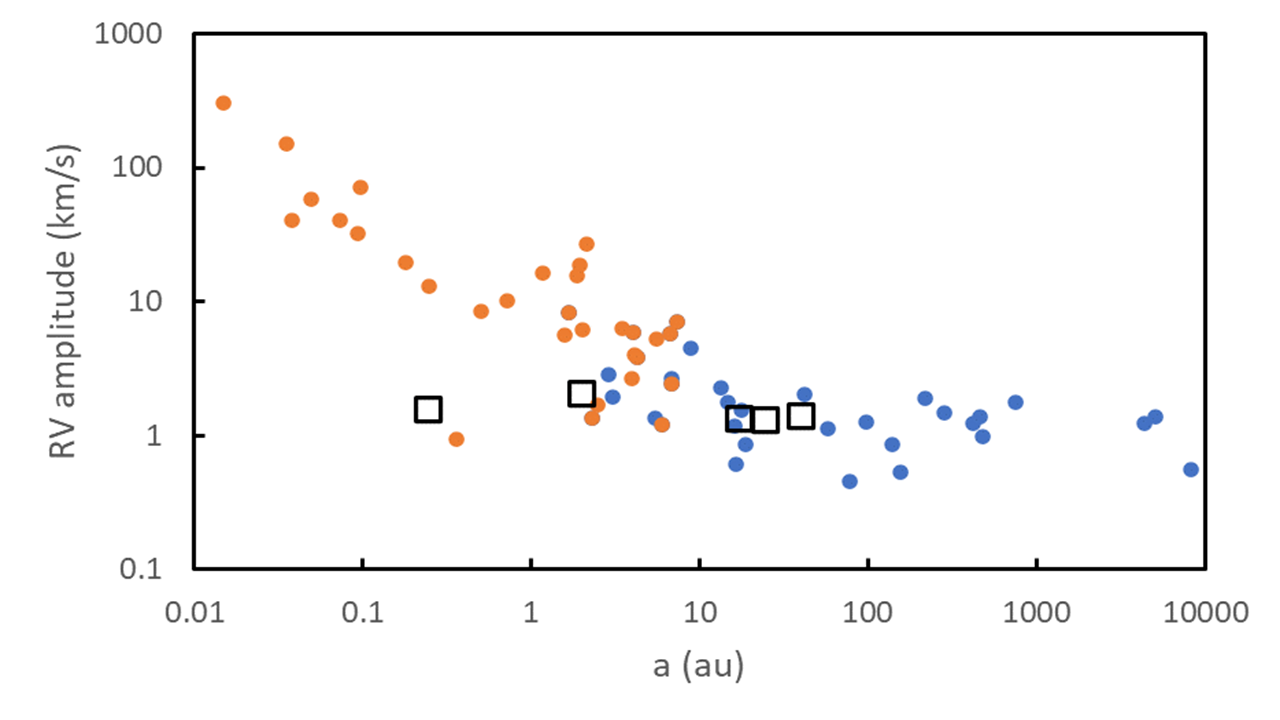}
    \includegraphics[width=6.5cm]{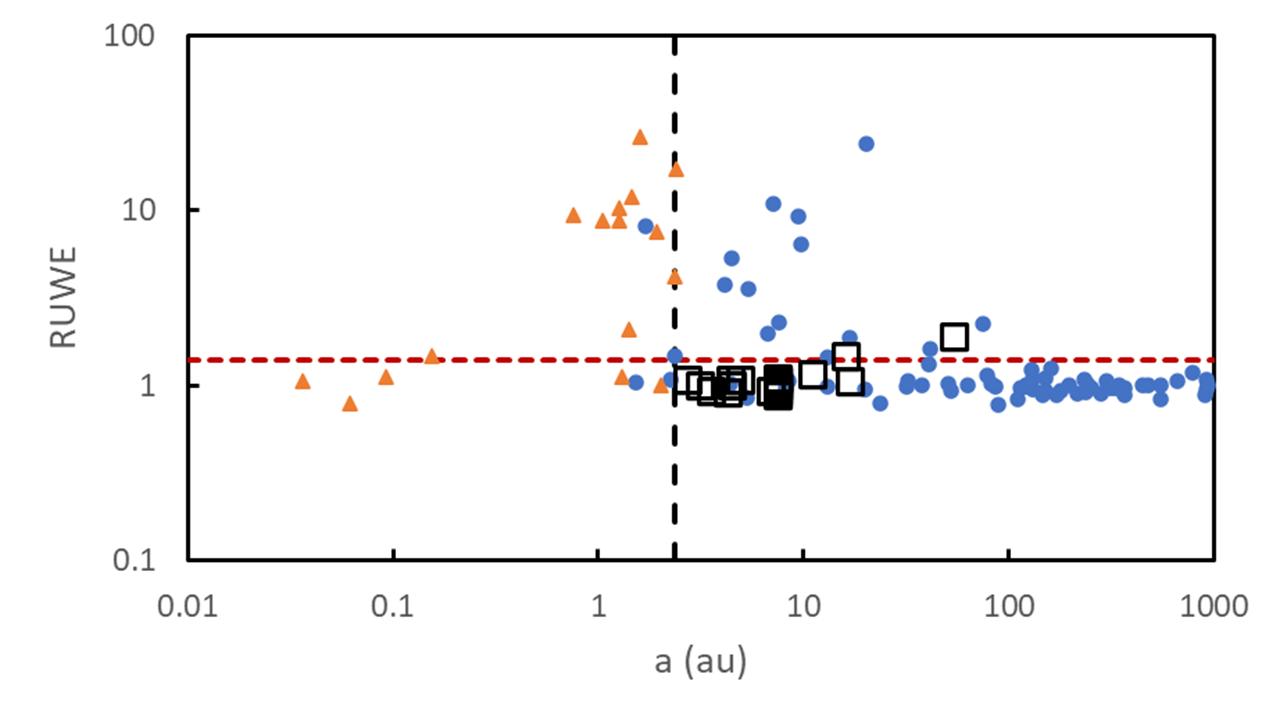}
    \includegraphics[width=6.5cm]{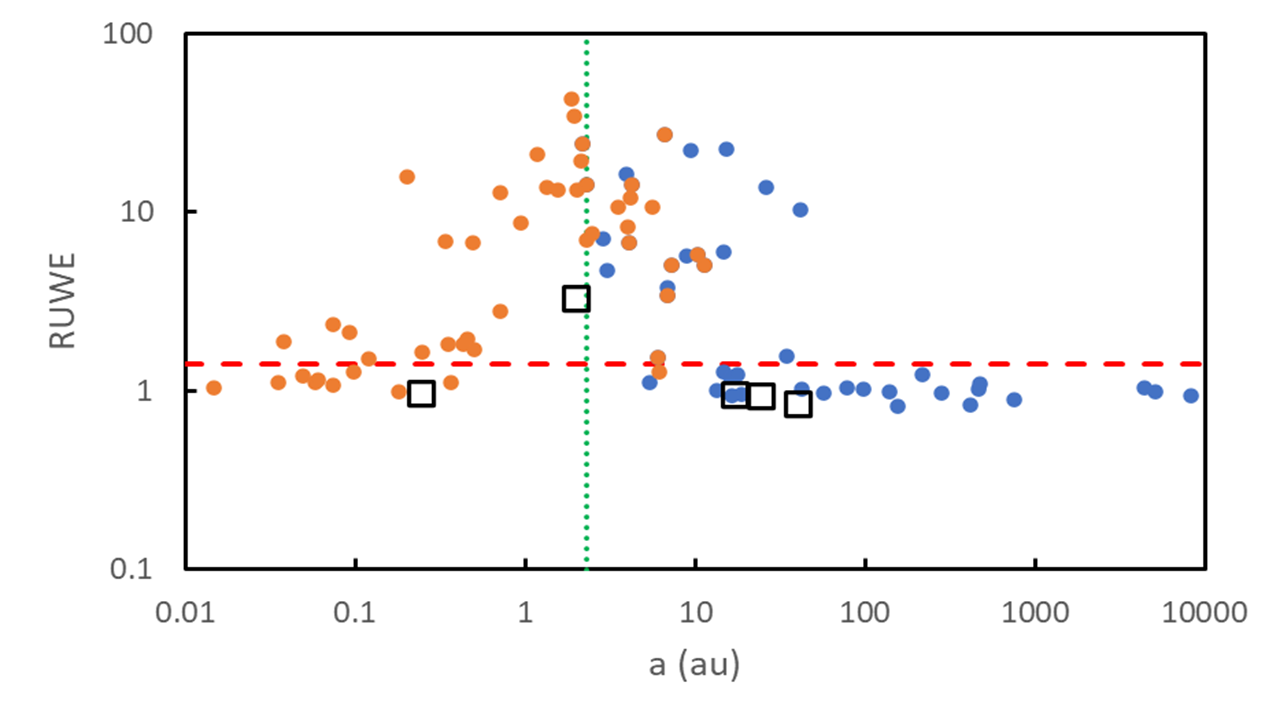}
    \includegraphics[width=6.5cm]{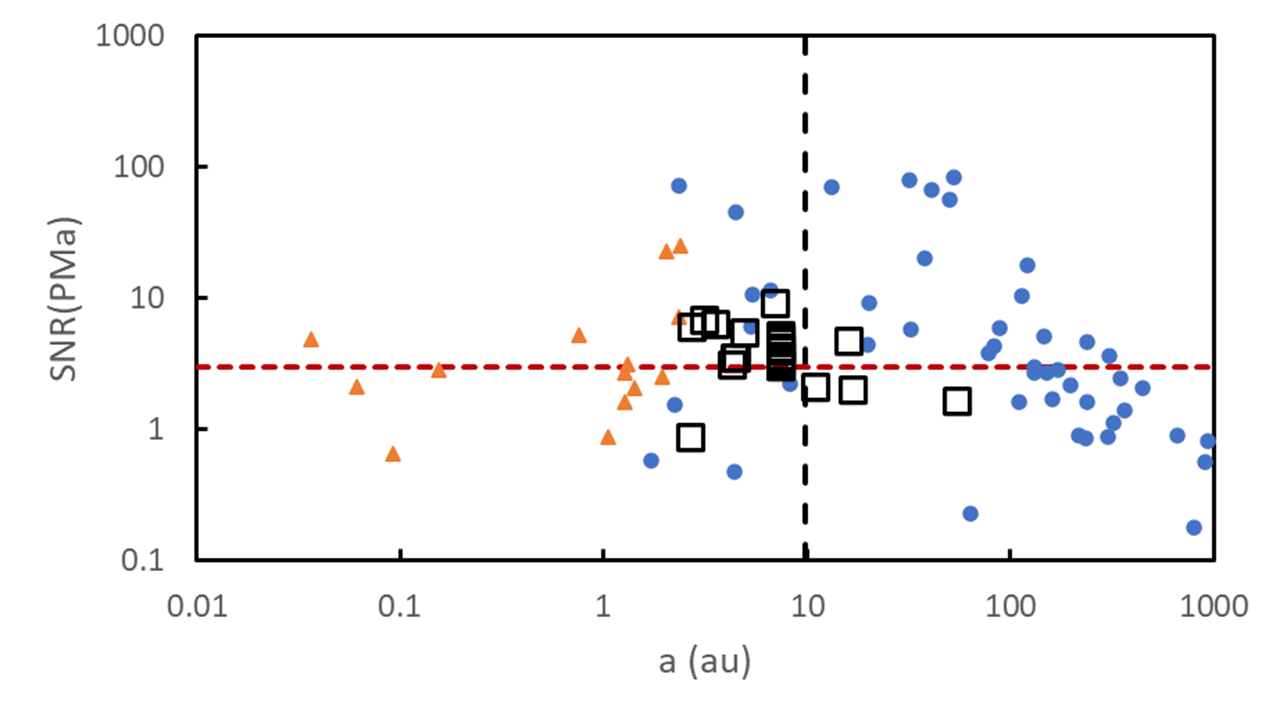}
    \includegraphics[width=6.5cm]{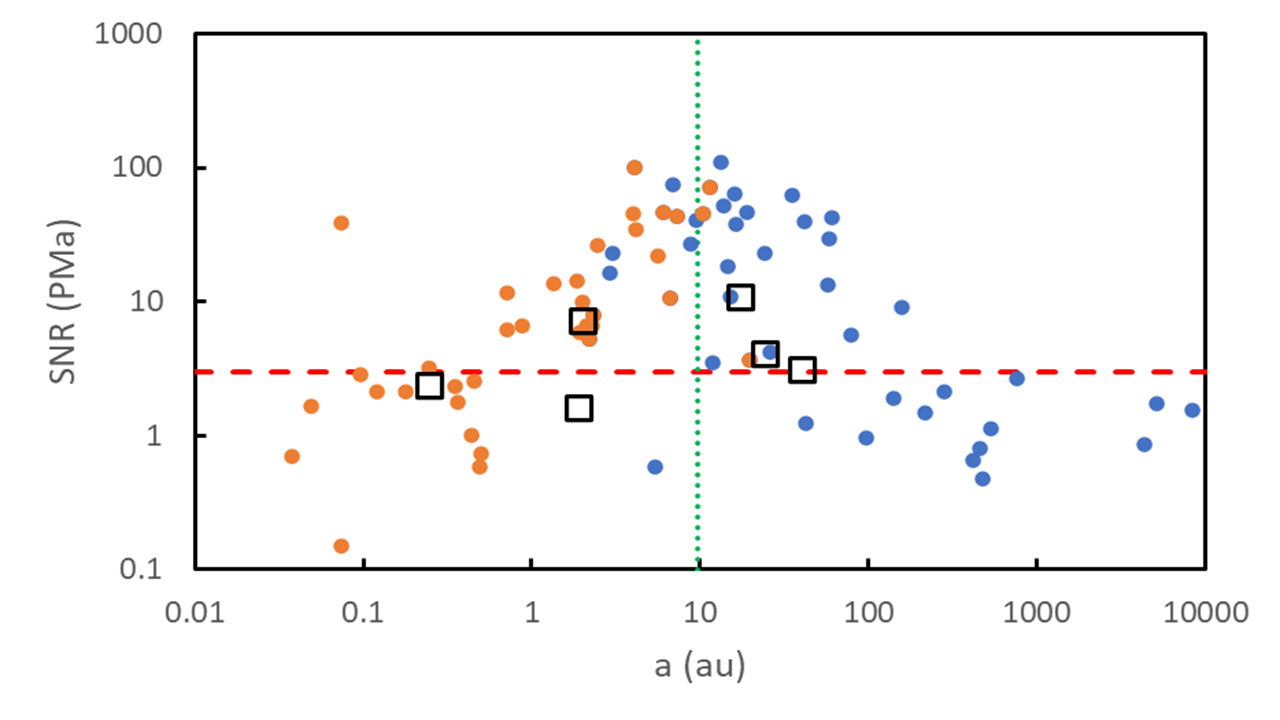}
    \caption{Relation between the semi-major axis and Gaia RV amplitude (upper row), $RUWE$ (middle row), and S/N(PMa) (lower row) for previously known binaries in the young association stars (left column) and in the Hyades (right column). Filled blue circles are visual binaries; filled orange triangles are spectroscopic binaries; open squares are substellar companions. Horizontal short dashed horizontal lines are the detection limits for binaries; vertical black long dashed lines mark semi-major axis corresponding to periods of 34 months (for $RUWE$) and 24.75 yr (for S/N(PMa)) where maximum sensitivity are expected for these two techniques.  }
    \label{fig:dyn}
\end{figure*}

\begin{table*}
\centering
\caption{Companion masses and semi-major axis from dynamical data (RVs, RUWE, and PMa)}
\scriptsize
\begin{tabular}{lccccccccl}
\hline
HIP	&	2MASS	&	RUWE	&	RV Ampl.	&	S/N(PMa)	& $dG$ &	$M_A$	&	$M_B$			&	$a$			&	Association	\\
	&		&		&	\kms	&		&	mag & \MSun	&	\MSun			&	au			&		\\
\hline
\multicolumn{10}{c}{Selected for RUWE$>$1.4}	\\	
\hline			
12635	&	J02422094+3837212	&	26.545	&	2.36	&	12.39	&	-0.031$\pm$0.038 & 0.756	&	0.267	$\pm$	0.014	&	2.159	$\pm$	0.125	&	AB Dor	\\
14551	&	J03075083-2749520	&	4.176	&	11.76	&	7.18	&	-0.176$\pm$0.041 & 1.608	&	0.091	$\pm$	0.019	&	2.346	$\pm$	0.114	&	Columba	\\
	&	J05044132+4024003	&	2.245	&	25.32	&		&	-0.009$\pm$0.034 & 0.725	&	0.203	$\pm$	0.168	&	2.726	$\pm$	4.043	&	AB Dor	\\
	&	J05301907-1916318	&	1.575	&	8.67	&		&	0.714$\pm$0.075 & 0.838	&	0.228	$\pm$	0.050	&	0.299	$\pm$	0.236	&	Columba	\\
	&	J05345923-2954041	&	1.912	&	5.91	&		&	0.106$\pm$0.034 & 0.756	&	0.144	$\pm$	0.035	&	0.369	$\pm$	0.309	&	Argus	\\
35564	&	J06582764+2215067	&	8.691	&	16.96	&	0.88	&	 & 1.386	&	0.683	$\pm$	0.200	&	0.365	$\pm$	0.109	&	Carina-Near	\\
37635	&	J07432149-5209508	&	17.534	&	2.91	&	45.54	&	-0.087$\pm$0.075 & 1.127	&	0.180	$\pm$	0.064	&	3.842	$\pm$	0.850	&	Carina-Near	\\
37855	&	J07453559-7940080	&	7.633	&	1.28	&	2.49	&	0.039$\pm$0.075 & 1.138	&	0.205	$\pm$	0.031	&	1.802	$\pm$	0.160	&	AB Dor	\\
	&	J08134777-5837139	&	1.661	&	5.08	&		&	-0.082$\pm$0.066 & 1.388	&	0.275	$\pm$	0.144	&	1.582	$\pm$	3.491	&	Volans	\\
	&	J08264964-6346369	&	11.879	&	16.58	&		&	-0.052$\pm$0.056 & 0.792	&	0.557	$\pm$	0.192	&	1.800	$\pm$	1.736	&	Volans	\\
	&	J08391155-5834281	&	2.527	&	29.38	&		&	0.520$\pm$0.050 & 1.003	&	0.658	$\pm$	0.113	&	0.253	$\pm$	0.092	&	Argus	\\
43783	&	J08550282-6038406	&	3.731	&	7.25	&	9.86	&	0.481$\pm$0.212 & 4.822	&	0.911			&	3.387			&	Volans/Crius 221	\\
46460	&	J09283051-6642067	&	3.192	&	72.03	&	1.69	&	0.681$\pm$0.057 & 2.227	&	0.851	$\pm$	0.627	&	0.818	$\pm$	0.625	&	Volans/Crius 221	\\
	&	J09534760-5453540	&	4.861	&	6.41	&		&	0.230$\pm$0.047 & 0.911	&	0.237			&	2.530			&	Carina	\\
50191	&	J10144416-4207189	&	2.873	&		&	0.97	&	0.428$\pm$0.066 & 2.110	&	0.559	$\pm$	0.360	&	0.285	$\pm$	0.179	&	Argus	\\
50567	&	J10194613-7133173	&	44.080	&	5.07	&	12.73	&	0.216$\pm$0.067 & 1.024	&	0.488	$\pm$	0.047	&	2.453	$\pm$	0.146	&	Volans/Crius 221	\\
	&	J12063292-4247508	&	4.253	&	22.38	&		&	0.432$\pm$0.039 & 0.910	&	0.635	$\pm$	0.113	&	0.471	$\pm$	0.473	&	Argus	\\
	&	J12203437-7539286	&	26.350	&	16.39	&		&	0.340$\pm$0.035 & 0.754	&	0.564	$\pm$	0.161	&	1.805	$\pm$	1.080	&	Argus	\\
108422	&	J21575146-6812501	&	10.369	&	35.72	&	2.72	&	0.633$\pm$0.039 & 0.948	&	0.421			&	0.746			&	Tuc-Hor	\\
\hline
\multicolumn{10}{c}{Selected for S/N(PMa)$>$3}	\\
\hline
560	&	J00065008-2306271	&	0.933	&	3.02	&	3.42	& -0.069$\pm$0.065 &	1.413	&	0.0017	$\pm$	0.0009	&	16.57	$\pm$	15.78	&	BPMG	\\
8233	&	J01460105-2720556	&	1.071	&	5.15	&	3.01	&	-0.037$\pm$0.065 & 1.431	&	0.0052	$\pm$	0.0060	&	22.07	$\pm$	18.26	&	Tuc-Hor	\\
11696	&	J02305064+5532543	&	1.061	&	16.12	&	4.18	&	0.093$\pm$0.070 & 1.326 &	0.0048	$\pm$	0.0030	&	7.68	$\pm$	7.08				&	AB Dor	\\
15201	&	J03155768-7723184	&	1.000	&	0.42	&	6.74	&	0.167$\pm$0.068 & 1.418	&	0.0135	$\pm$	0.0114	&	17.78	$\pm$	12.87	&	Carina	\\
16449	&	J03315364-2536509	&	1.079	&	3.94	&	3.48	&	0.007$\pm$0.046 & 1.762	&	0.0056	$\pm$	0.0052	&	12.54	$\pm$	10.06	&	Columba	\\
17675	&	J03471061+5142230	&	0.995	&	2.56	&	3.15	&	-0.193$\pm$0.063 & 1.445	&	0.0097	$\pm$	0.0099	&	24.09	$\pm$	19.02	&	Columba	\\
22152	&	J04460056+7636399	&	0.900	&	3.10	&	3.09	&	-0.169$\pm$0.072 & 1.221	&	0.0012	$\pm$	0.0006	&	17.11	$\pm$	15.72	&	Columba	\\
30034	&	J06191291-5803156	&	0.923	&	1.96	&	6.26	&	0.208$\pm$0.033 & 0.913	&	0.008	$\pm$	0.004	&	3.571	$\pm$	1.167	&	Carina	\\
45585	&	J09172755-7444045	&	1.070	&	1.59	&	5.37	&	0.041$\pm$0.088 & 2.072	&	0.0176	$\pm$	0.0084	&	6.85	$\pm$	2.46	&	Carina	\\
47391	&	J09392101-6119410	&		&	11.61	&	45.80	&	0.508$\pm$0.134 & 3.011	&	1.287	$\pm$	0.687	&	34.622	$\pm$	26.864	&	Carina	\\
60831	&	J12280445+4447394	&	0.905	&	1.02	&	3.12	&	-0.065$\pm$0.074 & 1.166	&	0.0034	$\pm$	0.0020	&	9.47	$\pm$	3.90	&	Carina-Near	\\
60832	&	J12280480+4447305	&	1.067	&	0.48	&	5.94	&	0.176$\pm$0.066 & 1.042	&	0.0074	$\pm$	0.0029	&	9.27	$\pm$	3.84	&	Carina-Near	\\
\hline
\multicolumn{10}{c}{Selected for RV Ampl.$>$100 \kms}	\\
\hline
47017	&	J09345645-6459579	&	0.971	&	122.40	&	1.46	&	0.487$\pm$0.073 & 1.376	&	0.322	$\pm$	0.242	&	2.925	$\pm$	24.901	&	Volans/Crius 221	\\
\hline
\normalsize
\end{tabular}
\label{tab:dyn}
\end{table*}

\begin{figure}[htb]
    \centering
    \includegraphics[width=8.5cm]{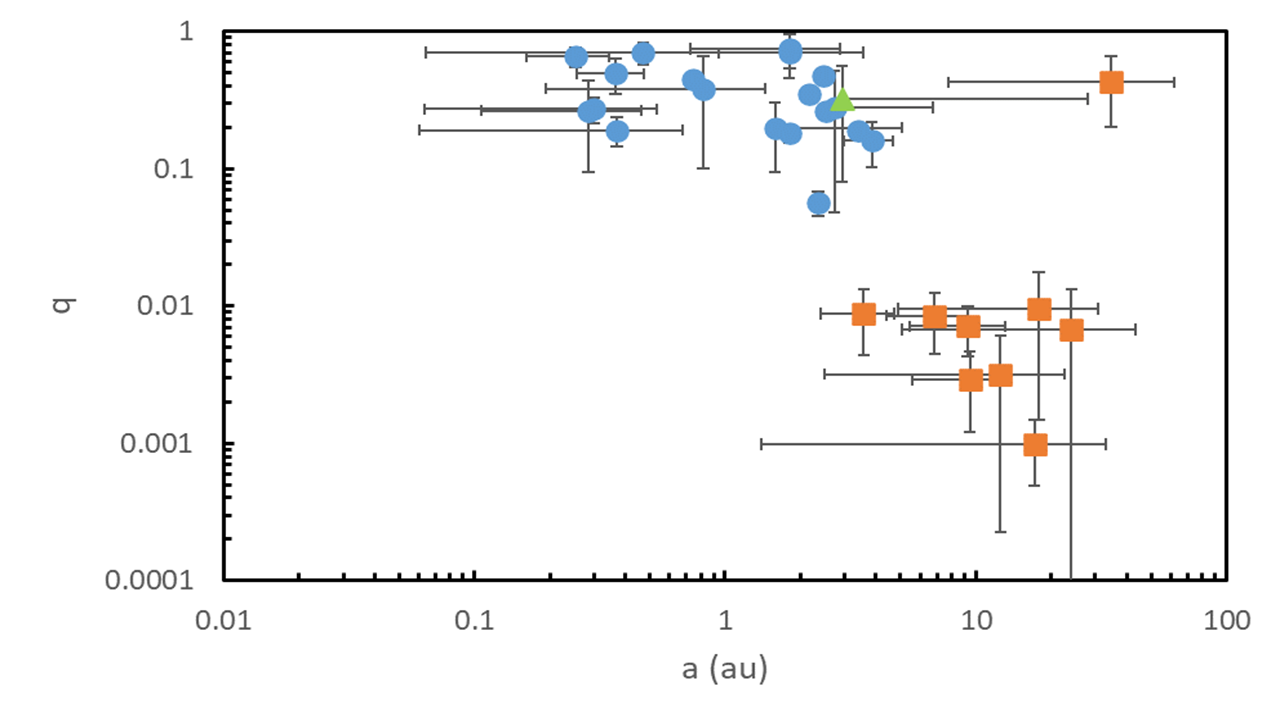}
    \caption{Relation between the semi-major axis and mass ratio for the companions of the young association stars discovered through Gaia PMa, RUWE, or a variation in RVs that are not visual binaries or do not have an orbit determination. Filled blue circles are companions selected on the basis of the value of $RUWE>1.4$; filled red squares are those selected from S/N(PMa)$>3$; filled green triangles are those selected from RVs }
    \label{fig:a_q_dyn}
\end{figure}

\begin{figure}[htb]
    \centering
    \includegraphics[width=8.5cm]{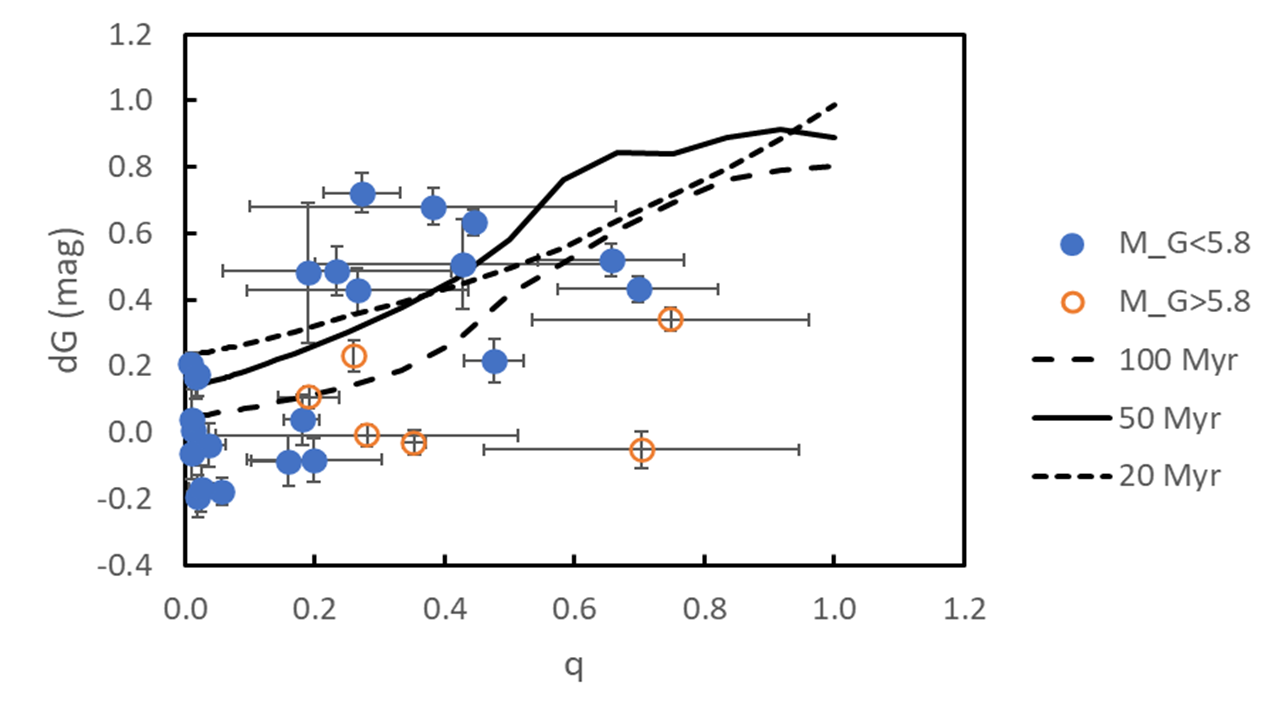}
    \caption{Run of the offset in Gaia $G$ magnitude $dG$ with respect to the single star main sequence as a function of the mass ratio $q$ for the stars with masses determined from a combination of Gaia PMa, $RUWE$, or a variation in RVs. Filled blue symbols are stars with $M_G<5.8$, open red symbols are for fainter stars. The lines are the expected runs of the magnitude offset as a function of $q$ for binaries with primaries of 1.3 \MSun and different ages.}
    \label{fig:q_dG}
\end{figure}

While other properties of the companions (e.g. the eccentricity of their orbits) are of high interest to determine their origin, in this paper we only considered the semi-major axis of the orbit and the mass of the companion because information about eccentricity could be derived only for a small fraction of the objects. We derived estimates of the mass and semi-major axis for all the companions found. Even though in a significant fraction of the cases they have rather large uncertainties, they are still useful for the statistical discussion done in this paper. In general, whenever these data were available from previous analysis, we used them. When this was not possible, for visual binaries we used the photometry of the individual stars to derive the masses using isochrones by \citet{Baraffe2015} and assumed that the semi-major axis is equal to the projected separation divided by the parallax. On average, this corresponds to the thermal eccentricity distribution considered by \citet{Ambartsumian1937} of $f(e)=2 e$ (see \citealt{Brandeker2006}). This last paper indicates that this assumption underestimates the semi-major axis by about 25\% in the case of circular orbits. This can be compared with the eccentricity distribution for wide binaries by \citet{Hwang2022}. These authors found that the eccentricity distribution ranges from uniform (that yields a mean value of $e=0.5$) for semi-major axis $a<100$ au, to thermal (that yields a mean value of $e=0.67$) and even suprathermal for $a>300$ au. 

For a number of objects, indication of binarity comes from PMa, RUWE, or RVs, but the secondary was not observed as a separate object and no period or semi-major axis were determined. These objects have different range of semimajor axis, depending on the technique used to detect them. This is illustrated by Figure \ref{fig:dyn} that shows the run of the Gaia robust RV amplitude, of $RUWE$, and of S/N(PMa) with semi-major axis of the orbit for previously known spectroscopic and visual binaries in the young associations and in the Hyades, for which we performed a similar analysis. They are compared with the detection limits, that we set at 10 \kms (this value is only indicative, however, for the reasons explained in Section 2.4 and Appendix A, and was not actually used to claim binarity), 1.4, and 3 for the robust RV amplitude, $RUWE$, and S/N(PMa), respectively. We notice that the Hyades are at a distance similar to that of the young associations considered in this paper, so they have a similar sensitivity of the astrometric indicators on the presence of companions. They are particularly useful in this context because thanks to efforts by many investigators in the last decades, they have been intensively studied looking for spectroscopic binaries. These plots indicate that binaries discovered through RVs have semimajor axis that typically are lower than about 1 au, those discovered through $RUWE$ have semimajor axis in the range 1-10 au, while those detected through PMa have a typical semimajor axis of around 10 au\footnote{In general, as discussed in Appendix D of \citet{Gratton2023} there is a range of possible solutions compatible with any observed value of $RUWE$, defining a V-shaped area in the $\log{a}$ vs. $\log{q}$ plane. The projected semimajor axis corresponding to the maximum sensitivity of $RUWE$ (that is the minimum mass of a secondary producing a given value of $RUWE$) is at about 2-3 au (depending on the primary mass), a value set by the length of the Gaia astrometric DR3 astrometric sequence (34 months); this is because a growing fraction of the astrometric trend is absorbed by the proper motion estimate for longer periods, producing smaller residuals around the best fit line (for this same reason we would expect even shorter periods in the case of Gaia DR2, as used by \citet{Wood2021}). Hence we expect that most binaries detected using $RUWE$ have periods of a few years and separation of a few au, with a peak-to-valley range of about one order of magnitude around this value. Stars with $RUWE>1.4$ should have periods that are systematically shorter than the stars with significant PMa (that typically have periods of decades corresponding to the time separation between Hipparcos and Gaia epochs), though there is some overlap between the two groups (that is, objects with both $RUWE>1.4$ and S/N(PMa)$>3$). This partial but not full overlap is indeed observed (see \citealt{Gratton2023}), but there are several objects with large $RUWE$ and low PMa and viz. There is also overlap with SB, that typically have shorter periods than those with significant $RUWE$. The plots of Figure 3 fully confirm our expectations. Rather, \citet{Wood2021} considered an empirical calibration of the $RUWE$ values against high contrast and interferometric measurements of companions; this calibration (A. L. Kraus et al. 2021, in preparation) is still unpublished so we cannot use it. Typical separations for binaries with significant $RUWE$ signals considered by \citet{Wood2021} are tens of au and then periods of several decades to a century. However, it is not fully clear that their empirical calibration provides an unbiased estimate of the appropriate parameters for binaries with significant $RUWE$ signals because only companions far enough to be visually detected can be considered in a similar calibration. As a consequence, the separation of the companions may be systematically overestimated. This bias is for instance present when we consider the stars in our sample and in the Hyades. Over a whole sample of about 350 stars, 79 have $RUWE>1.4$ and $G>5$; a visual companion has been detected only for 30 of these stars. We can attribute the lack of detections for 49 of the high $RUWE$ stars at least in part as due to companions that are closer than the sensitivity limit of visual observations. The median separation of the 30 visually detected companions among stars with high $RUWE$ is 0.10 arcsec, that corresponds to a semi-major axis of 5 au for a typical distance of 50 pc. While still likely biased towards high separation, this is substantially less than the median value estimated by \citet{Wood2021} (see their Figure 6, that suggests a value around 0.2 arcsec for a sample of stars that often are more distant than the stars considered in our paper).}. For the objects detected through astrometry these values are related to the maximum sensitivities achieved that depend on the baseline considered: the length of the Gaia DR3 observations (34 months) for the case of $RUWE$, and the separation between the Hipparcos and Gaia DR3 epochs (24.75 yr) for S/N(PMa). Of course we should consider the typical total mass of these binaries, that is about 1.5 \MSun. The different sensitivities of the various techniques indicate that masses and separation may be better constrained combining the different methods rather than considering only a single technique.

We listed in Table \ref{tab:dyn} the objects for which indication of binarity comes from PMa, RUWE, or RVs. For these objects, we looked for solutions that are compatible with the observed values of the RUWE, of the PMa, the scatter in RV, and the non-detection in HCI as done in \citet{Gratton2023}. This was done exploring the semi-major-axis -- mass ratio plane using a Monte Carlo code. We adopted eccentric orbits, with uniform priors between 0 and 1 on eccentricity (in agreement with \citealt{Hwang2022} for this range of separation), 0 and 180 degrees in the ascending node angle $\Omega$, and 0 and 360 degrees in the periastron angle $\omega$ and left the inclination and phase to assume a random value. In addition, whenever available, period was used to fix the solution.

The adopted final values are the mean of those for solutions compatible with observations within the errors, and the uncertainty is the standard deviation of this population. Data for the derivation of these masses are given in Tables \ref{tab:binary_abdor}-\ref{tab:binary_Volans} in the Appendix, while the values of the semimajor axis $a$ and of the masses of the stars are given in Table \ref{tab:dyn}. The distribution of the companions in the $a-q$ plane is shown in Figure \ref{fig:a_q_dyn}; they have $0.1<a<10$ au, but companions of those stars for which the PMa is available tend to be at slightly larger values of the semi-major axis $a$. For small mass objects with significant PMa, we considered the masses provided by the table by \citet{Kervella2022} and adopted a semi-major axis of 7.5 au.

We expect that in most cases close binaries unresolved by both Gaia and 2MASS should appear brighter in the colour-magnitude diagram than single stars of similar colour because of the contribution by the secondary. This happens when the secondary is also a main sequence star or a substellar object; the case is more complicate if the secondary is a stellar remnant, an occurrence that should be rare but that cannot be completely excluded. In addition, the secondary might itself be a close binary: this would cause it to be under-luminous with respect a single star with the same total mass. The magnitude offset $dG$ in the Gaia $G-$ magnitude between single stars and binaries should depend on the mass ratio between the two components, the maximum offset being expected for equal mass systems. The actual run with the mass ratio depends also on the colour that is considered (here the $G-K$ colour), on the age, and on the mass of the components. Although this relation is rather complex, we may use this offset as a check on the dynamical mass ratios indicated by the combination of PMa, $RUWE$, and RVs. This comparison is done in Figure \ref{fig:q_dG}. The agreement between the photometric and dynamical masses is not perfect. The most discrepant cases are 2MASS J05301907-1916318 (AG Lep) in Columba and 2MASS J08264964-6346369 in Volans/Crius 221. In the first case, the magnitude offset is $dG=0.714\pm 0.075$ mag, that would be consistent with a value of $q>0.6$, that is a mass of $M_B>0.5$ \MSun, while the value we obtained from dynamical arguments is only $M_B=0.228\pm 0.050$ \MSun. The second case is an SB according to \citealt{Steinmetz2020}; the magnitude offset $dG=-0.052\pm 0.056$ mag is negligible while the dynamical mass ratio ($q=0.70\pm 0.24$) indicates the presence of a rather massive companion ($M_B=0.557\pm 0.192$ \MSun). In this case a small mass ratio $q\sim 0.2$ (that is only 2-$\sigma$ off the preferred value and would correspond to a 0.15 \MSun star) would make dynamical and photometric results in fair agreement with each other. Given the fair agreement existing between dynamical and photometric masses for most cases, we however think that the adopted masses may be considered acceptable for the statistical purposes of this paper.


\subsection{Detection completeness}


\begin{figure*}[ht]
\centering
\includegraphics[width=\linewidth]{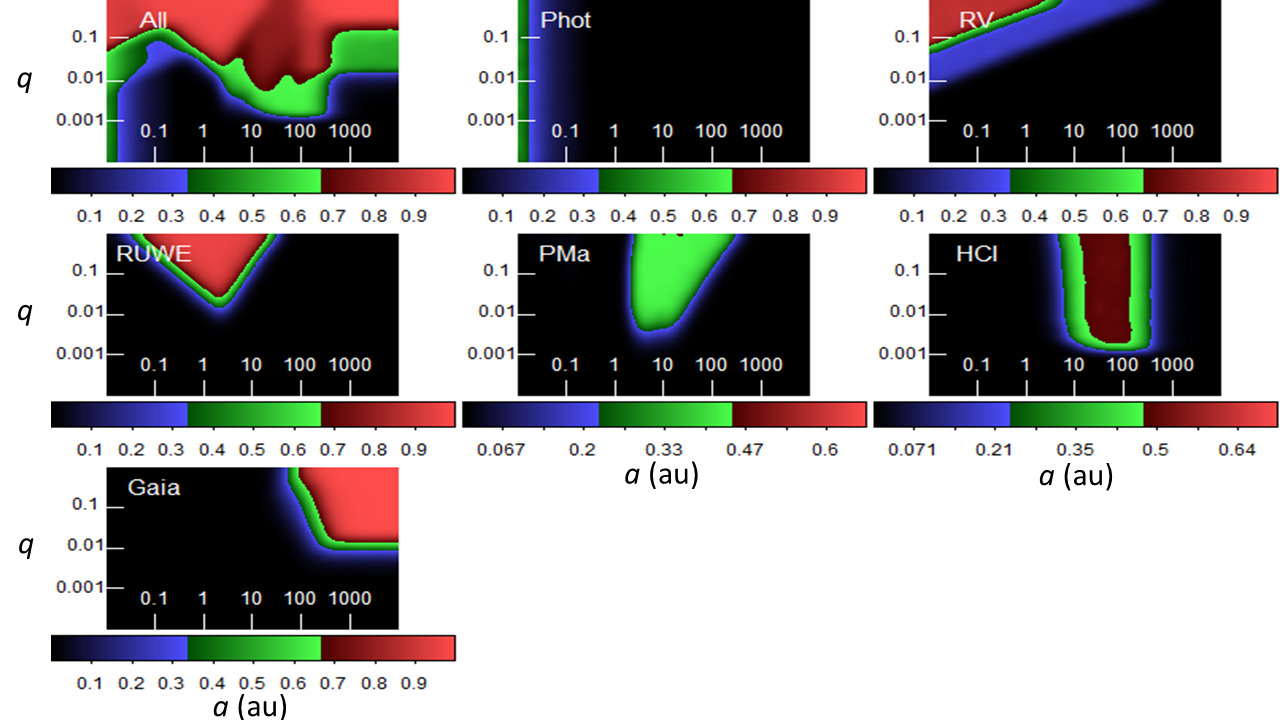}
    \caption{Completeness map of the search of companions in the semi-major axis $a$  (in au) - mass ratio $q$ plane. The upper left panel is the result obtained using all the techniques considered in this paper; the remaining panels are results for the individual techniques: Phot = Eclipsing binaries; RV = spectroscopic binaries; RUWE = Gaia goodness of fit RUWE parameter; PMa = Proper Motion Anomaly; HCI = high contrast imaging; Gaia = separate entry in the Gaia catalogue. Different level of completeness are shown as different colours; the colour scale used is shown on bottom of each panel. }
\label{fig:completeness}
\end{figure*}

\begin{figure*}[htb]
\centering
\includegraphics[width=\linewidth]{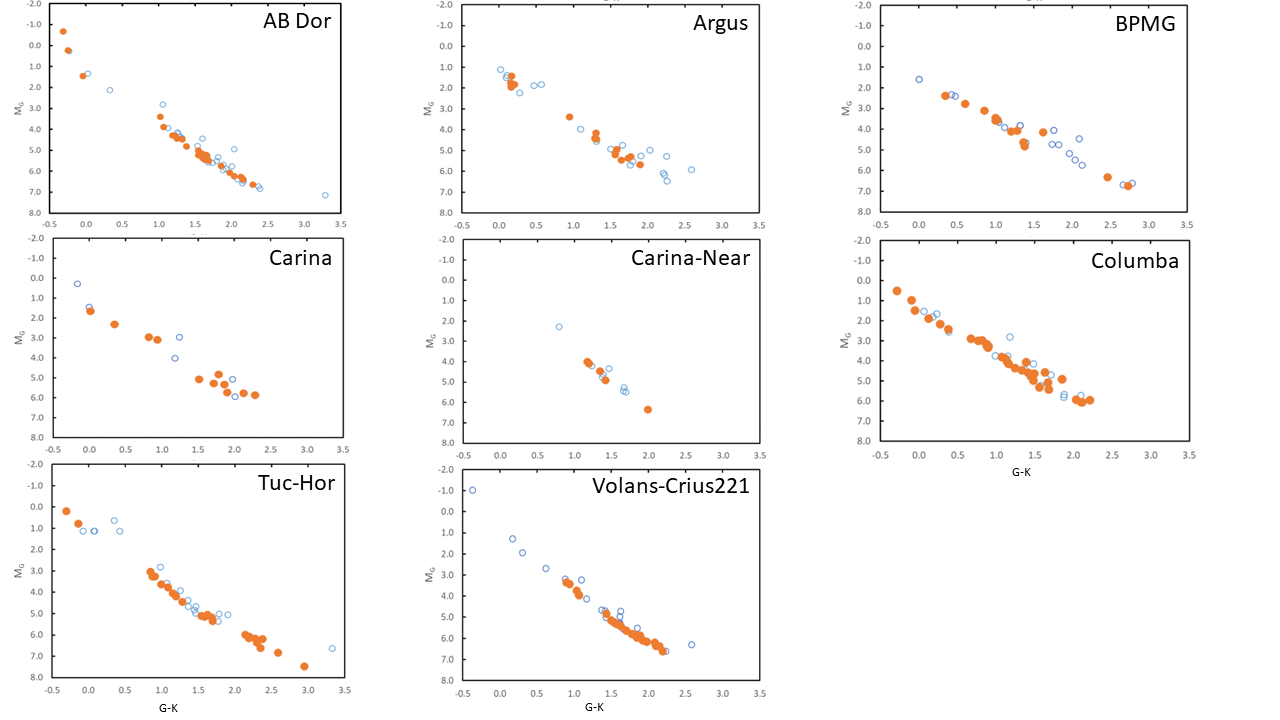}
\caption{$M_G-G-K$ colour-magnitude diagram for stars in the young associations considered in this paper. Orange filled circles are bona fide single stars; blue empty circles are multiple stars.
}
\label{fig:cmd}
\end{figure*}

\begin{figure}[ht]
\centering
\includegraphics[width=\linewidth]{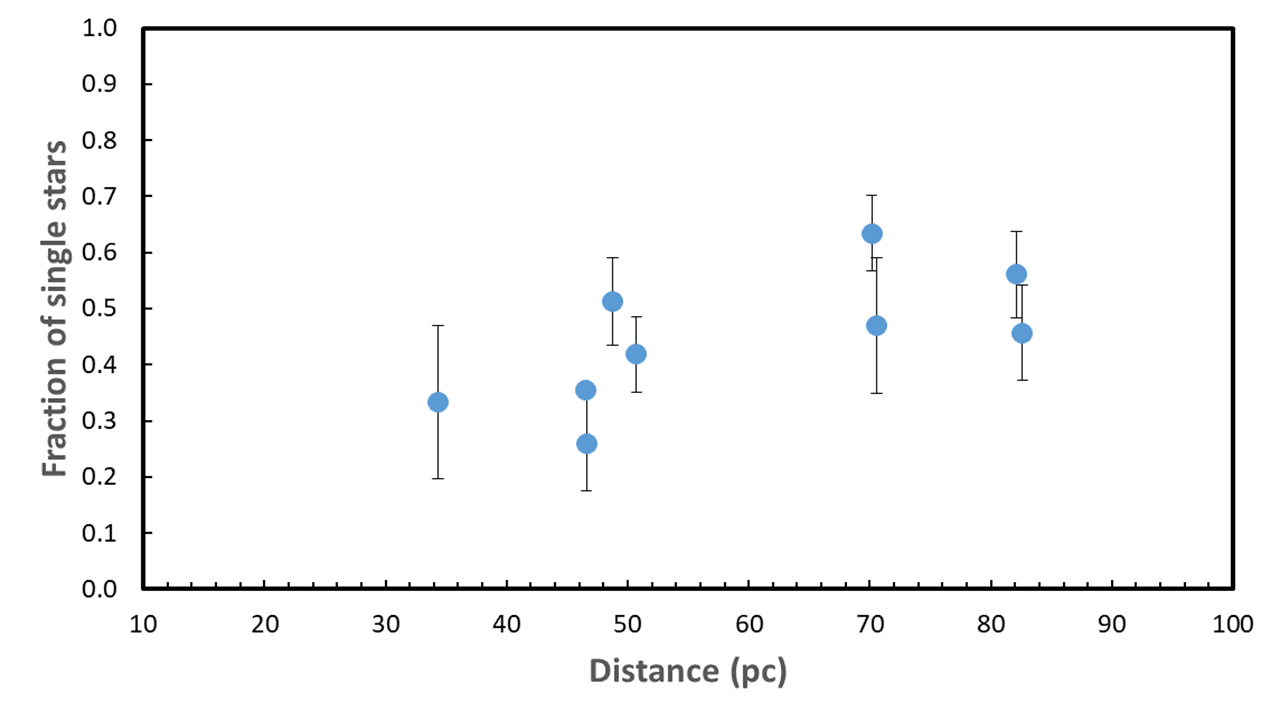}
    \caption{Fraction of single stars in individual associations as a function of the distance}
\label{fig:binfraction}
\end{figure}

Figure \ref{fig:completeness} illustrates the completeness of our survey in the semi-major axis $a$ vs. mass ratio $q$ plane.  This is shown for the individual techniques as well as for the combination of them. These completeness maps were obtained through simulations of 10,000 companions (stellar and planetary) for each of the stars in our sample (with appropriate values of the magnitude, mass, parallax and age) with random values of the orbital parameters. For each target and methodology we considered if the appropriate data sets were available. We then determined the signal expected for each simulated companion according to the various methods considered and compared them with the detection limits appropriate for each method. This method was already considered in \citet{Gratton2023}, where a full description is reported, save for errors and limits in RV; this procedure is actually similar to that proposed by \citet{Wood2021}. While the detection limits are possibly questionable, they were used consistently in the simulation and on the real data. This is of course not properly an injection/recovery process, in the sense that we did not start from real raw data adding a companion signal and recovering it. This full injection/recovery process was done to determine the detection limits in HCI; however, it was not possible to use it in other data sets, in particular for those provided by Gaia that are the most useful for the present purposes, because the raw data are not available. In these cases we simply considered the signal on the time series of Gaia positions and RVs relevant to each simulated companion, added a random Gaussian noise, extracted from these noisy simulated series the relevant signal (RUWE, RV variation, proper motion anomaly), and compared it with the observed values (both above and below threshold for detection). We notice that also signals below threshold can be useful to constrain the range of parameters where acceptable solutions can be found. For eclipses/transits the detection procedure is much more complicated and there is no simple threshold value available in the literature. For what concerns stellar companions, we simply assumed that all EBs with periods $<28$ days are recovered by TESS. TESS is in principle also able to recover some EB with longer periods depending on the number of sectors where the target was observed. However, the EB catalogues we used only refers to the early observations with TESS and very few stars were observed on a large number of sectors. For transiting planets the situation is much more complex and TESS is of course less efficient in discovering young transiting planets; we then discussed the case of transiting planets separately in Section 4.3, where we make explicit reference to literature papers that used an injection/recovery procedure \citep{Fernandes2022, Fernandes2023}.

For what concerns RVs, the precision achievable for solar-type stars is in principle much higher than for early-type stars, due to the slower rotation and much higher number of spectral lines available. On the other hand, we should take into account the concern related to the jitter due to the stellar activity. To evaluate the appropriate RV jitter, we considered in Appendix A the value of the robust RV amplitude in those stars that do not have any other clear signs of being close binaries (separation $<100$ au). We found that as expected for these objects, the robust RV amplitude is a function of both star absolute magnitude (a proxy for spectral type) and age. In addition, there are a number of young late-type stars that are very fast rotators (as indicated by the short period that can be derived by the TESS light curves) and for them the robust RV amplitude is not a reliable indicator of real RV variations, probably because of severe line blending not properly taken into account by the Gaia RV pipeline. This indicates that the Gaia RVs are sometime poor indicators of variation of the centre-of-mass RVs. In our procedure, we then considered the maximum between the internal error and the median RV jitter expected for the age and absolute magnitude of each star as an appropriate indicator of the errors in individual RVs and we did not consider as binaries young late-type star whose only indication of binarity is a high value of the Gaia robust RV amplitude, save for the case of HIP47017 that has a Gaia robust RV amplitude of 122.4 \kms. For the statistical purposes of this paper, we then considered detection thresholds of 1 km/s for those cases where there are high precision RV series, and of 10 km/s for those when only the Gaia RV sequence is available, respectively. These values are much larger than the typical internal errors, that are a few m/s from high precision RVs and $\sim 0.3$ km/s for those from Gaia. With these thresholds, the completeness level for separation in the range 0.1-1 au is however higher than obtained for B-stars in \citet{Gratton2023}. We notice that for what concerns eclipses, the detection efficiency considered preparing this figure includes the probability that the transit occurs.

For what concerns stellar companions, thanks to Gaia, the search for visual companions should be complete down to a projected separation of about 100 au. A fraction of the substellar companions is not detected by Gaia. The masses corresponding to a magnitude of $G=19$ (roughly the detection limit to be considered here in order to have reasonably accurate astrometry) for the age and average distance of each association are listed in Table \ref{tab:associations}. They are all in the BD regime, ranging from 0.013 \MSun for the BPMG to 0.040 \MSun for Volans/Crius 221. HCI imaging, available for (at least) 194 stars, should reveal all stellar companions with semi-major axis in the range 20-150 au. 

For the programme stars, the PMa (available for 178 stars) should have $S/N>3$ for all systems having a stellar companion in the range 3-100 au. There are 61 stars for which neither PMa nor HCI is available. Though additional data is available (for instance visual observations or speckle interferometry), for these stars (21\% of the total) there may be incompleteness in the 10-100 au region.  Stellar companions with separation in the range 0.2-10 au would produce a value of $RUWE>1.4$ and should be detected. RUWE is available for all but seven stars in the sample. RV variations with clearly significant amplitude would be detected for semi-major axis $<1$ au and for many others with larger radii. High precision RV sequences are available for 79 stars, while RV series from Gaia are available for 181 additional stars in the sample. Finally, TESS data about photometric variations are available for 269 stars. The search for eclipsing binaries with $a<0.2$ au (roughly a period of 27 days) should be then almost complete. However, we expect that only about 1/10 of the close binary systems should be eclipsing/transiting. We conclude that while some massive companions may still be missed, we think our search is almost complete for stellar companions and that few BDs are missing at separation greater than 10 au.

While completeness in the binary search is on average high, we find indications that a few binaries may still be missed. Figure \ref{fig:cmd} shows the $M_G-G-K$ colour-magnitude diagram for stars in the young associations considered in this paper. Different symbols are used for bona fide single stars and multiple stars. The single star sequences are generally very narrow. However, a few binaries may still be present among the sample of bona fide single stars. Furthermore, the search is likely less complete around stars that are more distant from the Sun. This is shown in Figure \ref{fig:binfraction}, where a trend for the fraction of single stars with distance of the individual associations is apparent. These data yield a Persson linear coefficient $r=0.63$ that is significant at 2.5\% level of confidence. This trend is due to the lack of PMa data for stars not included in the Hipparcos catalogue; in fact, the fraction of stars that have PMa data changes systematically with distance of the association, from about 100\% for the closest association (Carina Near) to about 40\% for the furthest ones (Argus and Volans/Crius 221). This highlights the importance of the distance limit in our analysis.

\subsection{All companions in the semi-major axis vs mass ratio plane}


\begin{figure*}[htb]
\centering
\includegraphics[width=\linewidth]{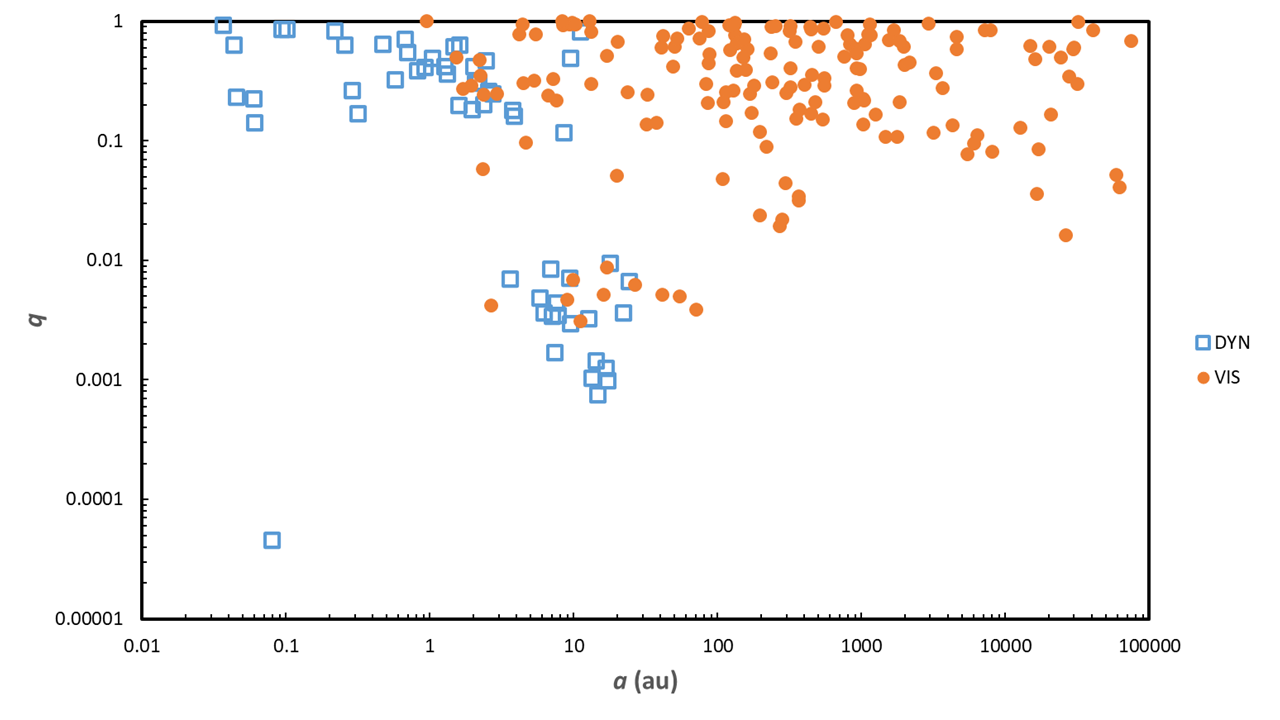}
\caption{Relation between semi-major axis $a$ (in au) and mass ratio $q$ between components for all companions found so far around stars that are member of the young associations considered in this paper. Filled orange circles are companions detected in imaging; open blue squares are those detected using dynamical methods or eclipses/transits.
}
\label{fig:all_companions}
\end{figure*}

\begin{figure*}[htb]
\centering
\includegraphics[width=\linewidth]{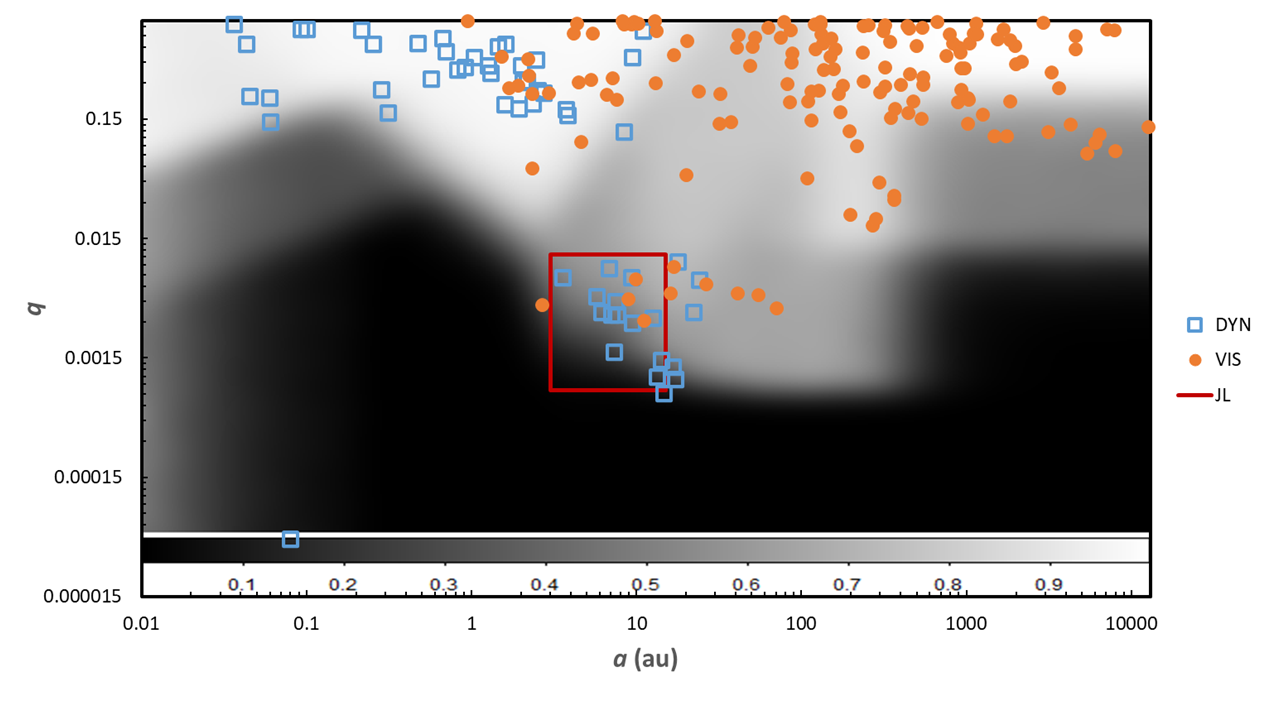}
\caption{Same as Figure \ref{fig:all_companions} but with the detection completeness in the background. The grey scale on the bottom of the figure indicates the level of completeness achieved in different region of the $a-q$ plane. The red rectangle marks the area covered by JL planets.
}
\label{fig:detection_vs_completeness}
\end{figure*}

Figure \ref{fig:all_companions} shows the correlation between semi-major axis $a$ and mass ratio $q$ for all companions considered in this paper. The distribution of the detected companions in this plane is the result of both the real distribution and of the selection effects due to completeness of the search, as shown in Figure \ref{fig:detection_vs_completeness}. As expected, detection efficiency of small mass companions ($q<0.01$) is rather low. After consideration of the selection effects, this diagram shows two main groups of objects: a wide distribution of companions with mass ratio $q$ from 0.015 to 1, and a second group of objects with semimajor axis from 1 up to about 100 au at $q<0.01$. This second group covers a region of the plane that is more extended in semi-major axis than that corresponding to the definition of JL planets. We should note that the buildup of dynamical companions at 7.5 AU is simply a consequence of the fact that lacking better information, we assumed this value to estimate their masses from the tables by \citet{Kervella2022}. Uncertainties in the semi-major axis are of at least a factor of three, so this value should only be taken as an approximation of the real value. We remind that our paper is only aimed to a statistical discussion of the frequency of companions of different masses and semi-major axis. This only requires order of magnitude estimates these quantities. Given this consideration, we will consider all members of this group as JL planets. Incompleteness is high in this region. If we strictly consider the area corresponding to the definition of JL-planets, the average completeness factor is 36\%. However, even within this restricted area there is a strong gradient with mass ratio and less extreme in semi-major axis, so more massive and further out planets are more easily detected than the close-by and smaller mass ones. Detection efficiency is as low as 0.5\% at the lower left corner of this range. If the mass function of JL-planets favours smaller mass object, as it is very likely, the completeness factor for the detection of JL-planets is well below the number given above. 

Finally, around the stars considered in this paper there is a 
hot-Neptune transiting DS Tuc A ($5.6\pm 0.2$ R$_{\rm Earth}$: \citealt{Newton2019, Benatti2019}) in the Tuc-Hor moving group, to which we have to add the mini-Neptune around HIP 94235 ($3.00^{+0.32}_{-0.28}$ R$_{\rm Earth}$ \citealt{Zhou2022}) for which no mass is available.

In the next two sections we consider first the properties of stellar companions and then those of the planetary companions, with emphasis on the JL ones.

\FloatBarrier

\section{Stellar companions}

\subsection{Statistics of stellar companions}


\begin{table}
\caption{Frequency of single stars for young nearby associations}
\begin{tabular}{lccccl}
\hline
Association & Primaries & Single stars & Frequency \\
\hline
AB Dor      & 55 & 23 & $0.42\pm 0.07$ \\
Argus       & 35 & 16 & $0.46\pm 0.08$ \\
$\beta$ Pic & 27 &  7 & $0.26\pm 0.08$ \\
Carina      & 17 &  8 & $0.47\pm 0.12$ \\
Carina Near & 12 &  4 & $0.33\pm 0.14$ \\
Columba     & 52 & 33 & $0.63\pm 0.07$ \\
Tuc-Hor     & 41 & 21 & $0.51\pm 0.08$ \\
Volans/Crius 221      & 41 &  23 & $0.56\pm 0.08$ \\
\hline
\end{tabular}
\label{tab:frequency}
\end{table}

Table \ref{tab:frequency} lists the total number of stars and the number of single stars in each association. The total number of single stars (we do not consider planetary companions here) is 135 over 280 systems surveyed\footnote{The total number of entries in our lists is 296 stars because there are several wide binaries with separate entries for the individual components.}, that is a fraction equal to $0.48\pm 0.03$\footnote{Throughout this paper, uncertainties in the fractions were computed as $\sqrt{q p/n}/n$, where $p$, $q$, and $n=q+p$ are the number of positive, negative, and total cases considered.}. The value is likely overestimated by a few per cent because our search for companions is not complete. This fraction is lower than the value of $0.54\pm 0.02$ obtained for field solar-type stars by \citet{Raghavan2010} (see upper panel of Figure \ref{fig:frequency}). This is not  an effect of a different mass distribution. In fact, the median mass of the primaries considered in this paper is 1.04 \MSun, that is only marginally different from the solar value. A high frequency of multiple stars in nearby young association was already noticed (CrA: \citealt{Kohler2008}; Taurus-Auriga: \citealt{Leinert1993, Kohler1998}); Sco-Cen: \citealt{Gratton2023}).

\begin{figure}[htb]
    \centering
    \includegraphics[width=8.5cm]{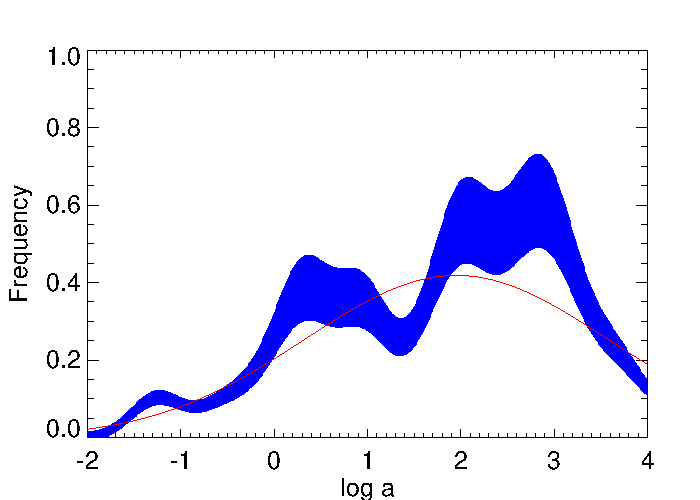}
    \includegraphics[width=8.5cm]{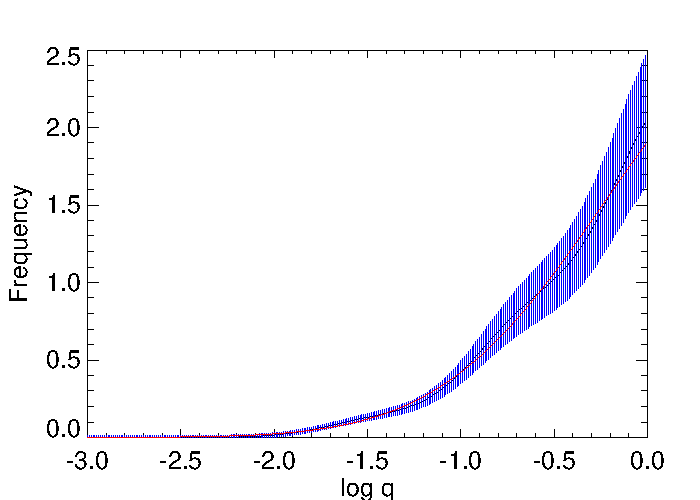}
    \caption{Distribution of companions with mass ratio and semi-major axis. Upper panel: v with $\log{q}>-1.08$ as a function of the logarithm of the semi-major axis $a$ in au. Lower panel: Distribution of companions with $0.5<\log{a}<3$ as a function of the logarithm of the mass ratio $q$. The red lines are best fit log-normal through the observed distributions. The shaded area in both panels corresponds to 1-$\sigma$ uncertainty.}
    \label{fig:a_dist}
\end{figure}

\begin{figure}[htb]
\centering
\includegraphics[width=8.5cm]{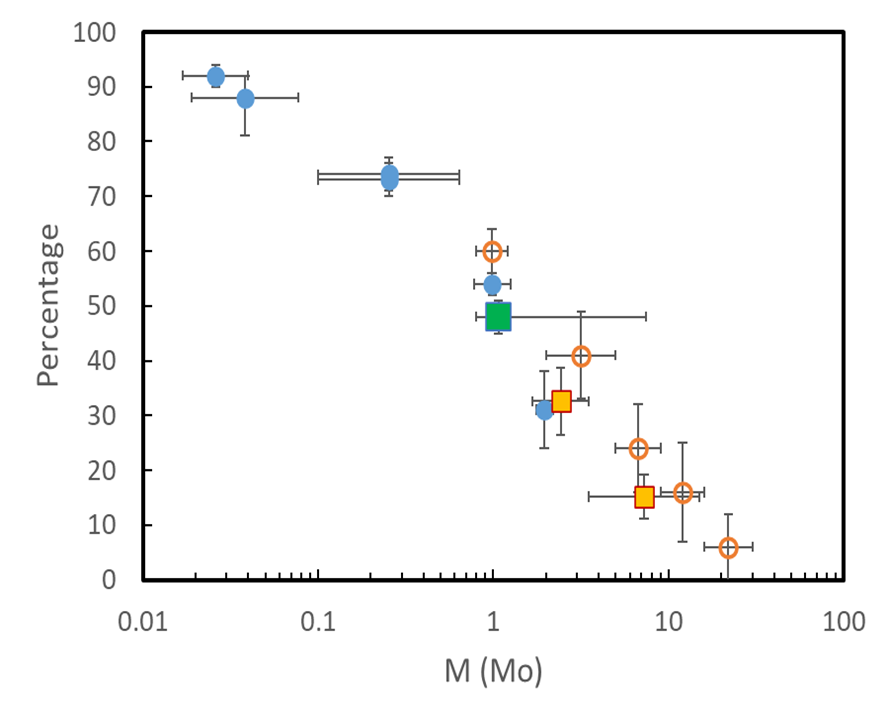}
\includegraphics[width=8.5cm]{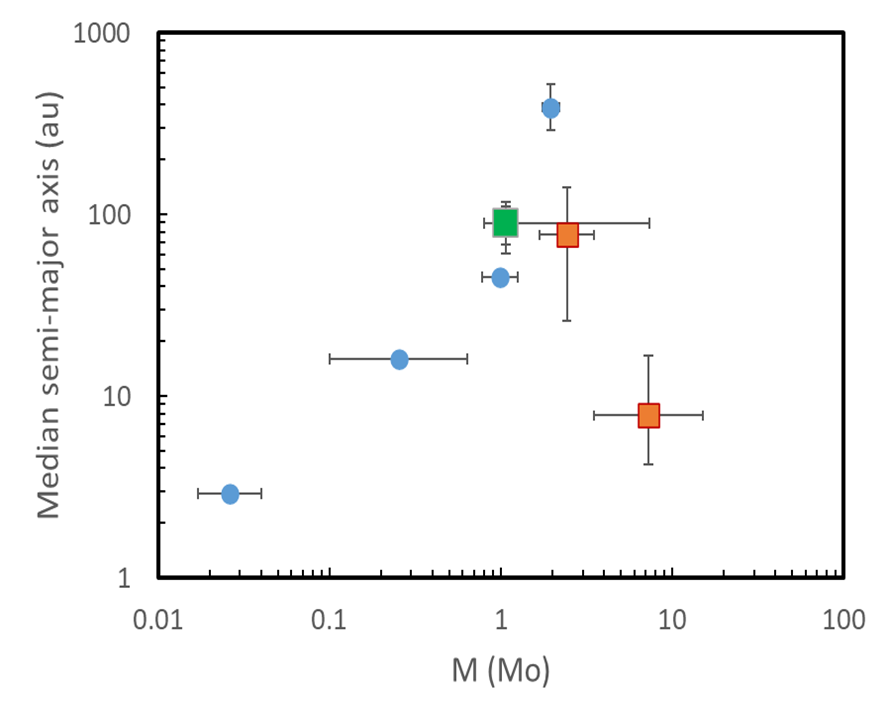}
\caption{Runs of statistical properties of binaries with the stellar mass. Upper panel: Run of the frequency of single stars from our data (green squares), the samples in Table~\ref{tab:frequency} (filled circles), from \citet{Moe2017} (opens circles), and the B stars in Sco-Cen \citep{Gratton2023}. Lower panel: Run of the median semi-major axis. Horizontal error bars reproduce the mass range of the different samples  }
\label{fig:frequency}
\end{figure}

The upper region of Figures \ref{fig:all_companions} and \ref{fig:detection_vs_completeness} contains stars and BDs (objects with $\log{q}>-1.8$). Our search is reasonably complete for these objects: the mean detection efficiency over the whole area is 74\%. The lowest efficiency is obtained for the search of BD companions with separation of the order of a few tenths au. These companions would be easily detected through high precision RVs, that are however available for only 79 stars, that is 28\% of the sample. Anyhow, the lack of any BD at small separation around these 79 stars shows that this region of the diagram is at most scarcely populated. On the other hand, the discovery efficiency for stellar companions (objects with $\log{q}>-1.15$) is 84\% and the minimum value is 31\%. This indicates that our sample is fairly complete so far stellar companions are considered. Overall, our lists include 182 stellar companions (mass $>0.075$ \MSun) and nine BDs (here objects with a mass $0.02<M<0.075$ \MSun) around 145 stars. We will consider here stellar and BD companions together. A log-normal fit to their semi-major axis $a$ yields an average value of $\log{a/au}=1.94$ (that is 88 au), with a sigma of 1.50. The companions with separation wider than 1000 au are 44, that is $23.0\pm 3.5$\% of the total. 

Within the upper strip we can still discern trends in the values of $q$ with separation. Namely, there is an empty region with $a<200$ au and $0.01<q<0.05$. As mentioned above we are not fully complete over this region, however, the lack of objects in this region is clearly significant. This is the well-known BD desert \citep{Marcy2000, Raghavan2010, Stevenson2023, Unger2023}). However, at separation $a>200$ au this region is populated by a number of objects, similarly to what found previously e.g. by \citet{Metchev2009}. We notice in fact that four out of ten of the BDs are at separation wider than 1000 au. This feature should be reproduced by mechanisms of formation of companions.

\subsection{Distribution with mass and semimajor axis of stellar companions}


Figure \ref{fig:a_dist} shows the distribution of stellar companions with semi-major axis $a$ and with the mass ratio $q$ corrected for completeness. The semi-major axis $\xi(\log{a/au})$ distribution can be described by a log-normal as:
\begin{equation}
\xi(\log{a/au})=0.418 \exp{[\frac{-0.5~(\log{a/au}-1.95)^2}{1.62^2}]}.
\end{equation}
The mean value made on the logarithm of $a$ yields a value of $a=89_{-21}^{+28}$ au. 
This value is close to that obtained for early-B stars in Sco-Cen by \citet{Gratton2023} and the peak of the run with stellar mass (see lower panel of Figure \ref{fig:frequency}.

The distribution with $\xi(\log{q})$\ of companions in the range 3-1000 au for $0.003<q<1$, again corrected for completeness, is reproduced by the tail of a log-normal law of the form:
\begin{equation}
\xi(\log{q})=2.453 \exp{[\frac{-0.5~(\log{q}-0.61)^2}{0.86^2}]}.
\end{equation}

\begin{figure}[htb]
\centering
\includegraphics[width=8.5cm]{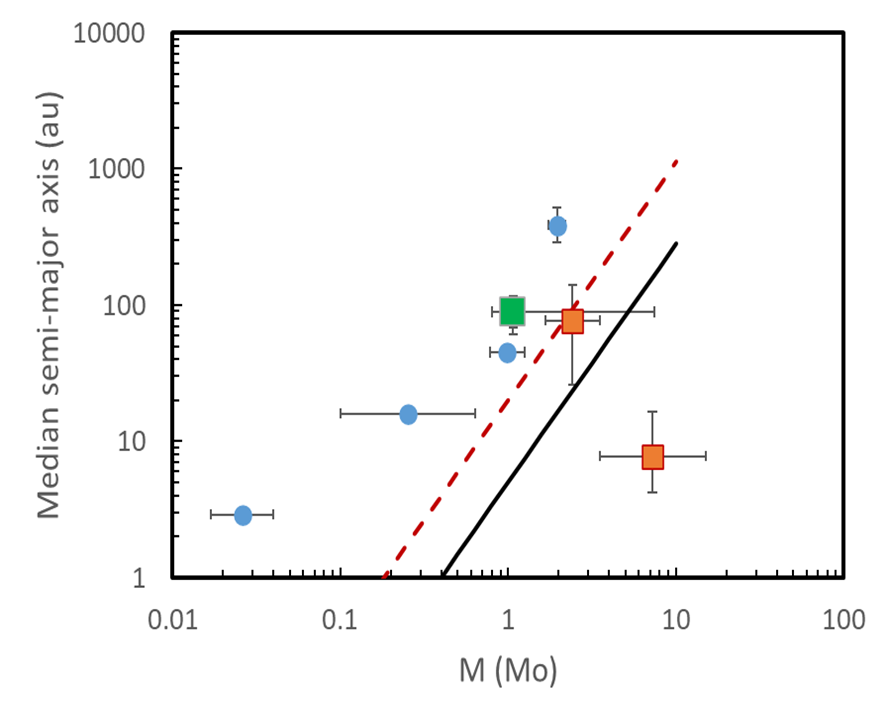}
\caption{Same as the lower panel of figure \ref{fig:frequency} but with added the location of the water-ice line (solid black line) and five times this distance (dashed red line) as a function of the mass of the primary star. Formation of JL planets by core accretion is expected in this range of separation from the star }
\label{fig:iceline}
\end{figure}

Figure \ref{fig:iceline} compares the median semi-major axis of the orbits of stellar companions with the location of the water ice-line (solid line) and five times this distance as a function of the mass of the primary star. Formation of JL planets by core-accretion mechanism is expected in this range of separation from the star. We notice that the peak of the distribution with semi-major axis for stellar companions is very close to the region where formation of JL planets is expected. Hence, such a formation will be prevented in a rather large fraction of the systems. This should be considered when estimating the frequency of JL planets in the context of the mechanisms for their formation. Also, this figure shows that for solar-type stars (as well for stars of lower mass), planets in binary systems are most likely to be found closer to the star than the companions (S-type planets), while the opposite should hold in the case of planets around B-stars (P-type planets).

\FloatBarrier

\section{Planetary companions}

\subsection{Detections and candidates JL planets}


\begin{figure*}[htb]
\centering
\includegraphics[width=5.0cm]{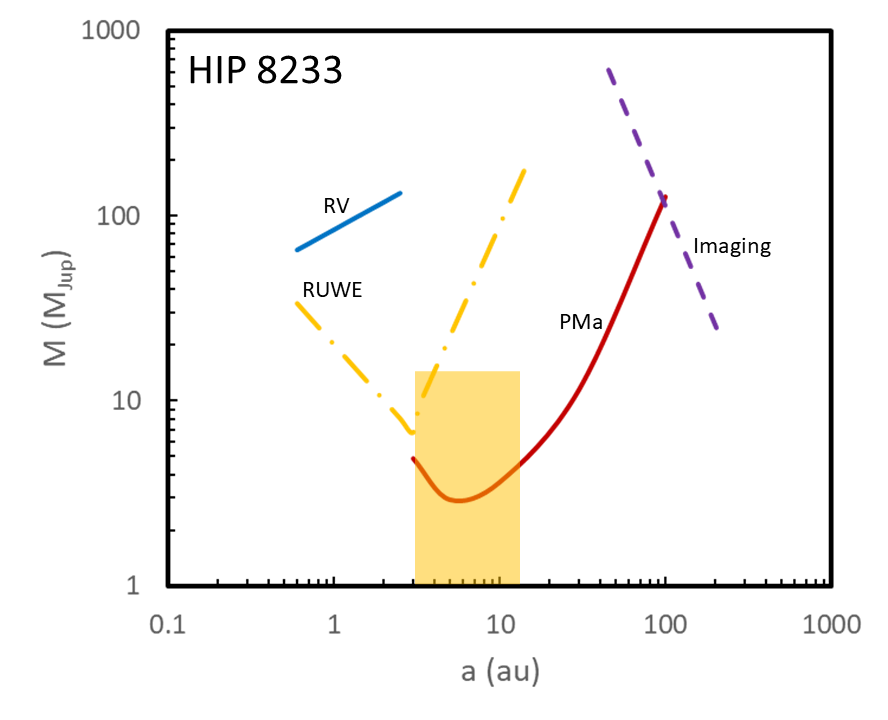}
\includegraphics[width=5.0cm]{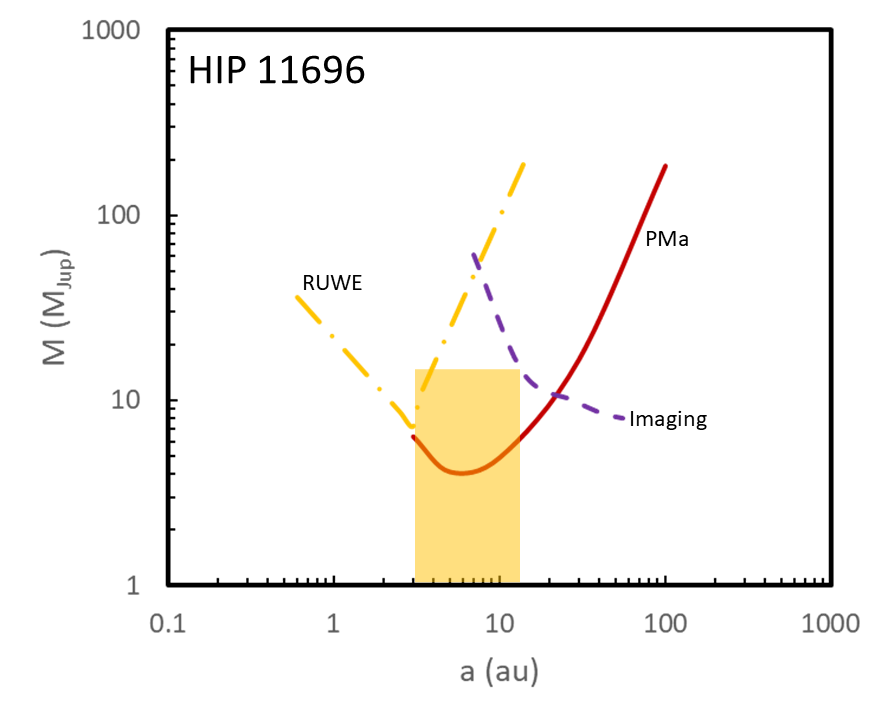}
\includegraphics[width=5.0cm]{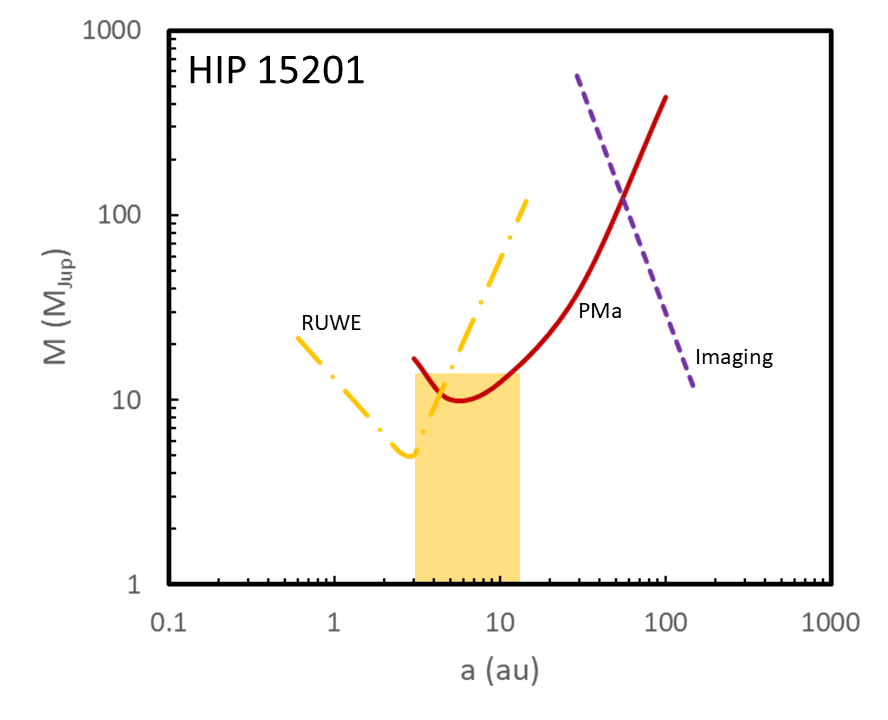}
\includegraphics[width=5.0cm]{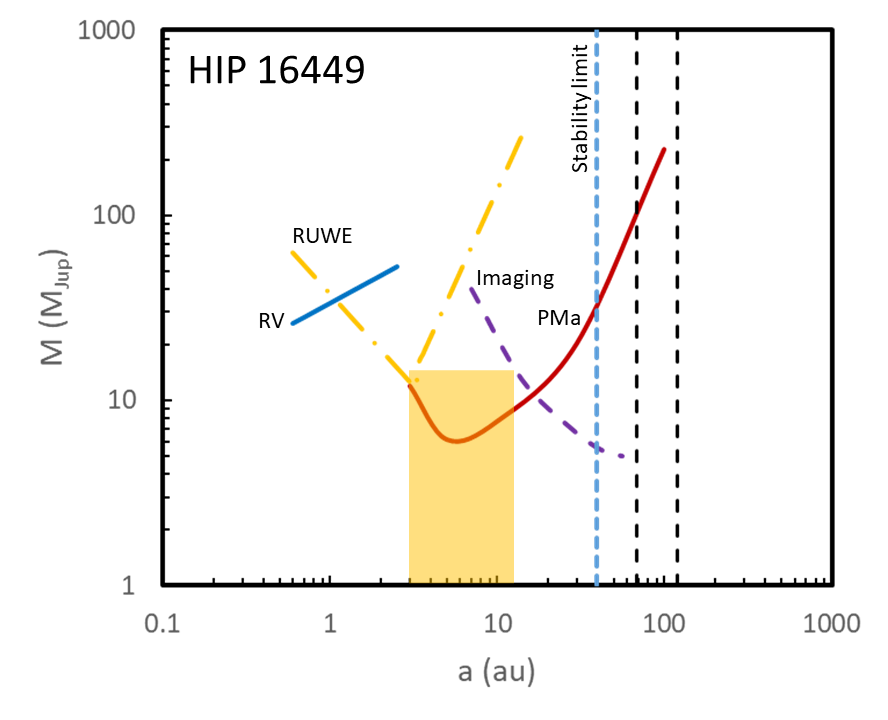}
\includegraphics[width=5.0cm]{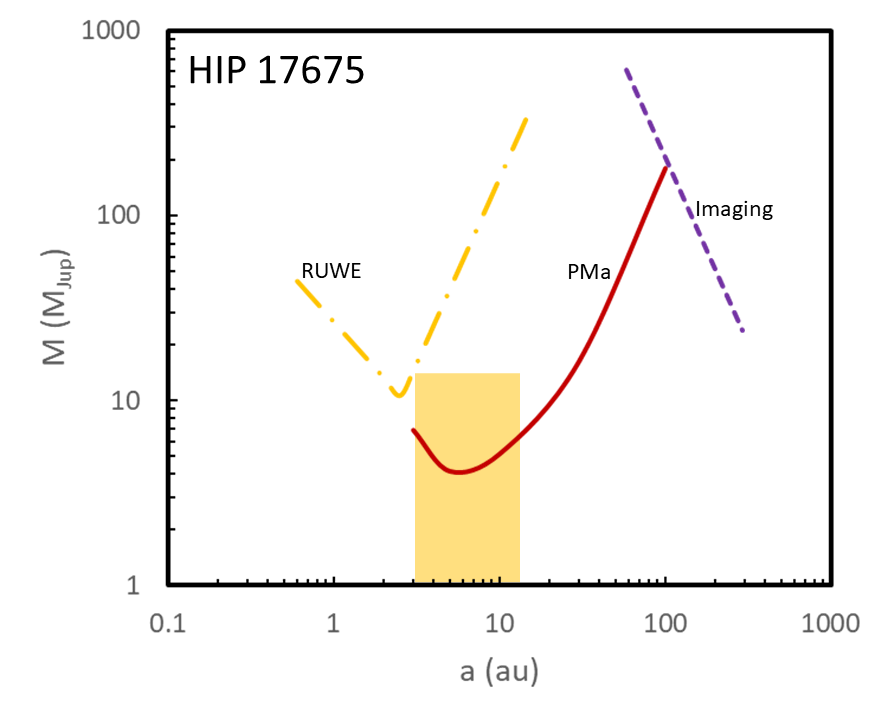}
\includegraphics[width=5.0cm]{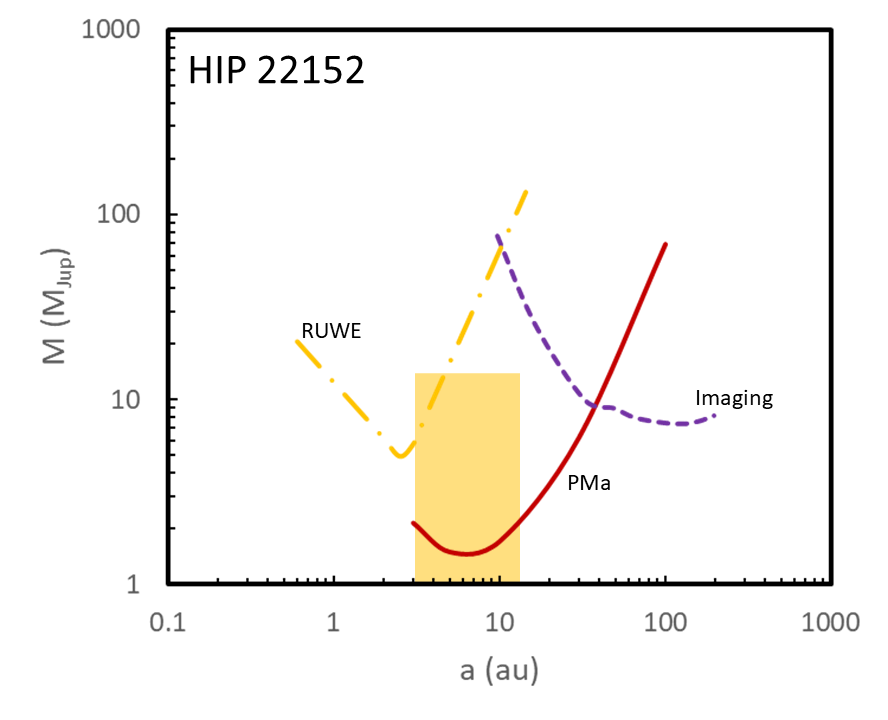}
\includegraphics[width=5.0cm]{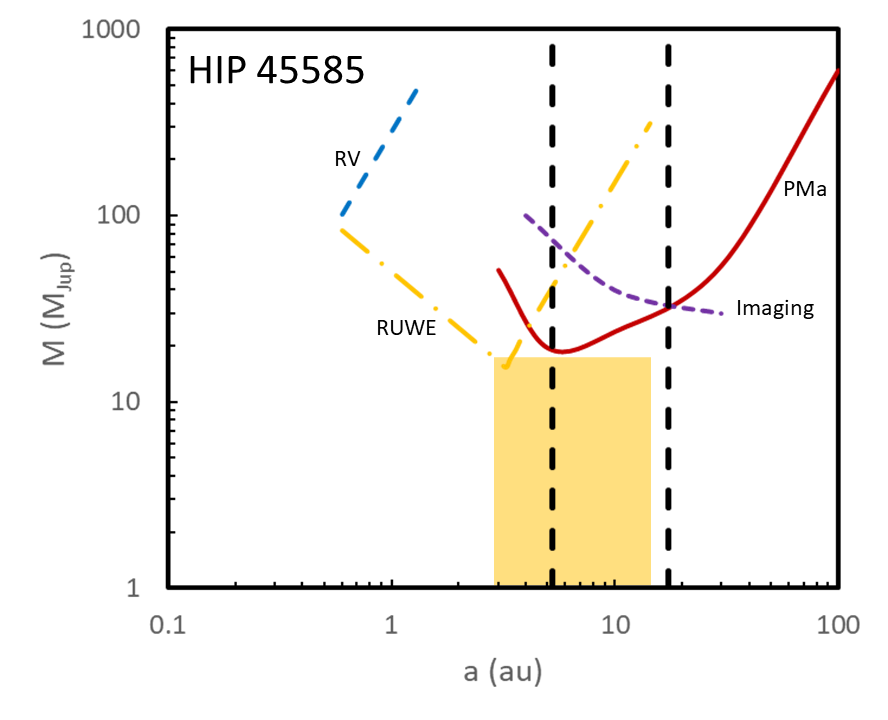}
\includegraphics[width=5.0cm]{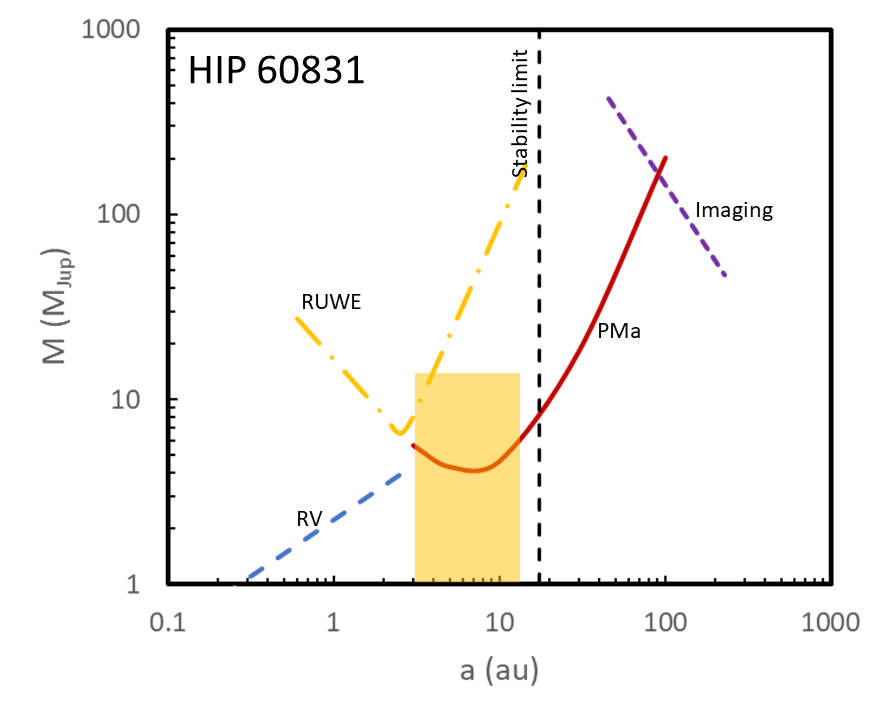}
\includegraphics[width=5.0cm]{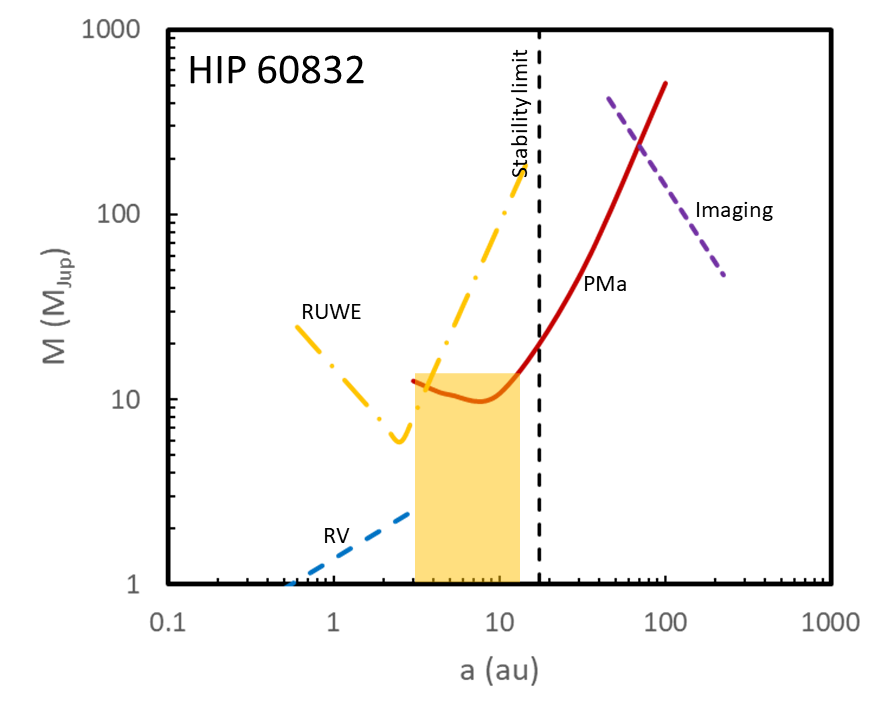}
\caption{Diagrams showing the mass of companions (in \MJup) that may be responsible for the observed proper motion anomaly (PMA) \cite{Kervella2022} as a function of semi-major axis $a$ in au (solid red line) for the stars discussed individually in Section 4.1. They are compared with the upper limits from imaging (short dashed violet line), RUWE (dash-dotted orange line) and RVs (dashed blue line). The area marked in orange is that occupied by JL-planets. When the star has debris disks, the vertical black dashed lines mark their position, and the vertical blue dashed line the stability limit. First row, from left to right: HIP~8233 in Tuc-Hor, HIP~11696 in AB Dor, HIP~15201 in Carina; second row; HIP~16449, HIP~17675 and HIP~22152 in Columba; thord row: HIP~45585 in Carina Near, and HIP~60831 and HIP~60832 in Carina Near}
\label{fig:HIP29568}
\end{figure*}

\begin{table}
\caption{Masses and semi-major axis for candidate planets from PMa}
\begin{tabular}{lccccl}
\hline
HIP       & $M_A$ &  $M_B$             &   $a$           &Frac(JL)\\
          & \MSun &  \MSun             &  au             &       \\
\hline
    11696 & 1.326 &  0.0048$\pm$0.0030 &  7.68$\pm$7.08  & 0.928 \\
    30314 & 1.067 &  0.0011$\pm$0.0004 & 13.28$\pm$13.93 & 0.507 \\
      560 & 1.413 &  0.0017$\pm$0.0009 & 16.57$\pm$15.78 & 0.596 \\
    10679 & 1.036 &  0.0054$\pm$0.0030 &  5.74$\pm$2.96  & 0.998 \\
    84586 & 2.234 &  0.0038$\pm$0.0012 &  7.28$\pm$4.28  & 0.984 \\
    88399 & 1.309 &  0.0043$\pm$0.0026 &  7.02$\pm$3.39  & 0.954 \\
    15201 & 1.391 &  0.0135$\pm$0.0114 & 17.78$\pm$12.87 & 0.634 \\
    45585 & 2.124 &  0.0176$\pm$0.0084 &  6.85$\pm$2.46  & 0.687 \\
    60831 & 1.166 &  0.0034$\pm$0.0020 &  9.47$\pm$3.90  & 0.940 \\
    60832 & 1.042 &  0.0074$\pm$0.0029 &  9.27$\pm$3.84  & 1.000 \\
    16449 & 1.762 &  0.0056$\pm$0.0052 & 12.54$\pm$10.06 & 0.780 \\
    17675 & 1.445 &  0.0097$\pm$0.0099 & 24.09$\pm$19.02 & 0.467 \\
    22152 & 1.221 &  0.0012$\pm$0.0006 & 17.11$\pm$15.72 & 0.452 \\
     1481 & 1.129 &  0.0016$\pm$0.0011 & 14.17$\pm$14.98 & 0.585 \\
     8233 & 1.431 &  0.0052$\pm$0.0060 & 22.07$\pm$18.26 & 0.444 \\
    52462 & 0.833 &  0.0007$\pm$0.0003 & 14.64$\pm$14.98 & 0.187 \\
    96334 & 1.000 &  0.0037$\pm$0.0023 &  6.15$\pm$3.77  & 0.935 \\
\hline
\hline
\end{tabular}
\label{tab:jl_masses}
\end{table}

Overall, we considered 24 stars hosting JL or candidates in the eight associations considered. JL planets have been already observed through HCI around seven stars. They are 51 Eri \citep{Macintosh2015, DeRosa2020, Samland2017}, AF Lep \citep{Mesa2023, DeRosa2023, Franson:2023arXiv}, and $\beta$ Pic \citep{Lagrange2009, Lacour2021} in the BPMG; HIP30034 (AB Pic) in Carina \citep{Chauvin2005, Vigan2021}; HR8799 \citep{Marois2008, Marois2010, Zurlo2022} and $\kappa$ And \citep{Carson2013, Uyama2020} in the Columba moving group; and b03 Cyg in the Argus moving group \citep{Currie2023}. 

In addition to these seven stars (with a total of 11 planets), indications for the presence of JL planets are obtained from PMa for 18 additional stars. The strong cases for JL companions being responsible of the observed PMa are already discussed for HIP30314 in AB Dor \citep{Mesa2022}; HIP560, HIP10679, HIP84586, and HIP88399 in the BPMG (\citealt{Mesa2022} and \citealt{Gratton2023}), for HIP1481 in Tuc-Hor \citep{Mesa2022}, and for HIP52462 \citep{Mesa2021} and HIP 96334 \citep{Mesa2022} in Volans/Crius 221 and will not be repeated here. We will now discuss the remaining ten cases.


{\bf HIP8233 (HR 517)}: member of Tuc-Hor with a mass of 1.42 \MSun. There is no HCI in the ESO archive and no high precision RV in the literature. The star has no infrared excess \citep{Pawellek2021}. The value of the PMa is consistent with the not significant RUWE and the non-detection of companions from Gaia over a wide range of semi-major axis (3-100 au) and mass (4-120 \MJup: see Figure \ref{fig:HIP29568}). Waiting confirmation from HCI data (an upper limit would also be adequate), we will assume here that the companion is a JL planet.

{\bf HIP11696}: member of AB Dor with a mass of 1.39 \MSun.  High contrast imaging has been obtained by Mesa et al. with LBT-SHARK (in preparation) with no detection. The observed PMa, the lack of significant RUWE and of a detection in HCI strongly suggest the presence of a JL companion.

{\bf HIP15201} (iot Hyi, 2MASS J03155768-7723184): member of Carina with a mass of 1.41 \MSun. There is unpublished Spectro-Polarimetric High-contrast Exoplanet REsearch instrument (SPHERE) observation in the ESO archive by De Rosa \& Nielsen, that is still not public. There is no indication of binarity in the literature and no high precision RV data. The value of the PMa is consistent with RUWE and the non-detection of companions from Gaia over a wide range of semi-major axis (3-55 au) and mass (10-100 \MJup: see Figure \ref{fig:HIP29568}). Waiting confirmation from the SPHERE data (an upper limit would also be adequate), we will assume here that the companion is a JL planet.

{\bf HIP16449 (HR 1082)}: member of Columba with a mass of 1.72 \MSun. High contrast imaging was obtained by \citet{Esposito2020, Dahlqvist2022, Pearce2022} with no detection. A debris disk was detected by ALMA with inner radius of 68 au and outer radius of 120 au \citet{Higuchi2020, Pearce2022}. The value of the PMa, the lack of a significant RUWE and of a HCI detection, and the location of the debris disk strongly suggest that the companion is a JL planet.

{\bf HIP17675} (2MASS J03471061+5142230): northern member of Columba with a mass of 1.44 \MSun. There is no high precision RV data. The value of the PMa is consistent with the not significant RUWE and the non-detection of companions from Gaia over a wide range of semi-major axis (3-100 au) and mass (4-180 \MJup: see Figure \ref{fig:HIP29568}). Waiting confirmation from HCI data (an upper limit would also be adequate), we will assume here that the companion is a JL planet.

{\bf HIP22152} (2MASS J04460056+7636399): northern member of Columba with a mass of 1.21 \MSun. HCI has been obtained within the International Deep Planet Survey \citep{Galicher2016} and there is Hubble Space Telescope imaging from Song et al. unpublished, reported in \citet{Vigan2017}, with no detection. 
Within the constraints given by the lack of detection in HCI and the lack of a significant RUWE signal, the PMa is only compatible with a planetary companion, over a range of separation from 3 to 40 au. We will assume here that the companion is a JL planet.

{\bf HIP45585} (2MASS J09172755-7444045): member of Carina with a mass of 2.07 \MSun. The object was observed with SPHERE by \citet{Dahlqvist2022} with no detection of a companion and an upper limit of about 30-100 \MJup over the separation range 3-300 au. There are HARPS High precision RV data available \citep{Trifonov2020} over about 74 d that show a rather large scatter and a possible trend of 23 m/s/day. However, since RUWE is not significant, this trend in RV (if real) can only be attributed to a small mass companion very close to the star and is certainly unrelated to the PMa. The star hosts a debris disk whose spectral energy distribution suggests two belts with \teff$\sim 290$ and 140 K \citep{Chen2014, Kennedy2014}. These temperatures correspond to semimajor axes of about 5 and 17 au. If these disks signal the presence of a planet, this should be at about 8-10 au from the star; this value corresponds to where the observed value of the PMa is consistent with the upper limits set by RUWE and imaging. The mass of the candidate companion (20-25 \MJup)  corresponds to a small BD. However, given the high mass of the star and hence the low value of the mass ratio $q$, we may consider this object as a JL planet.

{\bf HIP47115} (2MASS J09360518-6457011): wide binary with the HIP47017 system in Volans/Crius 221; the far companion (itself a ternary system) is at a projected separation of about 40,000 au. HIP47115 was observed with NACO\footnote{Nasmyth Adaptive Optics System (NAOS) – Near-Infrared Imager and Spectrograph (CONICA) at the ESO Very Large Telescope.} in the L'-band (Program 091.C-0154) on April 14, 2013. Since data are unpublished, we downloaded them from the ESO archive and found that they show the presence of a stellar companion with $\Delta L'=2.18\pm 0.04$ mag at $1.053\pm 0.006$ arcsec (projected separation of 89 au) and PA=$169.3\pm 0.3$ degree. This companion should then be a G6V star with a mass of 0.97 \MSun. This object is very likely responsible for the observed PMa. In fact its PA agrees very well with that of the PMa ($172.2\pm 9.6$ degree), as expected for long periods, and its mass is larger by only a factor of two than that expected for an object responsible for the PMa at the observed separation, that is a reasonable agreement given the approximations considered by \citet{Kervella2022}. The star also has a debris disk with \teff=190 K \citet{Cotten2016} that corresponds to a distance of about 7 au from the star. This is at the limit of the stability region due to the stellar companion (that however can be further from the star than its projected separation). We conclude that our analysis rejects a JL planet as responsible for the PMa for this star.

{\bf HIP60831} (HD108574, 2MASS J12280445+4447394): northern member of Carina Near with a mass of 1.15 \MSun. It is a wide companion of HIP60832, with a projected separation of 440 au. High precision RV data obtained with the SARG high resolution spectrograph at the Italian Telescopio Nazionale Galileo (TNG) by our group are available \citep{Carolo2011}. They do not show any significant periodicity or trend over four years time span and have an r.m.s. of about 20 m/s compatible with the expected jitter. The stellar companion cannot be responsible for the PMa because this is much larger than expected for the large separation and directed towards a different direction - in fact we expect that for very long periods, the PMa be directed towards the object responsible for it. The stability limit leaves only a JL solution for the companion responsible of the PMa.

{\bf HIP60832} (HD108575, 2MASS J12280480+4447305): northern member of Carina Near with a mass of 1.04 \MSun. It is a wide companion of HIP60831. The stellar companion cannot be responsible for the PMa because this is much larger than expected for the large separation and directed towards a different direction - in fact we expect that for very long periods, the PMa be directed towards the object responsible for it. High precision RV data from SARG \citep{Carolo2011} and Elodie are available. They do not show any any significant periodicity or trend over four years time span and have an r.m.s. of about 20 m/s compatible with the expected jitter. The stability limit leaves only a JL solution for the companion responsible of the PMa.

Summarising, among these ten stars there is clear indication for the presence of a JL companion in six and a weak one (the object may either be a planet or a small mass star) for three, while our reduction of unpublished data revealed that for the last object the companion is stellar. Nine candidates are then remaining. 

In Table \ref{tab:jl_masses} we give masses and semimajor axis derived in a uniform way for all the candidate JL planets suggested by PMa. The procedure we followed is similar to the Monte Carlo one described in Section 4.2; however we considered here that stars with significant PMa but not significant RUWE and RV variation (as it is the case for all the ones considered here) should have semimajor axis $1<a<300$ au. We also considered the upper limits due to HCI, whenever available, as well as stability limits indicated by other companions and the presence of debris disks. The best values and the error bars are the mean and standard deviation of the solutions compatible with the observational data. In the last column of this table we report the fraction of the solutions compatible with observational data that correspond to JL planets. Here we assumed that JL planets are those planets that have $3<a<20$ au and a mass $<20$ \MJup. This fraction actually depends on the assumed distributions of $a$ and $q$ in the Monte Carlo approach. Here we assumed uniform distributions in the logarithm of $a$ and $q$. In 13 out of 17 cases this fraction is higher than 0.5, and for three others higher than 0.4. The lower fraction is for HIP52462, where however most of the solutions are at masses below 1 \MJup, that is, they are still giant planets not far from the snow line.

On the other hand, we notice that in addition to false positive (objects with significant PMa but no real JL planet) we should also consider the presence of false negatives (stars hosting planets but that do not have a significant PMa, such as 51 Eri and $\beta$ Pic). We do not know how many false negatives are present in our sample. Our approach is to consider that all the seventeen PMa candidates considered here are indeed JL planets. Overall, the fraction of stars hosting only uncertain planet over the total of 24 is limited ($\sim 12$\%) and should not change much the discussion. We further notice that in total our list includes 29 companions with mass $M\leq 0.020$ \MSun and separation larger than 1 au because more than one planet has been detected around some stars.

\subsection{Statistics of JL planets}


\begin{figure}[htb]
\centering
\includegraphics[width=8.0cm]{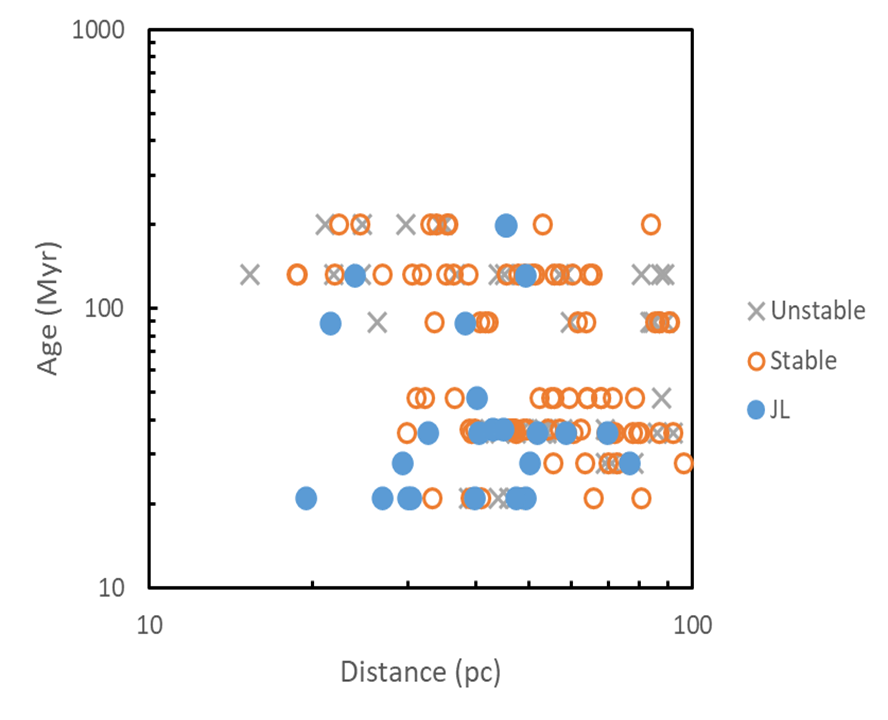}
\includegraphics[width=8.0cm]{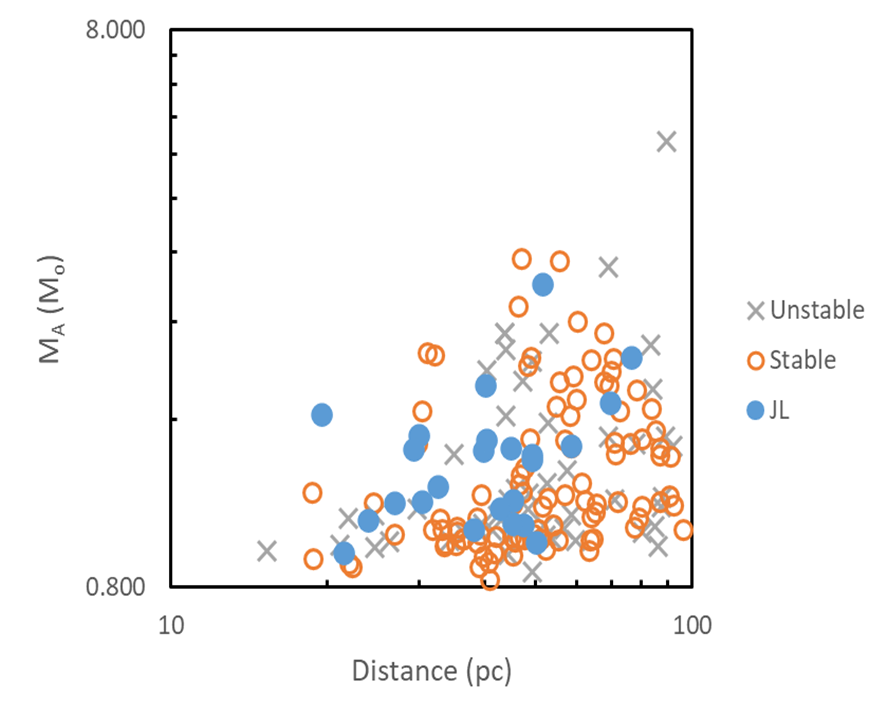}
\caption{Diagrams showing the stars found hosting with JL planets (blue filled circles) and those where there was no detection (open orange circles) in the distance vs age (upper panel) and distance vs mass of the primary (lower panel) planes. Only stars with PMA measures are plotted. In both panels, crosses mark stars that have stellar/BD companions that make the orbits of JL planets unstable. }
\label{fig:JL_companions}
\end{figure}

\begin{table*}
\centering
\caption{Frequency of JL planets}
\begin{tabular}{lcccccccc}
\hline
Association	& Stars PMA	&Stable	& Fraction	& Detected	&	Dyn	&	Total	&$N_{\rm eff}$ & Freq.		\\
	&		&		&	stable	&	JL comp	&		&	JL comp	&	&	\\
\hline
AB Dor	&	35	&	21	&	0.60	&	1	&	1	&	2	 &	8.41	&	0.24$\pm$0.16	\\
Argus	&	15	&	15	&	1.00	&	1	&	 	&	1	 &	2.65	&	0.38$\pm$0.38	\\
BPMG	&	22	&	15	&	0.68	&	3	&	4	&	7	 &	5.18	&	1.35$\pm$0.37	\\
Carina	&	12	&	9	&	0.75	&	1	&	2	&	3	&	2.03	&	1.48$\pm$0.70	\\
Carina Near&15	&	10	&	0.67	&		&	2	&	2	&	3.42	&	0.58$\pm$0.37	\\
Columba	&	30	&	21	&	0.70	&	2	&	3	&	5	 &	6.11	&	0.82$\pm$0.32	\\
Tuc-Hor	&	30	&	18	&	0.60	&		&	2	&	2	 &	7.36	&	0.27$\pm$0.18	\\
Volans/Crius 221&20	&10	&	0.50	&		&	2	&	2	&	6.61	&	0.30$\pm$0.19	\\
\hline
\end{tabular}
\label{tab:jl_frequency}
\end{table*}

Table \ref{tab:jl_frequency} gives the frequency of JL planets around the various associations. To derive this number we first considered the stars that have values of the PMa in the catalogue of \citet{Kervella2022}. This is a subset of the stars that are in the Hipparcos catalogue because for a few of them there is no astrometric solution in the Gaia DR3 catalogue. These few stars are likely binaries with large residuals in the astrometric solution. Likely most of these objects cannot have a JL planet in stable orbit because of the presence of massive nearby companions. We considered a total of 175 stars with this PMa available.

We then used the information about companions more massive than JL planets collected in the previous section to examine the possible stability of the orbits of JL planets. To this purpose we first estimated the inner and outer radii of the instability regions around these companions using the formulation by \citet{Holman1999}. We assumed here an orbit eccentricity given by equation (1) in \citet{Gratton2023b}. Driven by the results obtained this way, we assumed that the orbit of a JL planets would be unstable around any star for which $a_{\rm max}>1.5$ au or $a_{\rm min}<20$ au. Here, $a_{\rm min}$ and $a_{\rm max}$ are the inner and outer edges of the region where orbits are unstable due to the perturbations by a companion. We found a total of 121 stars (that is 69\% of the total) with PMa values that passed this criterion.

We found in the literature detection of JL planets through HCI around seven of these stars. In addition, we found additional systems where there is dynamical indication for the presence of a JL planet, in most cases from PMa (we considered meaningful those cases where S/N(PMA)$>3$). After culling cases in which the PMa can be attributed to other known companions, we finally have a total of 17 additional objects that likely are JL companions. So, in total there is indication of the presence of JL companions around 24 stars, that is $20\pm 4$\% of the stars that might potentially have a JL planet.

Most of the JL planets are detected using the PMa, with a few additional ones coming from HCI. However, only a fraction of the JL planets can be detected using existing data. This is clearly shown in Figure \ref{fig:JL_companions} that shows the stars found hosting with JL planets and those where there was no detection in the distance vs age and distance vs mass of the primary planes. Only stars with PMA measures are plotted; data for the remaining stars are too incomplete, and in fact there was no detection around them. In both panels, crosses mark stars that have stellar/BD companions that make the orbits of JL planets unstable. Detected planets are mainly around the closest, youngest, and most massive primaries. Distance is the most relevant parameter: while JL planets are found around 19 out of the 66 stars with distance lower than 50 pc where their orbits are stable, the similar fraction for stars beyond 50 pc is five out of 55. In addition, detected planets are generally massive, in most cases with a mass larger than 3 \MJup; only a fraction of the JL planets is expected to be so massive. The fraction of planets that can be detected this way depends on the limiting mass $M_{min,i}$ for this technique, that on turn depends on many factors, mainly distance from the Sun. An approximate value for this limiting mass for each star is given by the mass corresponding to a S/N(PMa)=3 at a separation in the range of JL planets given in the tables by \citet{Kervella2022}. We considered here an average of the values at 5 and 10 au, that should be representative for JL planets. We then considered the fraction of JL planets around each star that are expected to be above this limiting mass. This depends on the assumed mass function $f(m)$ for JL planets. We adopted a mass function that is a power law with a slope of -1.3 as proposed by \citet{Adams2021}, the same considered in \citet{Gratton2023b}. We then call $N_{\rm eff}$ the quantity:
\begin{equation}
N_{\rm eff} = \sum{\frac{\int_{M_{min,i}}^{13}f(m)dm}{\int_{1}^{13}f(m)dm}},
\end{equation}
where the index $i$ refers to each star and $m$ is in units of \MJup. $N_{\rm eff}$ represents the effective number of stars around which JL planets could be found with the PMa technique for each association. We assumed this is also the number of stars where we could detect JL planets including HCI too.

Summing all associations, in total we have $N_{\rm eff} =41.77$. So our estimate of the frequency of stars hosting JL planets in these association is 24/$N_{\rm eff}=0.57\pm 0.11$. This frequency is much higher than that obtained from RV surveys. It is however lower than that obtained for the case of the BPMG by \citet{Gratton2023b}. We also notice that for some of the associations, the nominal value of the frequency of JL planets is $>1$, though never in a statistically significant way. This may happen because in the actual realisation of the extraction of companions from the mass distribution, they may have randomly masses larger than the average value when the sample is small.

\begin{figure}[ht]
\centering
\includegraphics[width=\linewidth]{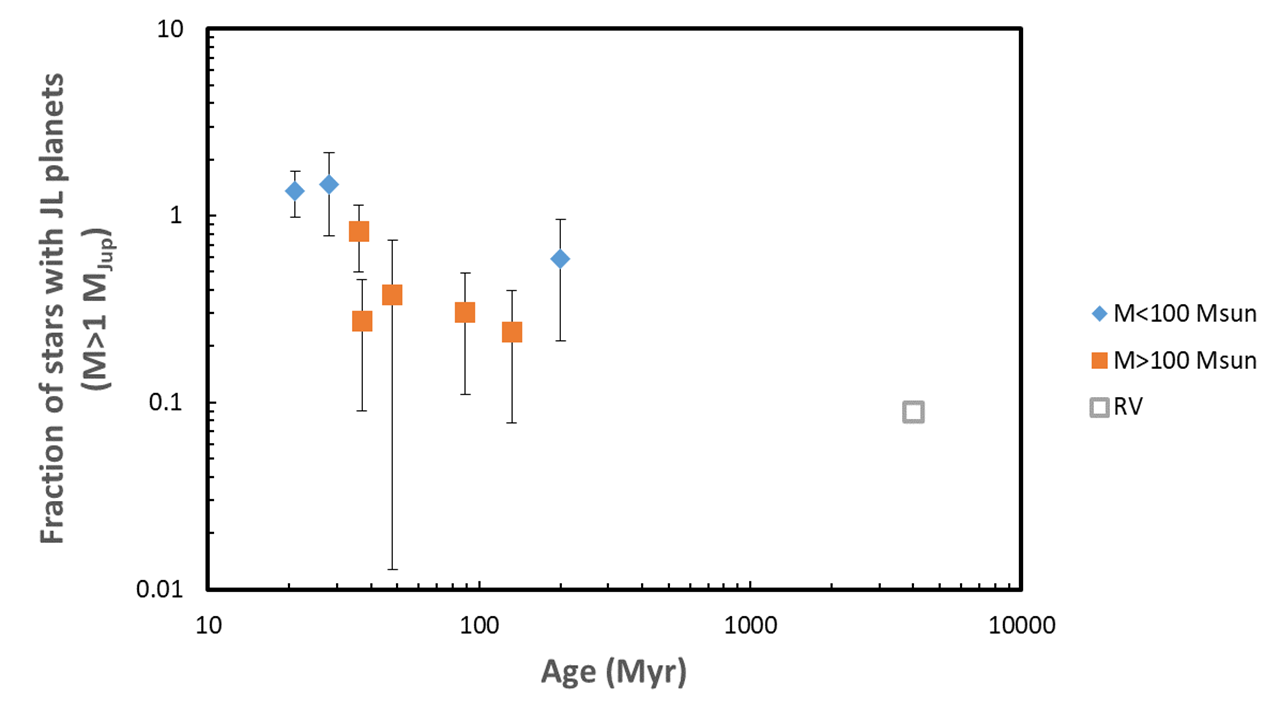}
\caption{Frequency of stars hosting JL planets (corrected for completeness) in individual associations as a function of age. Blue dots are associations with a total mass $<100$ \MSun; orange squares are associations with a total mass $>100$ \MSun. The open square represents the frequency of stars hosting JL planets from the RV surveys. We arbitrarily assumed an age of 4 Gyr as typical for the stars in these surveys.
}
\label{fig:frequency_jl_age}
\end{figure}

\begin{figure}[htb]
\centering
\includegraphics[width=\linewidth]{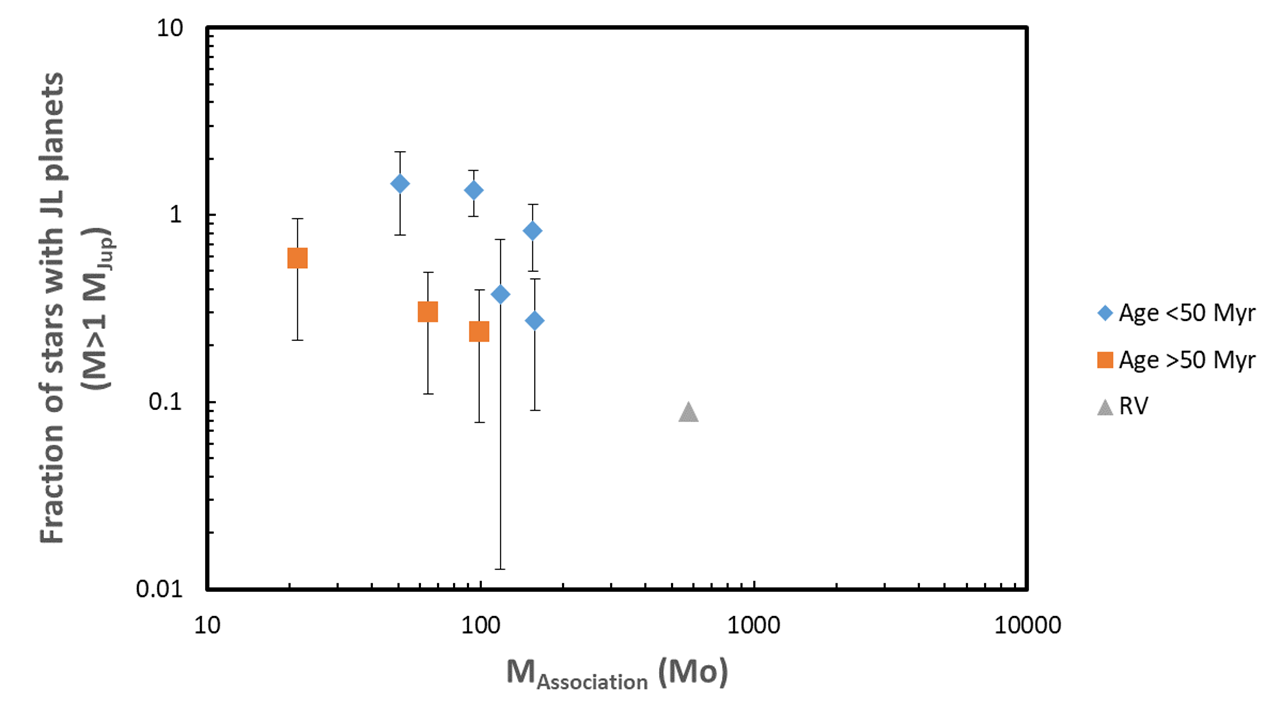}
\includegraphics[width=\linewidth]{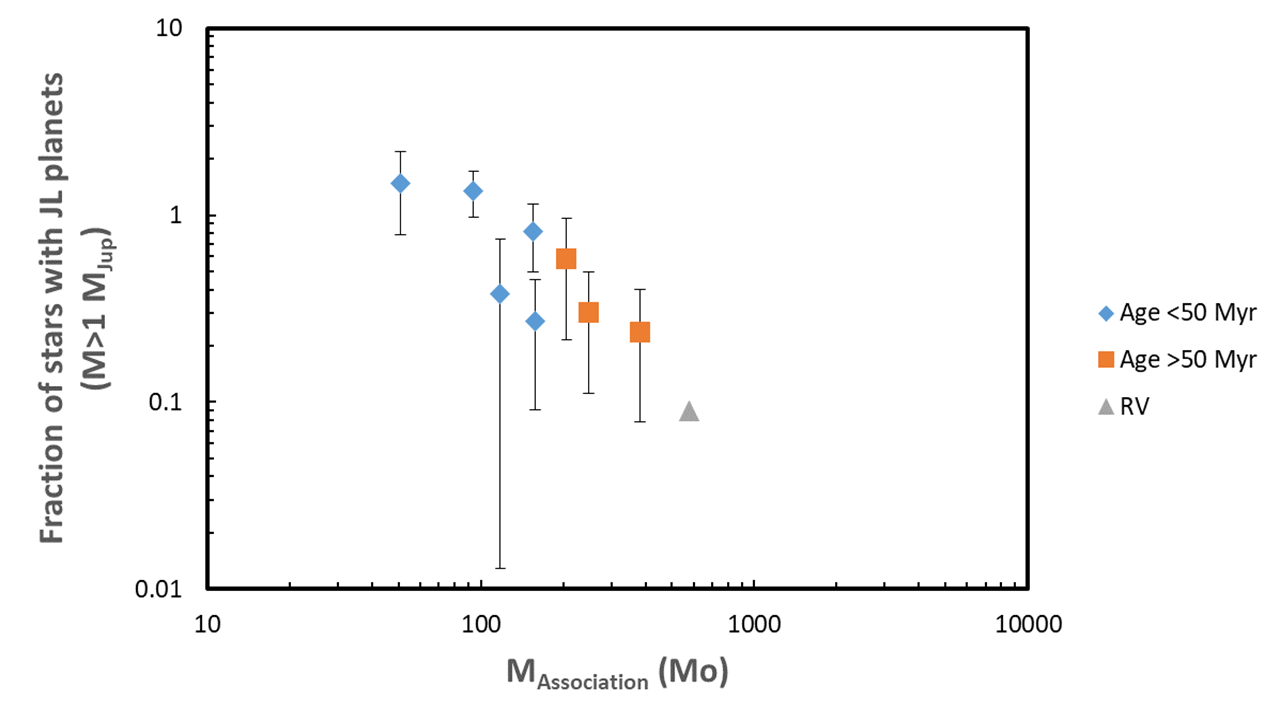}
\caption{Frequency of stars hosting JL planets (corrected for completeness) in individual associations as a function of association mass. Upper panel: Masses as obtained by summing the mass of the member stars. Bottom panel: Masses corrected for the selection along the galactic Y coordinate. Blue dots are associations with an age $<50$ Myr; orange squares are associations with a total mass $>50$ Myr. The open triangle represents the frequency of stars hosting JL planets from the RV surveys. We arbitrarily assumed a mass of 1000 \MSun for the typical birthplace of stars in these surveys.
}
\label{fig:frequency_jl_mass}
\end{figure}

In order to better explore this point, we plotted in Figures \ref{fig:frequency_jl_age} and \ref{fig:frequency_jl_mass} the runs of the frequency of JL planets for each individual associations as a function of age and mass. 

We obtain trends for decreasing frequency of JL planets with increasing age and mass. These trends are suggestive of real physical effects, that are even more obvious when compared to the results from RV surveys that refers to stars several Gyr old (we assumed an age of 4 Gyr in the plot) and likely mostly formed in rich groups with thousands of stars or more (we assumed a typical value of 1000 \MSun in the plot).

To quantify these hints, we obtained the best relation through the observed points in the frequency $f$ of JL planets vs. mass $m$ (in \MSun) plane. We did not consider the results obtained from RV surveys, because of their very different meaning. The best fit is:
\begin{equation}
\log{f} = -(0.05\pm 0.05)-( 0.70\pm 0.27) (\log{m}-2.0)
\end{equation}
The trend with mass is then significant at about 2.6$\sigma$. The reduced $\chi^2$ value is 1.69. If we assume higher masses by a factor of two for the associations with ages $>50$ Myr (AB Dor, Carina Near, and Volans/Crius 221) as proposed in Section 2.1.4 we obtain the equation:
\begin{equation}
\log{f} = (0.02\pm 0.03)-(1.01 \pm 0.14) (\log{m}-2.0)
\end{equation}
The trend is now highly significant at 7.5$\sigma$ and the reduced $\chi^2$ value is 2.78. 

The best relation through the observed points in the frequency $f$ vs. age $t$ (in Myr) plane is:
\begin{equation}
\log{f} = -(0.69\pm 0.13)-( 0.37\pm 0.07) (\log{t}-2.0).
\end{equation}
The trend with age is then significant at about 5.1$\sigma$ and the reduced $\chi^2$ value is 0.87. However, the dependence of the frequency of JL planets on age may be in part an artefact of observational biases because detection with direct imaging is easier for very young system. We then repeated the analysis but only considering those detections obtained through dynamics, that are insensitive to age. The planets around $\beta$ Pic, 51 Eri, and $\kappa$ And are only detected in imaging, the S/N of the PMa being $<3$ and should then be dropped from the analysis. In this case, the best fit when higher masses are considered for the older association is:
\begin{equation}
\log{f} = -(0.54\pm 0.14)-( 0.38\pm 0.07) (\log{t}-2.0).
\end{equation}
The trend in age would still be significant (at 4.0$\sigma$) and the reduced $\chi^2$ value is 0.86.

However, it is possible that there is some correlation between age and mass and that the trend with age is an artefact of this. We then further explored a possible linear combination of the dependencies on age and mass. In this case, the best relation is:
\begin{multline}
\log{f} = -(0.10\pm 0.14)-(0.78 \pm 0.30) (\log{m}-2.0)\\
-( 0.21 \pm 0.27) (\log{t}-2.0).
\end{multline}
The reduced $\chi^2$ value is now 0.62. We notice that the trend with mass is still significant at more than 2.5$\sigma$, while that on age is not larger than its error bar.

We conclude that the frequency of JL planets in young associations is much higher than obtained from RV surveys and that it depends on the total mass and possibly the age of the association. However data are not enough to precisely determine these dependencies.

\FloatBarrier

\subsection{Hot planets}


As mentioned in Section 2.5, around the stars considered in this paper there is no known hot-Jupiter, a hot-Neptune transiting DS Tuc A ($5.6\pm 0.2$ R$_{\rm Earth}$, period 8.1 d, and mass $<0.0453$ \MJup, that is $<14$ M$_{\rm Earth}$: \citealt{Newton2019, Benatti2019}) in the Tuc-Hor moving group, and a mini-Neptune around HIP 94235 ($3.00^{+0.32}_{-0.28}$ R$_{\rm Earth}$, period 7.7 d: \citealt{Zhou2022}) in the AB Dor moving group. The mass is not available for this last planet. It is not possible to draw firm conclusions about the frequency of these classes of objects in the sample from such sparse detections. However, we may add a few comments here.

We first have to discuss the detection efficiency. \citet{Mayor2011} estimated that there is a 70-80\%  probability that any hot-Jupiter would have been detected if present with high-precision RV series (available for 79 stars, 28\% of the stars surveyed). According to \citet{Fernandes2023} they would almost certainly be detected by TESS if transiting (2-minute cadence data available for 248 stars, 91\% of the stars surveyed), though transits are expected for about 1/10 of the hot Jupiters or less (depending on the period distribution). No detection over this sample is then not inconsistent with the overall frequency of hot Jupiters. In fact, \citet{Wright2012} estimated an overall frequency of $1.20\pm 0.38$\% for solar-type stars. More recently, \citet{Zhou2019} found lower occurrence rates of $0.26\pm 0.11$\% for A stars, $0.43\pm 0.15$\% for F-stars, and $0.71\pm 0.31$\% for G-stars. 

For what concerns the hot-Neptunes, they are very difficult to detect using RV in very active stars such as those considered here; even the TESS search is likely incomplete for this reason. The frequency of hot planets in 31 young clusters and associations with a median age of 45–50 Myr (including some of those considered in this paper) is discussed in \citet{Fernandes2022, Fernandes2023} who found a frequency of $90\pm 37$\% of sub-Neptunes/Neptunes (1.8–6 R$_{\rm Earth}$) with orbital periods less than 12.5 days for the the stars for which they could compute stellar properties. This frequency is higher than Kepler’s Gyr old occurrence rate of $6.8\pm 0.3$\% even when accounting for evaporated sub-Neptunes( \citep{Bergsten2022}. 

When comparing this occurrence rates with those for older populations, we should consider that young hot planets are likely inflated \citep{Benatti2019} and then of smaller mass than indicated by their radius when using standard relations such as that by \citet{Otegi2020}. For instance, in the case of DS Tuc b the radius corresponding to the upper limit of the mass ($<14$ M$_{\rm Earth}$) is 3.7 R$_{\rm Earth}$ using the appropriate relation by \citet{Otegi2020}, while the observed radius is $5.6\pm 0.2$ R$_{\rm Earth}$. The mass of the mini-Neptune around HIP 94235 is not known, but it may well be below 10 M$_{\rm Earth}$.

We may then notice that \citet{Kunimoto2021} obtained an analytic form for the distribution of (mainly old) planets with mass and periods combining Kepler transits and results from RV surveys. Using their equations, we derive an expected occurrence frequency of 24\% for planets with mass between 10 and 60 M$_{\rm Earth}$ and period less than 20 days. Considering the caveat given above, this value is well consistent with the current determination for young associations.

\section{discussion}

\subsection{Understanding the properties of stellar binaries in young associations}

\subsubsection{Comparison with binary formation model}


\begin{figure*}[htb]
    \centering
    \includegraphics[width=18cm]{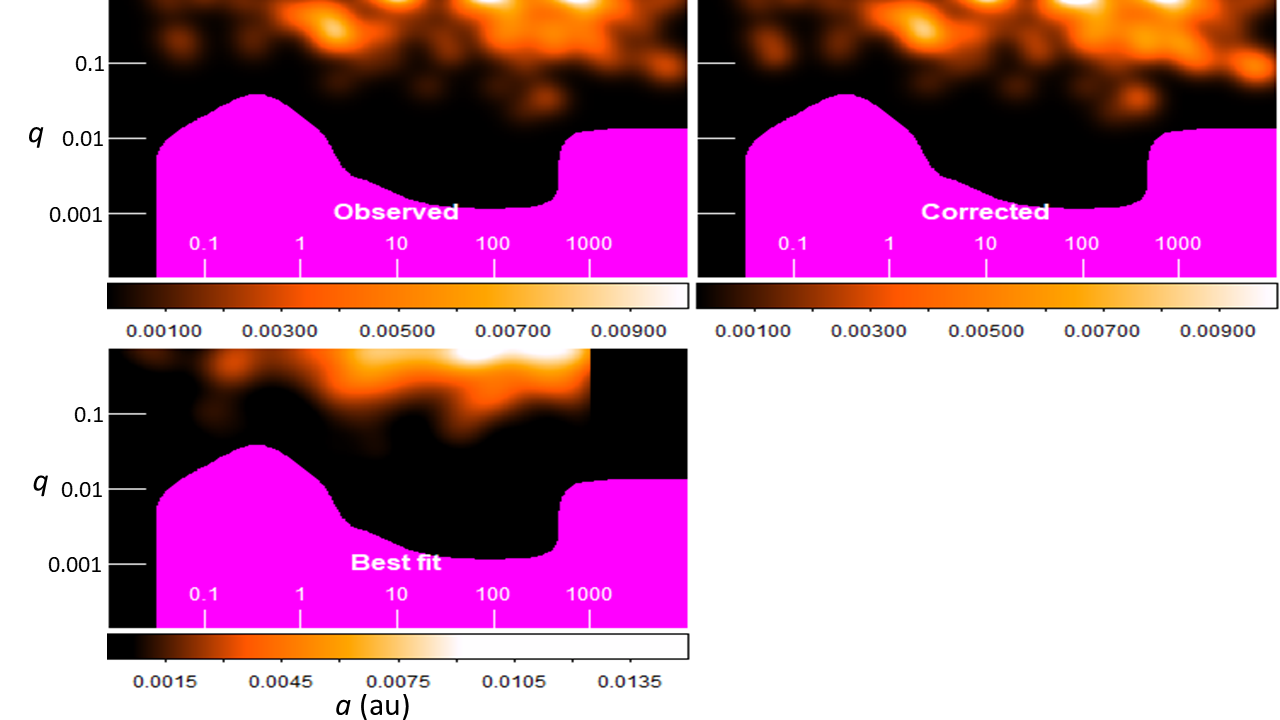}
  \caption{Comparison between observed (upper left panel), corrected for completeness (upper right panel) and model (bottom panel) maps of the smoothed distribution of companions in the separation (in au) vs mass ratio plane. We consider here the maps obtained considering only the closest companion, for consistency between the models and the observations. The magenta area marks the region with completeness $<0.2$, not used in the analysis}
    \label{fig:simu}
\end{figure*}

\begin{figure}[htb]
    \centering
    \includegraphics[width=8.5cm]{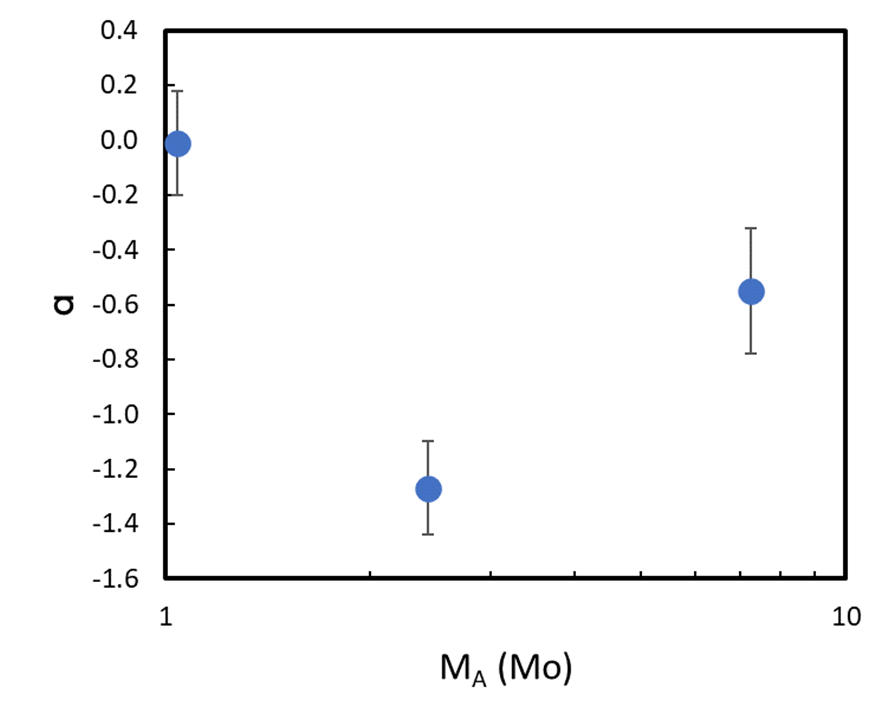}
    \includegraphics[width=8.5cm]{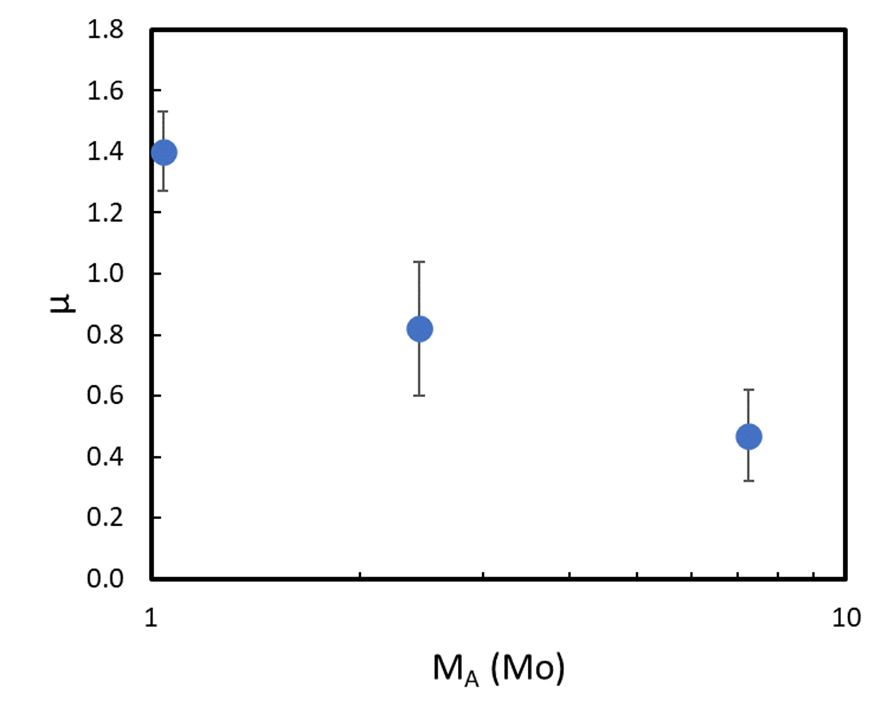}
    \caption{Upper panel: Run of the toy-model parameter $\alpha$ (exponent of the power-law of the mass of accreting events on the disk) as a function of the mass of the primaries $M_A$ from this paper and \citet{Gratton2023}. A lower value means that early episodes are more powerful. Lower panel: Same for the toy-model parameter $\mu$ (relevance of outward migration). A lower value means that outward migration is more relevant }
    \label{fig:mubeta}
\end{figure}

\begin{table}[htb]
  \caption[]{Best parameters for the toy model.}
  \label{t:model_parameter}
  \begin{tabular}{lcc}
  \hline
Parameter         & Prior Range &      All      \\
  \hline
$n_{\rm max}$      & [20, 60]    &   $37\pm 10$  \\
$\alpha$           & [-1.0, 1.0] &$-0.01\pm 0.19$\\
$d_{\rm min}$ (au) & [30, 100]    &  $62\pm 18$ \\
$d_{\rm max}$ (au) & [200, 1200]  & $750\pm 297$ \\
$\mu$              & [0.5, 2.0]   & $1.40\pm 0.13$\\
$\beta$            & [0.0, 1.0]   & $0.61\pm 0.27$\\
  \hline
  \end{tabular}
\end{table}

\begin{figure}[htb]
    \centering
    \includegraphics[width=8.5cm]{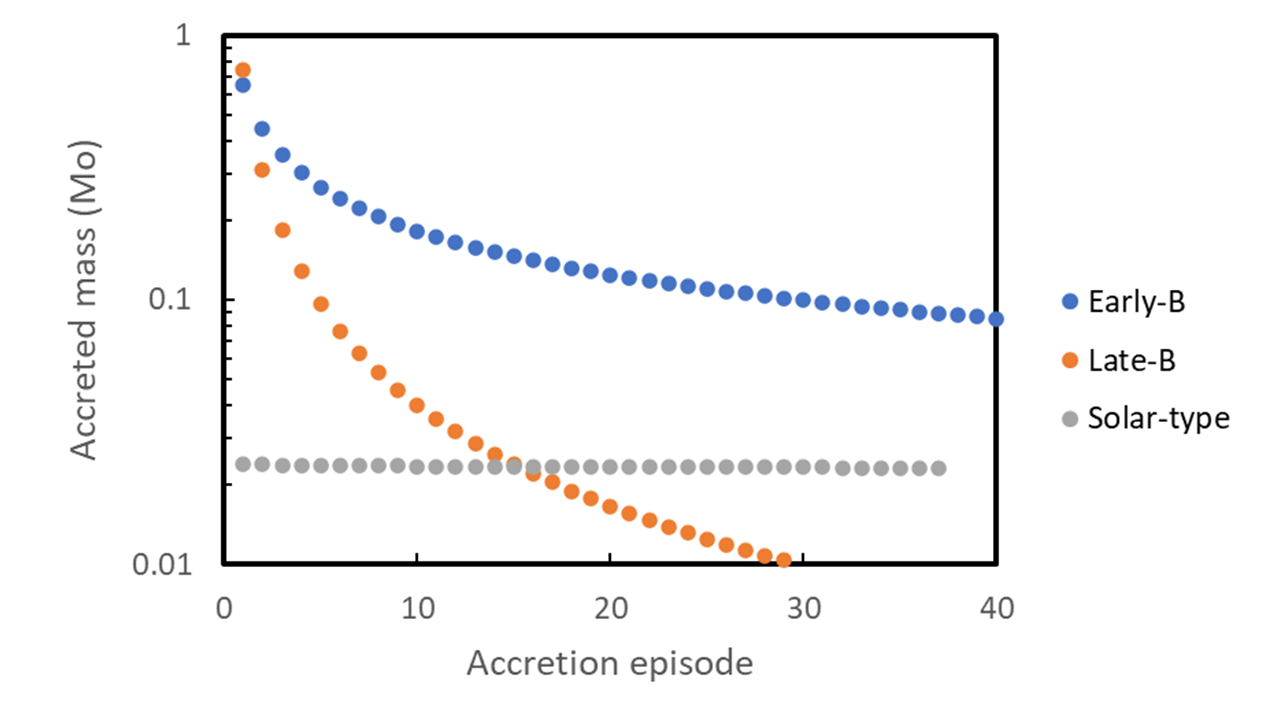}
    \caption{Run of the mass accreted from the interstellar matter to the disk with the episode of accretion for the best-fit parameters of the toy model for the formation of binaries by disk instability appropriate for early-B ($M_A=7.25$ \MSun), late-B ($M_A=2.43$ \MSun), and solar-type stars ($M_A=1.04$ \MSun). }
    \label{fig:accretion}
\end{figure}

\begin{figure}[htb]
    \centering
    \includegraphics[width=8.5cm]{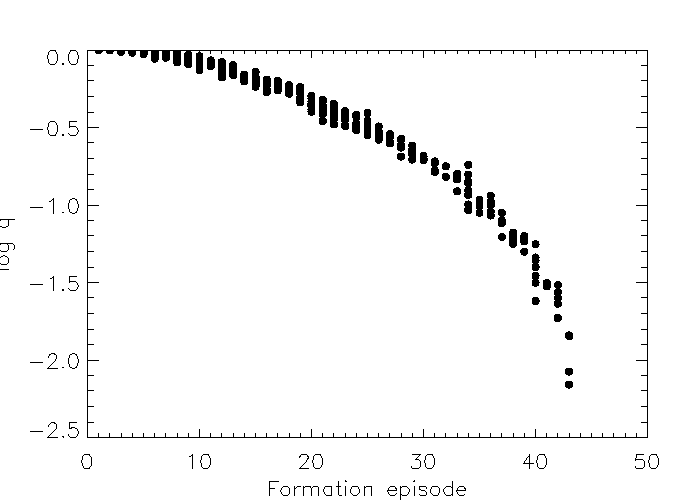}
    \caption{Run of the final mass ratio $q$ of a binary with the episode of accretion where they formed for the best-fit parameters of the toy model for the formation of binaries by disk instability appropriate for solar-type stars ($M_A=1.04$ \MSun). }
    \label{fig:formationq}
\end{figure}

In order to discuss the properties of the stellar companions found around the stars in young associations, we considered the same toy model considered by \citet{Gratton2023} for the binaries around B-stars in the Sco-Cen associations. As in that case, we may assume that this distribution mainly reflects the characteristics of binary systems at birth. The toy model considered by \citet{Gratton2023} is essentially similar to that formulated by \citet{Tokovinin2020} and it is appropriate to describe the formation of binaries by disk fragmentation. In this toy model the complex physics of binary formation is described by several distribution function whose parameters may be determined by matching the observed binary distribution with predictions by the models. \citet{Gratton2023} considered six different parameters: the total number $n_{\rm max}$  of accretion events onto the disk; the exponent $\alpha$  of the power-law of the mass of accreting events on the disk; the minimum $d_{\rm min}$ and maximum $d_{\rm max}$ radii of the disk; the exponent $\mu$ giving the relevance of outward migration; and the parameter $\beta$ of the relative accretion of secondary and primary. A Monte Carlo procedure was devised to select which values of the parameters allows a good fit between computed and observed distributions. The comparison was  done only considering the closest companion to the star that have $a<1000$ au, because widest binaries likely formed by other mechanism (cloud fragmentation). \citet{Gratton2023} found that only two of the model parameters really constrain the observed distributions: the exponent $\mu$ of the power-law of the mass of accreting events on the disk; and the parameter $\beta$ giving the relevance of outward migration. These parameters have been found to be systematically different for early and late-B stars in Sco-Cen.

The comparison (see Figure~\ref{fig:simu}) is made using maps of the smoothed distribution of companions in the $a - q$\ plane; these maps were obtained replacing the point relative to each companion with a Gaussian distribution with a sigma equal to 0.2 dex. The upper panels of Figure~\ref{fig:simu} show the maps we obtained in this way considering only the closest companions for our sample of 141 binary systems. This smoothed distribution is then compared with analogue distributions obtained using the model. For each set of parameters $n_{\rm max}$, $\mu$, $\alpha$, $d_{\rm min}$, $d_{\rm max}$, and $\beta$, we run the model 2,000 times, each time simulating the evolution of 10,000 systems, to define with a reasonable accuracy the distribution of companions. With these data we constructed maps of distributions in the $\log{a}-\log{q}$ plane, after applying the same smoothing to observing data. Once normalised to the total populations, we may compare these model maps with the observed ones and define a suitable goodness of fit parameter; in practice, we considered the mean quadratic residual between models and observation $r$. We considered acceptable those cases giving the best 1\% of the values of $r$. We assumed that the average values for each parameter over the acceptable solutions is its best guess and the $\sigma$ is its uncertainty.

Table \ref{t:model_parameter} gives the result of our analysis. Notice that this model produces a fraction of single stars of 41\%, close to the observed value of $<48\pm 3$\%. When compared to the results obtained by \citet{Gratton2023} for the B-stars in the Sco-Cen association, it should be considered that the typical mass of the stars considered in this paper is much lower than that of the B-stars, with a typical value of 1.04 \MSun. This is to be compared with typical values of about 2.43 \MSun for the late-B stars and about 7.25 \MSun for the early ones. We found that the properties of the companions of the solar-type stars in young associations yield higher values of $\alpha$ and $\mu$ than those of the B-stars (see Figure \ref{fig:mubeta}). The trend of $\alpha$ with mass implies that many episodes with similar accreted mass are needed for lower mass stars. This is illustrated by Figure~\ref{fig:accretion} that shows the run of the mass accreted from the interstellar matter to the disk with episode of accretion for the best fit parameters of the toy model for the formation of binaries by disk instability appropriate for early-B ($M_A=7.25$ \MSun), late-B ($M_A=2.43$ \MSun), and solar-type stars ($M_A=1.04$ \MSun). This run of the accretion onto the disk makes the probability of secondary formation in the early events for massive stars very high, which explains the very high frequency of binaries observed. This agrees very well with the concept that disk fragmentation is more efficient around massive stars because of the larger value of the accretion rate from the natal cloud and the larger expected disk-to-star mass ratio during early phases of formation, when binaries are likely to form \citep{Machida2010, Kratter2016, Elbakyan2023}. Binary that formed so early are much more likely to become close binaries due to migration and nearly equal mass binaries due to selective accretion than those formed later. The trend of $\mu$ with mass implies that inward migration is more pronounced in most massive binaries; this again can be due to the very large initial disk mass. On the other hand, less massive binaries should have only a small prevalence of inward with respect to outward migration. This agrees with the rather large value for the median semi-major axis of the binaries.

In this picture, low-mass companions (BDs) may originate due to a disk instability occurring in the very late episodes of accretion from the interstellar matter onto the disk, and then have no chance to further accrete mass and become a stellar companion. This is clearly shown in Figure~\ref{fig:formationq} that shows the run of the final mass ratio $q$ of a binary with episode of accretion for the best fit parameters of the toy model for the formation of binaries by disk instability appropriate for solar-type stars. BDs should be rare because the disk-to-star mass ratio and hence the probability of the onset of the instability is typically low in these late phases. In the model example shown above that produces a fraction of 59\% of multiples stars, only 1.2\% of the stars have a BD companion ($M_B<0.075$ \MSun), at a median separation of 90 au from the star. The fraction of BD companions expected to be detected by RV surveys is even lower; those with semi-major axis $a<10$ au is only 0.27\%. This agrees with the upper limit of 0.6\% for the frequency of BD companions around Sun-like stars obtained by \citet{Sahlmann2011}.

We conclude that a simple model of disk fragmentation such as the one we are considering allows to describe the main statistical properties of the observed distribution of those systems within 1000 au from the star. This model generally produces stellar companions, though in rare case massive BDs can be produced.

\subsubsection{ Comparison between the distribution at large separations with that for free floating objects}

Companions at separation farther than 500-1000 au are expected to form through instabilities in star-forming clouds. It is then expected that they have a different mass distribution from companions forming in the disk around primaries because in the last case the onset of the instability requires a larger minimum mass that is proportional to the stellar mass. Rather, it is expected that the mass function of these far companions resembles that of objects formed in isolation. We may test if this occurs for our sample.

The median mass of the 62 stellar and BD companions with semi-major axis $a>500$ au is 0.528 \MSun, that is very close to the value of 0.512 \MSun obtained for the 124 closer companions. The same result is obtained if we only consider companions with separation larger than 1000 au or even 10000 au. That is, in our sample there is no obvious trend of the mass of the stellar companions with separation. 

For comparison, using data by \citet{Luhman2022} we estimate that approximately $4750/6000\sim 79$\% of the stars of Sco-Cen are in the mass range $0.072<M<0.5$~M$_\odot$. This last result agrees with the expectation for the \citet{Chabrier2003, Chabrier2005} IMF's. If we then use this IMF and the detection efficiency of Gaia for these companions (see Figure~\ref{fig:detection_vs_completeness}), we find that the frequency of objects in this mass range should be 70\% in our sample, much more than the $50\pm 6$\%  observed. This indicates that there are less low-mass objects among far companions to the solar-type stars in these young associations than expected with the \citet{Chabrier2003, Chabrier2005} IMF's and observed for free-floating objects in Sco-Cen.

This result is quite unexpected and might signal that a large fraction of the far stellar companions of stars in young associations  formed within the disk and were lately ejected towards larger separations. Alternatively, it might indicate that a large fraction of the substellar and low mass objects found in the field  originated in multiple systems and it was later lost to the general field (see \citealt{2023Ap&SS.368...17M, 2022NatAs...6...89M} for similar conclusions). This is indeed a more palatable choice, given the higher typical density of the environments where most of the field stars likely formed. As for the case of the JL frequency discussed below, the young associations would then have characteristics that are more similar to those reproduced in model computations than the general field.

\subsection{Dependence of JL planet frequency on the mass and age of the association}

As discussed in \citet{Gratton2023b}, the higher frequency of JL planets in young associations with respect to that obtained from RV surveys may have different explanations. The frequency of giant planets is known to depend on stellar mass and metallicity (see e.g. \citealt{Sozzetti2009, Johnson2010, Santos2011, Mortier2012, Barbato2019}). However, the metallicity of the young associations is roughly solar \citep{VianaAlmeida2009, Biazzo2012, Barenfeld2013, Baratella2020} and the average mass of the stars considered within young associations (1.04 \MSun) is only mildly larger than that of the Sun. Hence the mass and metallicity dependencies are not likely to be the main reason of the observed difference. Rather, the dependencies on mass of the associations and on their age suggests that more JL form in smaller star forming regions, likely because of the presence of less disturbing neighbours, and that perhaps they are lost from the stars as the associations age. 

This suggests that in absence of disturbance by other stars, the natural outcome of the evolution of a protoplanetary disk is the formation of JL planets. They are not very frequent as given by this prediction because most stars do not form in isolation. As reviewed by \citet{Parker2020}, there are essentially three density regimes that pose a threat to planetary system formation and evolution. In the presence of massive stars (and then, depending on the overall mass of the association), far ultraviolet (FUV) and extreme ultraviolet (EUV) radiation can affect protoplanetary disks at stellar densities as low as 10 \MSun pc$^{-3}$. At stellar densities greater than 100 \MSun pc$^{-3}$, planetary orbits can be altered (which can include changes to the orbital eccentricity, inclination and semimajor axis of individual planets). At higher densities still (greater than 1000 \MSun pc$^{-3}$) protoplanetary disks can be physically truncated by encounters.

In addition, the associations considered in this paper are also young while the RV surveys target old stars with ages of billions of years. In the long run (hundreds of million to billions of years) a fraction of the planetary systems may be destabilised by close passage of other stars. In the following subsections we will consider separately these two mechanisms.

\subsubsection{Impact of nearby massive stars on the formation of JL planets} 

In the core accretion scenario \citep{Pollack1996, Ida2004, Mordasini2009}, the disk must survive for at least a few million of years in order to form giant planets. According to the planet formation simulations by \citet{Mordasini2012} there should be virtually no giant planet around solar mass stars if the disk lifetime is shorter than 1.5 Myr, while they may be present around 30-40\% of the stars for lifetimes of 5 Myr. 

Environment has likely a big impact for disk survival through various mechanism. The lifetime of disks is limited by encounters (around more massive stars) and photo-evaporation (around low-mass stars, $M<0.5$~M$_\odot$) by nearby massive stars. Various authors considered the impact of these effects \citep{Adams2001, Adams2004, Winter2018, Parker2021, ConchaRamirez2021, Winter2022, Wright2022}. \citet{Adams2001} proposed that encounters potentially disruptive of Solar Systems are frequent for very young clusters with an initial population larger than 1000 stars, while there should be virtually no effect if the population is as low as 100 stars. \citet{Winter2018} found a canonical threshold for the local stellar density ($n\geq 10^4$~pc$^{-3}$, that is however much larger than typical of associations, \citealt{Wright2022}) for which encounters can play a significant role in shaping the distribution of protoplanetary disk radii over a time-scale of about 3~Myr. \citet{ConchaRamirez2021} simulated the effects of photo-evaporation and estimated that a local stellar density lower than 100 stars pc$^{-2}$ is necessary for disks massive enough to form planets to survive for 2.0 Myr. In their simulations there is an order of magnitude difference in the disk masses in regions of projected density 100 versus 10$^4$ stars pc$^{-2}$. This agrees with observations. In fact, surveys of protoplanetary disks in star-forming regions show that disks closer to bright stars are less massive and more compact than their counterparts in sparser regions \citep{Mann2009,Fang2012,Mann2014,Ansdell2017,vanTerwisga2020}. 

We now briefly investigate how many massive disk-enemy stars exists in these associations. The mass function of the primaries for mass higher than 0.8 \MSun is well reproduced by a Salpeter mass function (slope of -2.24). However, there are few massive stars. In total there are 16 B-stars with a mass $M>2.4$ \MSun; this is to be compared with a total of about 1300 stars and an overall mass of about 1000 \MSun summing all the associations considered in this paper together. According to data available, the maximum mass of a star in these associations (the precursor of the massive WD GD50 in the AB Dor moving group, \citealt{Gagne2018e}) is 7.8 \MSun and there are only three additional stars above 5 \MSun. 
The frequency of B-stars (mass $>2.4$ \MSun: $1.2\pm 0.3$\%) is lower than that observed in the Sco-Cen association, where there are 182 B-stars over about 6000 stars (about $3.0\pm 0.2$\%) and 54 of them (about 30\%) with a primary more massive than 5 \MSun \citep{Gratton2023}. The most massive star in Sco-Cen might have been the precursor of PSR J1932+1059 with a mass of possibly 50 \MSun  \citep{Hoogerwerf2001}.

As a typical high density environment, we may consider the Orion Nebula Cluster. The total mass in stars is about 1800 \MSun \citep{Hillenbrand1998} and there are some 1750 members \citep{DaRio2012}. \citet{Hillenbrand1998} estimated that the core radius of the cluster is 0.16-0.21 pc and that the central stellar density approaches $2 \times 10^4$ stars pc$^{-3}$. There are 40 stars more massive than 5 \MSun \citep{Pflamm-Altenburg2006} and the maximum stellar mass is of 45.7 \MSun. 

There is a large difference in the high-mass content of the nearby young moving groups and the Orion Nebula Cluster. While the total mass is comparable to the sum of all nearby young associations, the number of massive stars and the mass of the most massive star are an order of magnitude larger. B-stars are about 1\% of the stars in the young moving groups, 3\% in Sco-Cen, and about 7\% in the Orion Nebula Cluster.

We may obtain a rough estimate of the relative impact of B-stars on disks around other stars in the same forming region considering the frequency of B-stars over the total number of stars and the density of stars in each environment. Initial density of the associations is not known. For very young systems, \citet{Parker2014} used the presence of structures to estimate this quantity, and gave values of 10000 \MSun pc$^{-3}$ for the Orion Nebula Cluster and 1000 \MSun pc$^{-3}$ for Upper Scorpius (that we may assume to be representative of the whole Sco-Cen association). No value is available for the associations considered in this paper because they are too old for the application of this method. However, the low frequency of B-stars may suggests that at birth these associations looked similar to Tau or CrA, that according to \citet{Parker2014} have densities of a few tens \MSun pc$^{-3}$. We further assume that the impact scales down with the square of the typical distance from a B-star. We then obtain that the impact of the B-stars is 340 times higher in the Orion Nebula Cluster with respect that in nearby young moving groups, with the Sco-Cen association in between (a factor of about 30 higher). While this estimate is very rough, we may conclude that the nearby young moving groups are quiet environments where disk evolution can proceed rather undisturbed by nearby stars while the Orion Nebula Cluster is an hostile environment where extended disks are rapidly destroyed (see also the discussion in \citealt{Parker2020} and references therein). Interestingly, in star-forming regions where massive stars are present, there are very few detections of gas in protoplanetary disks \citep{Ansdell2017, Mann2014, Mann2015} suggesting that they were subject to intense photoevaporation.

On the other hand, RV surveys average over a wide range of environments. Most stars in the disk of our Galaxy likely formed in regions similar to the Sco-Cen association or the Orion Nebula Cluster. \citet{Miller1978} estimated that about 20\% of the stars should form in T-associations (that evolve into the young moving groups such as those considered in this paper), about 60\% in more massive OB-associations (most massive stars with a mass $>10$~M$_\odot$), 10\% in R-associations  (most massive stars in the range 3-10~M$_\odot$), and the remaining 10\% in open clusters (see also \citet{Lamers2006}). \citet{Lada2003} spell that the vast majority (90\%) of stars that form in embedded clusters form in rich clusters of 100 or more members with masses in excess of 50~M$_\odot$, and embedded clusters account for a significant (70-90\%) fraction of all stars formed in giant molecular clouds. 

Most of these environments are hostile to long-term survival of disks. Depending on the actual fraction of stars that formed in such hostile environments, the fraction of stars hosting JL planets in RV surveys may well be much lower than observed in the nearby young moving groups.

\subsubsection{Loss of JL planets}

The possible dependence of the frequency of JL planets on age suggests that solar-like system may be disrupted on a typical timescale of the order of several hundred million years even in the low density environments of associations. Taken at face value, Eq. (8) indicates that about 50\% of the JL companions are lost in the first 200 Myr, that is the range covered by our sample, but this value is highly uncertain and it is not yet sure that this dependence on age really exists. In this Section we considered various mechanisms that may cause disruption of these systems.

The associations considered here probably had densities of the order of a few tens star/pc$^3$ at birth but the density rapidly drops due to expansion. Using formulas by \citet{Maraboli2023} we found that close encounters with single stars are by far too rare for these young associations to produce ejection of single planets. Even considering close encounters with binaries, that have much larger cross sections, only a tiny fraction of single JL planets are lost in such low density environments as the young associations considered here \citep{Li2020}. However, even a passage at moderate separation can perturb enough the orbits in a multiple packed planetary system to produce ejection of some planet though in general we do not expect ejection of a strict Jupiter analogue unless the passage is very close (a few tens au, see \citealt{Cai2017,Flammini2019, Li2019,Li2020b}). If we limit ourselves to the stars with imaging detections, two out of seven of the planetary systems are rather packed ($\beta$ Pic and especially HR8799) and can possibly become unstable as a consequence of a close passage \citep{Gotberg2016, Zurlo2022}. This data is likely incomplete and suggests that this evolution is possible for a fraction of systems.

Alternatively, we can consider the effect of wide stellar companions. We already considered as unstable all solar-like systems for which the inner radius of the unstable region due to a stellar companion is $a_{max}<20$ au. This practically means that orbits of all JL planets in systems where there is a stellar companion with $a<185$ au are considered unstable. This roughly corresponds to the separation where the characteristics of the planet population in binary systems begins to be statistically different from that of single stars \citep{Desidera2007}. In addition, we should consider the disruptive effect of long-term instability that may be induced by the presence of far stellar companions. \citet{Kaib2013} considered the disruptive effect of the galactic tidal field on solar-like systems residing in wide binaries (separations greater than 1,000 au). They found that between a third to a half of the solar-like systems residing in wide binaries are disrupted over a timescale of a few hundred million of years, that is indeed the timescale we are interested in. In our sample, about 25\% of the stars have a similar wide companion. We then expect that about 10\% of the original JL planets can be lost due to the presence of far stellar companions.

We conclude that the combination of the perturbations induced by wide binaries whose orbits are modified by the galactic tidal field and of close encounters of packed planetary systems with multiple stars may possibly produce an age dependence of the frequency of JL planets, but that the expected depletion is only some 10-20\% over the age range 20-200 Myr covered by the associations considered in this paper. 

\section{Conclusions}


To discuss the mechanisms of formation of companions (both stellar and substellar) to solar-type stars ($M>0.8$ \MSun), we have presented an analysis of the existing data for eight nearby (distance $<100$ pc) young (age $<200$ Myr) associations. The limits in distance and age are important because completeness rapidly declines for further and older associations. In total the sample includes 296 stars in 280 systems. Data clearly indicate two separate populations of companions; a massive sequence (stellar binaries, but also massive BDs) with mass ratios $q>0.05$ covering the whole range of separation surveyed here, and a low-mass sequence ($q<0.015$) that is concentrated at the semi-major axis $1<a<100$ au (JL planets). There is an empty region separating these two groups: this region is the BD desert \citep{Marcy2000}. We tentatively identified the first sequence as the product of disk fragmentation, mainly in the very early phases of disk evolution, and the second one as the product of the core accretion mechanism that requires long-living disks.

Once survey completeness was considered, we found that both groups of companions are very frequent around solar-type stars in these young associations. The survey for stellar companions is almost complete. Only $48\pm 4$\% of the stars are single in our analysis; this value is likely overestimated by a few per cent because our search is not complete. The distribution of the companions in the separation mass ratio plane can be reproduced well within the framework of disk fragmentation. To explain the observations, the parameters used in the model have a trend with stellar mass corresponding to a few, massive episodes of accretion from the interstellar matter onto the disk in B stars, and a more uniform and prolonged accretion over time for solar-type stars. This agrees with expectations. In addition, inward migration is much more efficient in early B stars, a fact that can be explained by the higher mass of the disks in the early phases.

Again, once detection limits are considered, we found that also JL companions are very frequent around stars in young associations, being present in $0.57\pm 0.11$ of the stars that do not have stellar companions that make their orbit unstable. The frequency of JL seems to depend on the mass and possibly the age of the associations. This can be explained by a combination of lower formation rate in environments richer in early-type stars - that might photoionise the proto-planetary disk before formation of the JL planets - and of the destruction of planetary systems related to long-term dynamical instabilities - that more easily occur in systems with several giant planets. 

Confirmation of this result might possibly be obtained in the near future both by exploiting the next release of Gaia data (foreseen in a couple of years from now) and with HCI with the Extremely Large Telescope (ELT, in a few more years). The longer high precision astrometric series provided by Gaia DR4 and the smaller inner working angle possible with ELT should allow for a much higher fraction of the JL planets to be discovered in young associations. Progress is also expected thanks to longer time series of high-precision RV.

Modelling the disk and multiple planetary system destruction by nearby stars may also provide a better understanding of the relevant processes. For what concerns disks, a systematic investigation of disk survival in different environments with, for example, ALMA is also very important. Some studies already exist \citep{Ansdell2017, Mann2014, Mann2015}. We then anticipate significant progress in the next few years.

\begin{acknowledgements}
This work has made use of data from the European Space Agency (ESA) mission {\it Gaia} (\url{https://www.cosmos.esa.int/gaia}), processed by the {\it Gaia} Data Processing and Analysis Consortium (DPAC, \url{https://www.cosmos.esa.int/web/gaia/dpac/consortium}). Funding for the DPAC has been provided by national institutions, in particular, the institutions participating in the {\it Gaia} Multilateral Agreement.
This research has made use of the SIMBAD database, operated at CDS, Strasbourg, France. 
D.M., R.G., and S.D. acknowledge the PRIN-INAF 2019 'Planetary systems at young ages (PLATEA)' and ASI-INAF agreement n.2018-16-HH.0. A.Z. acknowledges support from ANID -- Millennium Science Initiative Program -- Center Code NCN2021\_080. S.M.\ is supported by the Royal Society as a Royal Society University Research  Fellowship (URF-R1-221669). V.S. acknowledges support from the European Research Council (ERC) under the European Union’s Horizon 2020 research and innovation programme (COBREX; grant agreement n◦ 885593). 
SPHERE is an instrument designed and built by a consortium consisting of IPAG (Grenoble, France), MPIA (Heidelberg, Germany), LAM (Marseille, France), LESIA (Paris, France), Laboratoire Lagrange (Nice, France), INAF-Osservatorio di Padova (Italy), Observatoire de Gen\`eve (Switzerland), ETH Zurich (Switzerland), NOVA (Netherlands), ONERA (France) and ASTRON (Netherlands), in collaboration with ESO. SPHERE was funded by ESO, with additional contributions from CNRS (France), MPIA (Germany), INAF (Italy), FINES (Switzerland) and NOVA (Netherlands). 

\end{acknowledgements}

\bibliographystyle{aa} 
\bibliography{main} 

\begin{appendix}

\section{Gaia RV jitter}

\begin{figure}[htb]
    \centering
    \includegraphics[width=8.5cm]{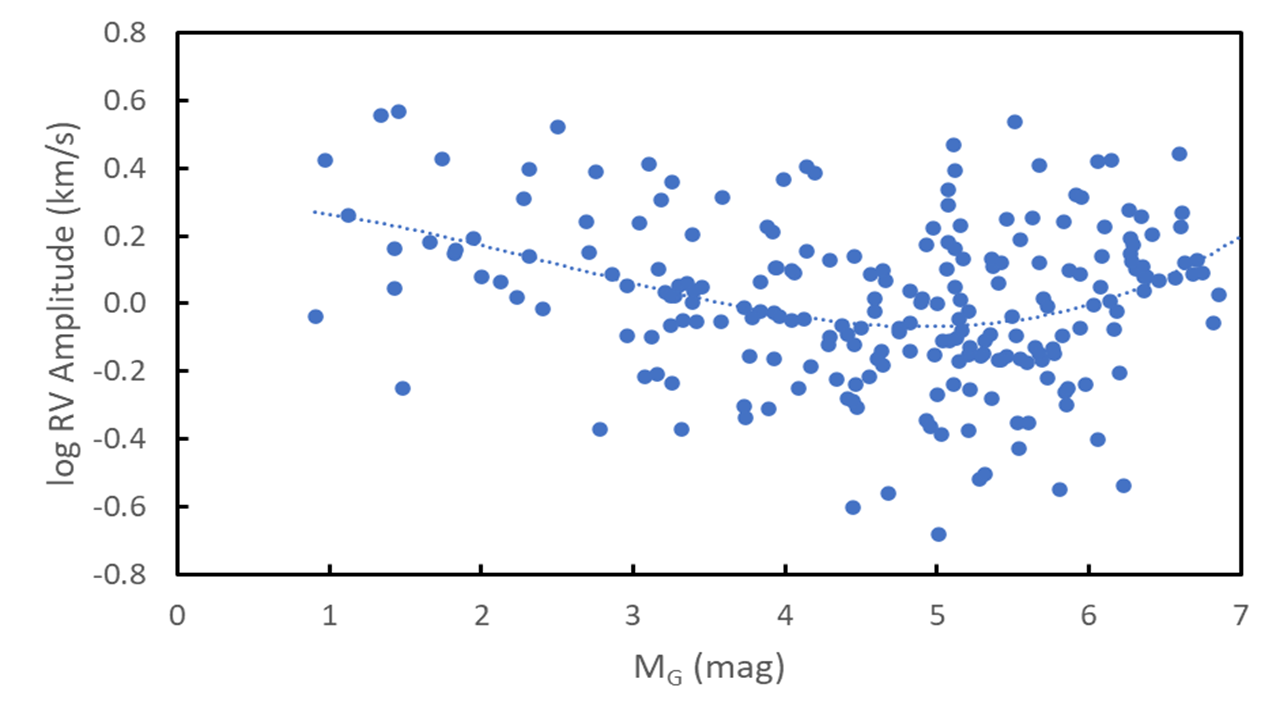}
    \caption{Run of the ratio between Gaia robust RV amplitude and its median value for each association with the absolute magnitude of the stars. Stars in the young associations considered in this paper and the Hyades are plotted. Only stars without any companion within 100 au are considered. The dashed line is a polynomial best fit through data. }
    \label{fig:jitter-magnitude}
\end{figure}

\begin{figure}[htb]
    \centering
    \includegraphics[width=8.5cm]{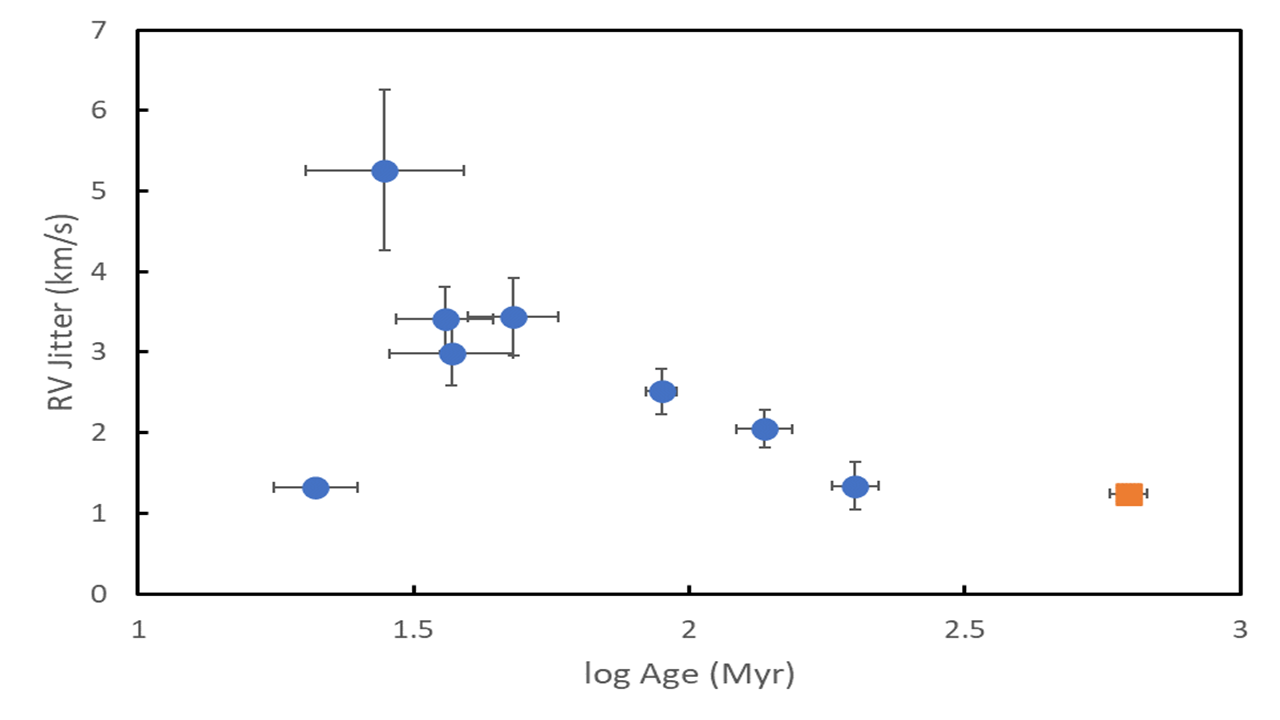}
    \caption{Run of the median Gaia robust RV amplitude with the age for young associations (blue filled symbols) and the Hyades (open red square). Only stars without any companion within 100 au are considered }
    \label{fig:jitter-age}
\end{figure}

The Gaia robust RV amplitude may signal a variation of the RV of the centre of mass of stars. However, this quantity is also influenced by activity and rotation. In order to consider it properly, these dependencies should be taken into account. In practice, we expect trends with the absolute magnitude and age of the stars. To show this, we constructed two plots. The first one (Figure \ref{fig:jitter-magnitude}) shows the run of the ratio between Gaia robust RV amplitude and its median value for each association with absolute magnitude of the stars. Only stars without any companion within 100 au were considered. As expected, the faster rotation of the most luminous stars causes a higher value of the Gaia robust RV amplitude; however, also the less luminous stars considered in this paper (that is however limited to masses $>0.8$ \MSun) have on average a value of the Gaia robust RV amplitude higher than for stars with solar masses. Once this trend with magnitude is considered, we can derive median values of the Gaia robust RV amplitude for each different association and consider the trend with age. This is shown in Figure \ref{fig:jitter-age}. As expected, the median value of the Gaia robust RV amplitude decreases with age. However, the youngest association considered here, the BPMG, has a rather low value of the median Gaia robust RV amplitude.

In practice, we find that the expected value of the Gaia robust RV amplitude $J$ (in \kms) for stars that should have constant RV may be represented by using the following relations:
\begin{equation}
J=1.3*f
\end{equation}
if the age of the association $t$ (in Myr) is higher than 300, and:
\begin{equation}
J=(1.7327*\log{t} +5.813)*f
\end{equation}
if it is lower. The quantity $f$ (in \kms) is given by:
\begin{equation}
\log{f}=0.0068732 M_G^3 - 0.053288 M_G^2 + 0.028749 M_G + 0.25273
\end{equation}
where $M_G$ is the stellar absolute magnitude in the Gaia $G-$band. The r.m.s. scatter around this relation is about 0.2 dex. We will adopt these relations as estimate of the jitter in Gaia RVs. We may consider as SBs those star deviating more than 2.5 $\sigma$ from the mean value; that is, those stars with robust RV amplitude larger than $3.16 J$, that is roughly 11.3 and 5.8 \kms for solar type stars of 20 and 200 Myr, respectively. 

\section{Companions of stars in young associations}
\label{sec:companions}

This Appendix contains tables giving the relevant data about companions around the programme stars. Tables \ref{tab:photometry_abdor} - \ref{tab:photometry_Volans} contain the basic data for the members of the various moving groups (a table for each moving group). The first four columns contain the Hipparcos, HD, and 2MASS numbers as well as alternative names for each star. Column 5 gives the membership probability obtained using the Banyan-$\Sigma$ code \citep{Gagne2018}. Columns 6 and 7 give the coordinates. Column 8 the parallax. Whenever possible this was from Gaia DR3 \citet{Gaia_DR3}; else it was taken from Gaia DR2 or alternatively from the Hipparcos catalogue \citep{1997ESASP1200.....E}. Columns 9 and 10 give the Gaia $G$ magnitude of the primary $G_A$ and if existing of the secondary $G_B$, respectively. Column 11 gives the 2MASS $K$ magnitude \citep{2mass}. Columns 12 and 13 give the the absolute $M_G$ magnitude of the primary and secondary ($M_{GB}$). Column 14 gives the absolute $M_K$ magnitude of the primary.

Tables \ref{tab:info_abdor} - \ref{tab:info_Volans} explains if relevant information about binarity are available and Tables \ref{tab:binary_abdor} - \ref{tab:binary_Volans} give the data relevant (a table for each moving group). In these second group of tables, the first four columns are as described above. Column 5 gives the Gaia $RUWE$ parameter; columns 6-10 give the Gaia mean RV, its error, the number of epochs considered in its derivation, the probability that variation around the mean value is due to random noise, and the robust RV amplitude. Columns 11 and 12 give the separation and position angle of visual companions. Column 13 gives the signal-to-noise ratio (S/N) of the Proper Motion anomaly (PMa) as given by \citet{Kervella2022}.

Tables \ref{tab:mass_abdor} - \ref{tab:mass_Volans} give the derived values for the stars and their companions (a table for each moving group). The first four columns are as described above. Column 5 and 6 gives the mass of the primary and of companions. Here, mass of the primary is the mass of all components within the semimajor axis of the companion considered. Column 7 gives the semi-major axis $a$. Column 8 gives the mass ratio $q$. Column 9 gives the detection method: either visual binary VIS if the secondary was detected with some imaging mode and a separate magnitude could be obtained; or dynamical binary DYN if the secondary is only detected from the dynamical impact on the primary. Finally, Column 10 gives a reference appropriate for each object. In many cases this is a reference to a paper describing the detection and characterisation of the companions. In addition: Gaia means that the companion was detected as a separate entry in the Gaia catalogue; RUWE/RV/PMa and RUWE/RV means that the companion was detected from a high RUWE value ($>1.4$), a high scatter in RV ($>10$ \kms) or from a high S/N PMa ($>3$). All available data were used to derive a consistent solution, using the method described in \citet{Gratton2023}.


\begin{table*}
\caption{Photometry of stars in the AB Dor moving group}
\scriptsize
\begin{tabular}{cccccccccccccc}
\hline
HIP	&	HD	&	2MASS	&	Others/WDS	&	Prob	&	RA	&	Dec	&	$\pi$	&	$G_A$	&	$G_B$	&	$K_A$	&		$M_G$	&	$M_{GB}$	&	$M_K$	\\
&&&&&degree&degree&mas&mag&mag&mag&mag&mag&mag\\
\hline
	&	1405	&	J00182085+3057224	&	PW And	&	99.0	&	4.5870	&	30.95615	&	35.43	&	8.512	&		&	6.387	&	6.259	&		&	4.134	\\
3589	&	4277	&	J00455088+5458402	&	BD+54 144	&	99.3	&	11.46204	&	54.97783	&	20.61	&	7.627	&	9.927	&	6.361	&	4.197	&	6.497	&	2.931	\\
5191	&	6569	&	J01062614-1417468	&		&	99.0	&	16.60897	&	-14.29642	&	21.99	&	9.226	&	13.714	&	7.340	&	5.938	&	10.426	&	4.052	\\
6276	&		&	J01203226-1128035	&	BD-12 243	&	99.7	&	20.13445	&	-11.46770	&	28.33	&	8.234	&		&	6.549	&	5.495	&		&	3.810	\\
	&	236717	&	J01242709+5751062	&		&	9.8	&	21.11297	&	57.85171	&	25.8598	&	8.384	&		&	6.754	&	5.447	&		&	3.817	\\
	&		&	J02051802+5217027	&	TYC 3293-1912-1	&	63.3	&	31.32513	&	52.28408	&	12.9464	&	10.826	&		&	8.67	&	6.387	&		&	4.231	\\
\hline
\end{tabular}
\normalsize
\label{tab:photometry_abdor}
\end{table*}

\begin{table*}
\caption{Photometry of stars in the Argus moving group}
\scriptsize
\begin{tabular}{cccccccccccccc}
\hline
HIP	&	HD	&	2MASS	&	Others/WDS	&	Prob	&	RA	&	Dec	&	$\pi$	&	$G_A$	&	$G_B$	&	$K_A$	&		$M_G$	&	$M_{GB}$	&	$M_K$	\\
&&&&&degree&degree&mas&mag&mag&mag&mag&mag&mag\\
\hline
7345	&	9672	&	J01343776-1540348	&	49 Cet	&	96.6	&	23.6574	&	-15.6764	&	16.84	&	5.611	&		&	5.458	&	1.743	&		&	1.590	\\
	&		&	J05345923-2954041	&	CD-29 2360	&	99.3	&	83.7468	&	-29.9011	&	17.42	&	10.253	&		&	7.994	&	6.458	&		&	4.199	\\
	&		&	J07015339-4227561	&	CD-42 2906	&	99.7	&	105.4725	&	-42.4656	&	11.37	&	10.412	&	14.28	&	8.648	&	5.691	&	9.559	&	3.927	\\
	&		&	J07282203-4908377	&	CD-48 2972	&	99.9	&	112.0918	&	-49.1438	&	11.47	&	9.639	&		&	8.058	&	4.936	&		&	3.355	\\
36312	&	59659	&	J07283006-4908589	&		&	99.9	&	112.1253	&	-49.1497	&	11.42	&	8.678	&	9.64	&	7.585	&	3.966	&	4.927	&	2.873	\\
	&		&	J07311991-5447027	&	CPD-54 1295	&	98.1	&	112.8330	&	-54.7841	&	9.33	&	10.311	&	15.82	&	8.749	&	5.160	&	10.665	&	3.598	\\
\hline
\end{tabular}
\normalsize
\label{tab:photometry_argus}
\end{table*}

\begin{table*}
\caption{Photometry of stars in the BPMG}
\scriptsize
\begin{tabular}{cccccccccccccc}
\hline
HIP	&	HD	&	2MASS	&	Others/WDS	&	Prob	&	RA	&	Dec	&	$\pi$	&	$G_A$	&	$G_B$	&	$K_A$	&		$M_G$	&	$M_{GB}$	&	$M_K$	\\
&&&&&degree&degree&mas&mag&mag&mag&mag&mag&mag\\
\hline

560	&	203	&	J00065008-2306271	&		&	95.7	&	1.7087	&	-23.1075	&	25.16	&	6.094	&		&	5.240	&	3.098	&		&	2.244	\\
	&		&	J01373940+1835332	&	TYC1208-468-1	&	76.6	&	24.4141	&	18.5924	&	18.95	&	10.032	&	10.181	&		&	6.420	&	6.569	&		\\
10679	&	14082B	&	J02172472+2844305	&		&	98.7	&	34.3531	&	28.7418	&	25.23	&	7.623	&		&	6.262	&	4.633	&		&	3.272	\\
10680	&	14082	&	J02172527+2844423	&		&	99.3	&	34.3554	&	28.7450	&	25.28	&	6.904	&	7.623	&	5.787	&	3.918	&	4.637	&	2.801	\\
11437	&		&	J02272924+3058246	&	AG Tri	&	99.4	&	36.8719	&	30.9735	&	24.42	&	9.744	&	11.434	&	7.080	&	6.683	&	8.373	&	4.019	\\
21547	&	29391	&	J04373613-0228248	&	51 Eri	&	9.1	&	69.4006	&	-2.4735	&	33.44	&	5.141	&		&	4.537	&	2.762	&		&	2.158	\\
\hline
\end{tabular}
\normalsize
\label{tab:photometry_bpmg}
\end{table*}

\begin{table*}
\caption{Photometry of stars in the Carina moving group}
\scriptsize
\begin{tabular}{cccccccccccccc}
\hline
HIP	&	HD	&	2MASS	&	Others/WDS	&	Prob	&	RA	&	Dec	&	$\pi$	&	$G_A$	&	$G_B$	&	$K_A$	&		$M_G$	&	$M_{GB}$	&	$M_K$	\\
&&&&&degree&degree&mas&mag&mag&mag&mag&mag&mag\\
\hline
15201	&	21024	&	J03155768-7723184	&	iot Hyi	&	58.5	&	48.9903	&	-77.3885	&	34.26	&	5.405	&		&	4.470	&	3.079	&		&	2.144	\\
	&		&		&		&		&		&		&		&		&	13.855	&		&		&	11.529	&		\\
26144	&	37402	&	J05342614-6006151	&		&	99.6	&	83.6089	&	-60.1042	&	14.06	&	8.266	&		&	7.088	&	4.006	&		&	2.828	\\
	&		&		&		&		&		&		&		&		&	14.132	&		&		&	9.872	&		\\
	&	42270	&	J05532917-8156532	&		&	97.4	&	88.3722	&	-81.9481	&	17.00	&	8.915	&	11.971	&	6.944	&	5.067	&	8.123	&	3.096	\\
29314	&	43199	&	J06105289-6129594	&		&	96.4	&	92.7205	&	-61.4997	&	13.10	&	7.370	&		&	6.554	&	2.956	&		&	2.140	\\
\hline
\end{tabular}
\normalsize
\label{tab:photometry_carina}
\end{table*}


\begin{table*}
\caption{Photometry of stars in the Carina-Near moving group}
\scriptsize
\begin{tabular}{cccccccccccccc}
\hline
HIP	&	HD	&	2MASS	&	Others/WDS	&	Prob	&	RA	&	Dec	&	$\pi$	&	$G_A$	&	$G_B$	&	$K_A$	&		$M_G$	&	$M_{GB}$	&	$M_K$	\\
&&&&&degree&degree&mas&mag&mag&mag&mag&mag&mag\\
\hline
35564	&	57852	&	HIP35564	&	HR 2814	&	99.9	&	110.08926	&	-52.31152	&	28.63	&	5.899	&		&		&	3.183	&		&		\\
	&		&		&		&		&		&		&		&		&	6.464	&		&		&	3.748	&		\\
	&	57853	&		&		&		&		&		&		&	6.464	&		&		&	3.748	&		&		\\
36414	&	59704	&	J07293141-3807217	&		&	98.7	&	112.38089	&	-38.12266	&	18.87	&	7.613	&		&	6.442	&	3.991	&		&	2.820	\\
37563	&	62850	&	J07423606-5917507	&		&	99.9	&	115.65025	&	-59.29742	&	30.44	&	7.041	&		&	5.701	&	4.458	&		&	3.118	\\
37635	&	62848	&	J07432149-5209508	&		&	94.8	&	115.83953	&	-52.16401	&	33.67	&	6.552	&		&	5.321	&	4.188	&		&	2.957	\\
\hline
\end{tabular}
\normalsize
\label{tab:photometry_carinanear}
\end{table*}

\begin{table*}
\caption{Photometry of stars in the Columba moving group}
\scriptsize
\begin{tabular}{cccccccccccccc}
\hline
HIP	&	HD	&	2MASS	&	Others/WDS	&	Prob	&	RA	&	Dec	&	$\pi$	&	$G_A$	&	$G_B$	&	$K_A$	&		$M_G$	&	$M_{GB}$	&	$M_K$	\\
&&&&&degree&degree&mas&mag&mag&mag&mag&mag&mag\\
\hline
3210	&	3888	&	J00405159-5312357	&		&	25.7	&	10.2149	&	-53.2099	&	22.57	&	7.249	&		&	6.113	&	4.017	&		&	2.881	\\
6165	&	8077	&	J01190560-5351020	&		&	2.2	&	19.7734	&	-53.8505	&	11.50	&	8.753	&		&	7.615	&	4.056	&		&	2.918	\\
7805	&	10472	&	J01402406-6059566	&		&	77.8	&	25.1003	&	-60.9991	&	14.05	&	7.530	&		&	6.632	&	3.269	&		&	2.371	\\
	&	TYC 8047-232-1	&	J01521461-5219332	&	CD-52 381	&	45.8	&	28.0609	&	-52.3259	&	11.59	&	10.614	&		&	8.405	&	5.934	&		&	3.725	\\
11029	&	14691	&	J02220152-1046390	&	HR692	&	99.5	&	35.5064	&	-10.7775	&	33.51	&	5.335	&		&	4.522	&	2.961	&		&	2.148	\\
12413	&	16754	&	J02394796-4253300	&	HR789	&	73.1	&	39.9499	&	-42.8918	&	24.71	&	4.685	&		&	4.460	&	1.649	&		&	1.424	\\
\hline
\end{tabular}
\normalsize
\label{tab:photometry_columba}
\end{table*}

\begin{table*}
\caption{Photometry of stars in the Tucana-Horologium moving group}
\scriptsize
\begin{tabular}{cccccccccccccc}
\hline
HIP	&	HD	&	2MASS	&	Others/WDS	&	Prob	&	RA	&	Dec	&	$\pi$	&	$G_A$	&	$G_B$	&	$K_A$	&		$M_G$	&	$M_{GB}$	&	$M_K$	\\
&&&&&degree&degree&mas&mag&mag&mag&mag&mag&mag\\
\hline
490	&	105	&	J00055255-4145109	&		&	99.9	&	1.4689	&	-41.7531	&	25.75	&	7.395	&		&	6.464	&	4.449	&		&	3.171	\\
1113	&	987	&	J00135300-7441178	&		&	99.9	&	3.4709	&	-74.6883	&	21.80	&	8.654	&		&	7.406	&	5.347	&		&	3.655	\\
1481	&	1466	&	J00182612-6328389	&		&	99.9	&	4.6088	&	-63.4775	&	23.36	&	7.339	&		&	6.462	&	4.181	&		&	2.991	\\
2484	&	2884	&	J00313267-6257297	&	bet1 Tuc	&	93.4	&	7.8861	&	-62.9582	&	22.84	&	4.329	&		&	4.664	&	1.122	&		&	1.274	\\
2487	&	2885	&	J00313347-6257560	&	bet2 Tuc	&	0.0	&	7.8893	&	-62.9656	&	22.84	&	4.329	&	4.564	&	4.320	&	1.122	&	1.357	&	0.901	\\
2578	&	3003	&	J00324391-6301533	&	bet3 Tuc	&	99.9	&	8.1829	&	-63.0315	&	22.84	&	4.329	&	4.529	&	5.061	&	1.122	&	1.322	&	1.778	\\
\hline
\end{tabular}
\normalsize
\label{tab:photometry_tuchor}
\end{table*}

\begin{table*}
\caption{Photometry of stars in the Volans/Crius 221 moving group}
\scriptsize
\begin{tabular}{cccccccccccccc}
\hline
HIP	&	HD	&	2MASS	&	Others/WDS	&	Prob	&	RA	&	Dec	&	$\pi$	&	$G_A$	&	$G_B$	&	$K_A$	&		$M_G$	&	$M_{GB}$	&	$M_K$	\\
&&&&&degree&degree&mas&mag&mag&mag&mag&mag&mag\\
\hline
	&		&	J00301560-8613552	&	TYC 9494-799-1	&	0.0	&	7.5722	&	-86.2321	&	18.31	&	9.460	&		&	7.685	&	5.77	&		&	4.00	\\
	&	67928	&	J08075272-5135016	&	TYC 8148-343-1	&	0.0	&	121.9696	&	-51.5836	&	9.36	&	9.857	&	12.357	&	8.232	&	4.71	&	7.21	&	3.09	\\
	&	69406	&	J08134777-5837139	&	CPD-58 1080B	&	0.0	&	123.4489	&	-58.6204	&	9.18	&	8.364	&		&	7.488	&	3.18	&		&	2.30	\\
	&		&			&		&		&		&		&		&		&	11.716			&		&	6.53	&		\\
	&		&	J08363202-6024116	&	TYC 8926-337-1	&	0.0	&	129.1333	&	-60.4030	&	9.29	&	11.113	&		&	9.269	&	5.95	&		&	4.11	\\
45096	&	299896	&	J09111005-5501535	&	TYC 8587-323-1	&	0.0	&	137.7915	&	-55.0312	&	24.54	&	9.248	&		&	7.167	&	6.20	&		&	4.12	\\
\hline
\end{tabular}
\normalsize
\label{tab:photometry_Volans}
\end{table*}

\begin{table}
\caption{Information about multiplicity for stars in the AB Dor moving group. TESS=1 means that there is TESS data, 0 that there is not. RV=1 means that there is Gaia RV data; RV=2 means that there is high precision RV data; RV=0 means that there is no RV data. HCI=1 means that there is HCI data, 0 that there is not. RUWE=1 means that there is RUWE data, 0 that there is not. PMA=1 means that there is PMA data, 0 that there is not.}
\scriptsize
\begin{tabular}{lcccccc}
\hline
2MASS	& Age&TESS &RV&HCI&RUWE&PMA\\
        &Myr &&&&&\\
\hline
J00182085+3057224	&	137	&	1	&	1	&	0	&	1	&	0	\\
J00455088+5458402	&	137	&	1	&	1	&	0	&	1	&	1	\\
J01062614-1417468	&	137	&	1	&	2	&	1	&	1	&	1	\\
J01203226-1128035	&	137	&	1	&	2	&	1	&	1	&	1	\\
J01242709+5751062	&	137	&	1	&	1	&	0	&	1	&	0	\\
J02051802+5217027	&	137	&	1	&	1	&	0	&	1	&	0	\\
\hline
\end{tabular}
\normalsize
\label{tab:info_abdor}
\end{table}

\begin{table}
\caption{Information about multiplicity for stars in the Argus moving group. TESS=1 means that there is TESS data, 0 that there is not. RV=1 means that there is Gaia RV data; RV=2 means that there is high precision RV data; RV=0 means that there is no RV data. HCI=1 means that there is HCI data, 0 that there is not. RUWE=1 means that there is RUWE data, 0 that there is not. PMA=1 means that there is PMA data, 0 that there is not.}
\scriptsize
\begin{tabular}{lcccccc}
\hline
2MASS	& Age&TESS &RV&HCI&RUWE&PMA\\
        &Myr &&&&&\\
\hline
J01343776-1540348	&	48	&	1	&	2	&	1	&	1	&	1	\\
J05345923-2954041	&	48	&	1	&	1	&	0	&	1	&	0	\\
J07015339-4227561	&	48	&	1	&	1	&	1	&	1	&	0	\\
J07282203-4908377	&	48	&	1	&	1	&	1	&	1	&	0	\\
J07283006-4908589	&	48	&	1	&	1	&	1	&	1	&	1	\\
J07311991-5447027	&	48	&	1	&	1	&	0	&	1	&	0	\\
J07354747-3212141	&	48	&	1	&	2	&	1	&	1	&	1	\\
J07472597-4902511	&	48	&	2	&	1	&	0	&	1	&	0	\\
J07484982-4327055	&	48	&	2	&	1	&	1	&	1	&	0	\\
J07541770-5711131	&	48	&	2	&	1	&	1	&	1	&	0	\\
J07535548-5710073	&	48	&	1	&	1	&	1	&	1	&	0	\\
J08391155-5834281	&	48	&	1	&	1	&	0	&	1	&	0	\\
J08500812-5745593	&	48	&	1	&	1	&	1	&	1	&	0	\\
J09020394-5808497	&	48	&	1	&	1	&	0	&	1	&	0	\\
J09133027-6259092	&	48	&	1	&	1	&	1	&	1	&	0	\\
J09361783-7820417	&	48	&	1	&	2	&	1	&	1	&	1	\\
J09395143-2134175	&	48	&	1	&	1	&	0	&	1	&	0	\\
J09424742-7239499	&	48	&	1	&	1	&	0	&	1	&	0	\\
J09471986-4003098	&	48	&	2	&	2	&	1	&	1	&	0	\\
HD85151	&	48	&	1	&	0	&	0	&	1	&	0	\\
J09555832-6721218	&	48	&	1	&	1	&	0	&	1	&	0	\\
J10144416-4207189	&	48	&	1	&	2	&	1	&	1	&	1	\\
J10272528-6542167	&	48	&	1	&	1	&	1	&	1	&	1	\\
J11411978-4305442	&	48	&	1	&	2	&	1	&	1	&	1	\\
J11490366+1434197	&	48	&	1	&	2	&	1	&	1	&	0	\\
J12063292-4247508	&	48	&	1	&	1	&	0	&	1	&	0	\\
J12203437-7539286	&	48	&	1	&	1	&	1	&	1	&	0	\\
J13491293-7549475	&	48	&	1	&	1	&	0	&	1	&	0	\\
J14072929-6133441	&	48	&	1	&	1	&	1	&	1	&	1	\\
J14462144-6746158	&	48	&	1	&	1	&	0	&	1	&	0	\\
J15122343-7515156	&	48	&	1	&	2	&	1	&	1	&	1	\\
J15401155-7013403	&	48	&	1	&	1	&	1	&	1	&	1	\\
J16170540-6756285	&	48	&	1	&	2	&	1	&	1	&	1	\\
J18205694+2951317	&	48	&	1	&	0	&	0	&	1	&	1	\\
J19561425+1125254	&	48	&	1	&	1	&	0	&	1	&	1	\\
J20003558-7254380	&	48	&	1	&	2	&	1	&	1	&	1	\\
J20143203+3648225	&	48	&	1	&	1	&	0	&	1	&	1	\\
\hline
\end{tabular}
\normalsize
\label{tab:info_argus}
\end{table}

\begin{table}
\caption{Information about multiplicity for stars in the BPMG. TESS=1 means that there is TESS data, 0 that there is not. RV=1 means that there is Gaia RV data; RV=2 means that there is high precision RV data; RV=0 means that there is no RV data. HCI=1 means that there is HCI data, 0 that there is not. RUWE=1 means that there is RUWE data, 0 that there is not. PMA=1 means that there is PMA data, 0 that there is not.}
\scriptsize
\begin{tabular}{lcccccc}
\hline
2MASS	& Age&TESS &RV&HCI&RUWE&PMA\\
        &Myr &&&&&\\
\hline
J00065008-2306271	&	21	&	1	&	1	&	1	&	1	&	1	\\
J01373940+1835332	&	21	&	1	&	1	&	1	&	1	&	0	\\
J02172472+2844305	&	21	&	1	&	1	&	1	&	1	&	1	\\
J02172527+2844423	&	21	&	1	&	1	&	1	&	1	&	1	\\
J02272924+3058246	&	21	&	1	&	2	&	1	&	1	&	1	\\
J04373613-0228248	&	21	&	1	&	2	&	1	&	1	&	1	\\
J05004928+1527006	&	21	&	1	&	1	&	1	&	1	&	0	\\
J05270477-1154033	&	21	&	1	&	2	&	1	&	1	&	1	\\
J05471708-5103594	&	21	&	1	&	2	&	1	&	1	&	1	\\
J06182824-7202416	&	21	&	1	&	2	&	1	&	1	&	1	\\
J15385757-5742273	&	21	&	1	&	1	&	1	&	1	&	1	\\
J17150362-2749397	&	21	&	0	&	0	&	1	&	0	&	0	\\
J17172550-6657039	&	21	&	1	&	1	&	1	&	1	&	1	\\
J17295506-5415487	&	21	&	1	&	1	&	1	&	1	&	0	\\
J17414903-5043279	&	21	&	1	&	1	&	1	&	1	&	1	\\
J17483374-5306433	&	21	&	1	&	1	&	1	&	1	&	0	\\
J18030341-5138564	&	21	&	1	&	2	&	1	&	1	&	1	\\
J18064990-4325297	&	21	&	1	&	0	&	1	&	1	&	1	\\
J18195221-2916327	&	21	&	0	&	0	&	1	&	1	&	1	\\
J18452691-6452165	&	21	&	1	&	2	&	1	&	1	&	1	\\
J18453704-6451460	&	21	&	1	&	1	&	1	&	1	&	0	\\
J18480637-6213470	&	21	&	1	&	0	&	0	&	0	&	0	\\
J18530587-5010499	&	21	&	1	&	2	&	1	&	1	&	1	\\
J19114467-2604085	&	21	&	0	&	0	&	1	&	1	&	0	\\
J19225122-5425263	&	21	&	1	&	2	&	1	&	1	&	1	\\
J19225894-5432170	&	21	&	1	&	2	&	1	&	1	&	1	\\
J20041810-2619461	&	21	&	0	&	1	&	0	&	1	&	1	\\
J20090521-2613265	&	21	&	0	&	2	&	1	&	1	&	1	\\
J20554767-1706509	&	21	&	0	&	0	&	1	&	1	&	1	\\
J21475527-5255502	&	21	&	1	&	1	&	0	&	1	&	1	\\
\hline
\end{tabular}
\normalsize
\label{tab:info_bpmg}
\end{table}

\begin{table}
\caption{Information about multiplicity for stars in the Carina moving group. TESS=1 means that there is TESS data, 0 that there is not. RV=1 means that there is Gaia RV data; RV=2 means that there is high precision RV data; RV=0 means that there is no RV data. HCI=1 means that there is HCI data, 0 that there is not. RUWE=1 means that there is RUWE data, 0 that there is not. PMA=1 means that there is PMA data, 0 that there is not.}
\scriptsize
\begin{tabular}{lcccccc}
\hline
2MASS	& Age&TESS &RV&HCI&RUWE&PMA\\
        &Myr &&&&&\\
\hline
J03155768-7723184	&	28	&	1	&	1	&	1	&	1	&	1	\\
J05342614-6006151	&	28	&	1	&	1	&	0	&	1	&	1	\\
J05532917-8156532	&	28	&	1	&	2	&	1	&	1	&	0	\\
J06105289-6129594	&	28	&	1	&	1	&	0	&	1	&	1	\\
J06191291-5803156	&	28	&	1	&	2	&	1	&	1	&	1	\\
J06434625-7158356	&	28	&	1	&	2	&	1	&	1	&	1	\\
J07003046-7941459	&	28	&	1	&	2	&	1	&	1	&	1	\\
J08210046-5213406	&	28	&	1	&	1	&	0	&	1	&	1	\\
J08225514-5207254	&	28	&	1	&	2	&	1	&	1	&	1	\\
J09172755-7444045	&	28	&	1	&	2	&	1	&	1	&	1	\\
J09233498-6111359	&	28	&	1	&	1	&	1	&	1	&	1	\\
J09303148-7041479	&	28	&	2	&	1	&	0	&	1	&	0	\\
J09312500-7344496	&	28	&	1	&	0	&	1	&	1	&	1	\\
J09392101-6119410	&	28	&	1	&	1	&	0	&	1	&	1	\\
J09534760-5453540	&	28	&	1	&	1	&	0	&	1	&	0	\\
J10131476-5230540	&	28	&	1	&	2	&	1	&	1	&	0	\\
J10300355-7432320	&	28	&	1	&	1	&	0	&	1	&	0	\\
\hline
\end{tabular}
\normalsize
\label{tab:info_carina}
\end{table}

\begin{table}
\caption{Information about multiplicity for stars in the Carina Near moving group. TESS=1 means that there is TESS data, 0 that there is not. RV=1 means that there is Gaia RV data; RV=2 means that there is high precision RV data; RV=0 means that there is no RV data. HCI=1 means that there is HCI data, 0 that there is not. RUWE=1 means that there is RUWE data, 0 that there is not. PMA=1 means that there is PMA data, 0 that there is not.}
\scriptsize
\begin{tabular}{lcccccc}
\hline
2MASS	& Age&TESS &RV&HCI&RUWE&PMA\\
        &Myr &&&&&\\
\hline
HIP35564	&	200	&	1	&	1	&	1	&	1	&	1	\\
J07293141-3807217	&	200	&	1	&	1	&	1	&	1	&	1	\\
J07423606-5917507	&	200	&	1	&	1	&	1	&	1	&	1	\\
J07432149-5209508	&	200	&	1	&	1	&	1	&	1	&	1	\\
J07461480-5948506	&	200	&	1	&	1	&	1	&	1	&	1	\\
J07461694-5948341	&	200	&	2	&	2	&	1	&	1	&	1	\\
J08123271-5408473	&	200	&	1	&	1	&	0	&	1	&	1	\\
J08191905+0120198	&	200	&	1	&	2	&	1	&	1	&	1	\\
J08343166-0043339	&	200	&	1	&	1	&	1	&	1	&	1	\\
J11564232-3216053	&	200	&	1	&	1	&	1	&	1	&	1	\\
J11564379-3216026	&	200	&	1	&	2	&	1	&	1	&	1	\\
J12280445+4447394	&	200	&	1	&	2	&	0	&	1	&	1	\\
J12280480+4447305	&	200	&	1	&	2	&	0	&	1	&	1	\\
J12300479-1323350	&	200	&	1	&	1	&	0	&	1	&	1	\\
J13554999+1403234	&	200	&	1	&	1	&	1	&	1	&	1	\\
\hline
\end{tabular}
\normalsize
\label{tab:info_carinanear}
\end{table}

\begin{table}
\caption{Information about multiplicity for stars in the Columba moving group. TESS=1 means that there is TESS data, 0 that there is not. RV=1 means that there is Gaia RV data; RV=2 means that there is high precision RV data; RV=0 means that there is no RV data. HCI=1 means that there is HCI data, 0 that there is not. RUWE=1 means that there is RUWE data, 0 that there is not. PMA=1 means that there is PMA data, 0 that there is not.}
\scriptsize
\begin{tabular}{lcccccc}
\hline
2MASS	& Age&TESS &RV&HCI&RUWE&PMA\\
        &Myr &&&&&\\
\hline
J00405159-5312357	&	36	&	1	&	1	&	0	&	1	&	1	\\
J01190560-5351020	&	36	&	1	&	1	&	1	&	1	&	1	\\
J01402406-6059566	&	36	&	1	&	2	&	1	&	1	&	1	\\
J01521461-5219332	&	36	&	1	&	2	&	1	&	1	&	0	\\
J02220152-1046390	&	36	&	1	&	1	&	0	&	1	&	1	\\
J02394796-4253300	&	36	&	1	&	0	&	1	&	1	&	1	\\
\hline
\end{tabular}
\normalsize
\label{tab:info_columba}
\end{table}

\begin{table}
\caption{Information about multiplicity for stars in the Tuc-Hor moving group. TESS=1 means that there is TESS data, 0 that there is not. RV=1 means that there is Gaia RV data; RV=2 means that there is high precision RV data; RV=0 means that there is no RV data. HCI=1 means that there is HCI data, 0 that there is not. RUWE=1 means that there is RUWE data, 0 that there is not. PMA=1 means that there is PMA data, 0 that there is not.}
\scriptsize
\begin{tabular}{lcccccc}
\hline
2MASS	& Age&TESS &RV&HCI&RUWE&PMA\\
        &Myr &&&&&\\
\hline
J00055255-4145109	&	37	&	1	&	2	&	1	&	1	&	1	\\
J00135300-7441178	&	37	&	1	&	2	&	1	&	1	&	1	\\
J00182612-6328389	&	37	&	1	&	2	&	1	&	1	&	1	\\
J00313267-6257297	&	37	&	1	&	1	&	1	&	1	&	1	\\
J00313347-6257560	&	37	&	1	&	0	&	1	&	0	&	0	\\
J00324391-6301533	&	37	&	1	&	2	&	1	&	1	&	1	\\
\hline
\end{tabular}
\normalsize
\label{tab:info_tuchor}
\end{table}

\begin{table}
\caption{Information about multiplicity for stars in the Volans/Crius 221 moving group. TESS=1 means that there is TESS data, 0 that there is not. RV=1 means that there is Gaia RV data; RV=2 means that there is high precision RV data; RV=0 means that there is no RV data. HCI=1 means that there is HCI data, 0 that there is not. RUWE=1 means that there is RUWE data, 0 that there is not. PMA=1 means that there is PMA data, 0 that there is not.}
\scriptsize
\begin{tabular}{lcccccc}
\hline
2MASS	& Age&TESS &RV&HCI&RUWE&PMA\\
        &Myr &&&&&\\
\hline
J00301560-8613552	&	89	&	1	&	1	&	0	&	1	&	0	\\
J08075272-5135016	&	89	&	2	&	0	&	0	&	1	&	0	\\
J08134777-5837139	&	89	&	2	&	1	&	0	&	1	&	0	\\
J08363202-6024116	&	89	&	1	&	1	&	0	&	1	&	0	\\
J09111005-5501535	&	89	&	1	&	2	&	0	&	1	&	1	\\
J09134337-4642014	&	89	&	1	&	1	&	0	&	1	&	0	\\
\hline
\end{tabular}
\normalsize
\label{tab:info_Volans}
\end{table}

\begin{table*}
\caption{Binary data for stars in the AB Dor moving group}
\scriptsize
\begin{tabular}{ccccccccccccc}
\hline
HIP	&	HD	&	2MASS	&	Others/WDS	&	RUWE	&	RV	&	RV err	&	Nb	&	pvalue	&	RV ampl	&	sep	&	PA	&	S/N PMA	\\
&&&&& \kms & \kms &&& \kms & arcsec & degree & \\
\hline
	&	1405	&	J00182085+3057224	&	PW And	&	0.989	&	-11.91	&	0.33	&	17	&	7.00678E-06	&	3.87	&		&		&		\\
3589	&	4277	&	J00455088+5458402	&	BD+54 144	&	0.959	&	-16.99	&	0.17	&	71	&	0	&	4.99	&	2.72	&	68.9	&	3.00	\\
5191	&	6569	&	J01062614-1417468	&		&	0.956	&	7.87	&	0.21	&	15	&	0.016022373	&	1.73	&	22.56	&	331.12	&	1.03	\\
6276	&		&	J01203226-1128035	&	BD-12 243	&	0.984	&	10.21	&	0.14	&	37	&	7.83559E-06	&	1.87	&		&		&	2.06	\\
	&	236717	&	J01242709+5751062	&		&	0.965	&	-11.94	&	0.13	&	29	&	0.36065376	&	0.91	&		&		&		\\
	&		&	J02051802+5217027	&	TYC 3293-1912-1	&	0.872	&	-10.52	&	0.48	&	14	&	0.005912767	&	4.38	&		&		&		\\
\hline
\end{tabular}
\normalsize
\label{tab:binary_abdor}
\end{table*}

\begin{table*}
\caption{Binary data for stars in the Argus moving group}
\scriptsize
\begin{tabular}{ccccccccccccc}
\hline
HIP	&	HD	&	2MASS	&	Others/WDS	&	RUWE	&	RV	&	RV err	&	Nb	&	pvalue	&	RV ampl	&	sep	&	PA	&	S/N PMA	\\
&&&&& \kms & \kms &&& \kms & arcsec & degree & \\
\hline
7345	&	9672	&	J01343776-1540348	&	49 Cet	&	0.782	&	8.76	&	0.50	&	23	&	1.61E-01	&	9.18	&		&		&	1.69	\\
	&		&	J05345923-2954041	&	CD-29 2360	&	1.912	&	25.45	&	0.38	&	28	&	3.87717E-05	&	5.91	&		&		&		\\
	&		&	J07015339-4227561	&	CD-42 2906	&	0.833	&	23.51	&	0.31	&	13	&	1.92E-01	&	2.35	&	6.25	&	109.97	&		\\
	&		&	J07282203-4908377	&	CD-48 2972	&	0.817	&	19.46	&	0.96	&	20	&	1.11E-16	&	14.58	&		&		&		\\
36312	&	59659	&	J07283006-4908589	&		&	0.982	&	18.38	&	0.36	&	15	&	6.88E-08	&	3.15	&	81.64	&	285.06	&	0.55	\\
	&		&	J07311991-5447027	&	CPD-54 1295	&	0.977	&	19.46	&	0.37	&	9	&	0.5561346	&	2.86	&	3.45	&	9.55	&		\\
\hline
\end{tabular}
\normalsize
\label{tab:binary_argus}
\end{table*}

\begin{table*}
\caption{Binary data for stars in the BPMG}
\scriptsize
\begin{tabular}{ccccccccccccc}
\hline
HIP	&	HD	&	2MASS	&	Others/WDS	&	RUWE	&	RV	&	RV err	&	Nb	&	pvalue	&	RV ampl	&	sep	&	PA	&	S/N PMA	\\
&&&&& \kms & \kms &&& \kms & arcsec & degree & \\
\hline
560	&	203	&	J00065008-2306271	&		&	0.933	&	6.06	&	0.49	&	18	&	7.13E-03	&	3.42	&		&		&	3.02	\\
	&		&	J01373940+1835332	&	TYC1208-468-1	&	1.371	&	-2.15	&	1.74	&	7	&	6.08E-05	&	9.62	&	1.635	&	22.805	&		\\
10679	&	14082B	&	J02172472+2844305	&		&	0.919	&	5.00	&	0.16	&	12	&	3.93E-02	&	0.95	&		&		&	5.21	\\
10680	&	14082	&	J02172527+2844423	&		&	1.008	&	5.54	&	0.27	&	12	&	8.29E-10	&	2.15	&		&		&	2.04	\\
11437	&		&	J02272924+3058246	&	AG Tri	&	0.892	&	5.55	&	0.23	&	10	&	8.89E-01	&	1.62	&	22.002	&	315.683	&	0.57	\\
21547	&	29391	&	J04373613-0228248	&	51 Eri	&	1.145	&		&		&		&		&		&		&		&	2.08	\\
\hline
\end{tabular}
\normalsize
\label{tab:binary_bpmg}
\end{table*}

\begin{table*}
\caption{Binary data for stars in the Carina moving group}
\scriptsize
\begin{tabular}{ccccccccccccc}
\hline
HIP	&	HD	&	2MASS	&	Others/WDS	&	RUWE	&	RV	&	RV err	&	Nb	&	pvalue	&	RV ampl	&	sep	&	PA	&	S/N PMA	\\
&&&&& \kms & \kms &&& \kms & arcsec & degree & \\
\hline
15201	&	21024	&	J03155768-7723184	&	iot Hyi	&	1.000	&	14.16	&	0.13	&	9	&	0.8163255	&	0.42	&		&		&	6.74	\\
	&		&		&		&		&		&		&		&		&		&	275.57	&	77.14	&		\\
26144	&	37402	&	J05342614-6006151	&		&	8.692	&	21.79	&	1.22	&	16	&	0	&	11.45	&	0.01	&	166.00	&	1.61	\\
	&		&		&		&		&		&		&		&		&		&	291.84	&		&		\\
	&	42270	&	J05532917-8156532	&		&	0.990	&	17.03	&	0.90	&	8	&	1.75E-14	&	6.66	&	2.08	&	285.32	&		\\
29314	&	43199	&	J06105289-6129594	&		&	0.978	&	14.47	&	0.53	&	20	&	8.48E-05	&	5.94	&		&		&	0.85	\\
\hline
\end{tabular}
\normalsize
\label{tab:binary_carina}
\end{table*}

\begin{table*}
\caption{Binary data for stars in the Carina Near moving group}
\scriptsize
\begin{tabular}{ccccccccccccc}
\hline
HIP	&	HD	&	2MASS	&	Others/WDS	&	RUWE	&	RV	&	RV err	&	Nb	&	pvalue	&	RV ampl	&	sep	&	PA	&	S/N PMA	\\
&&&&& \kms & \kms &&& \kms & arcsec & degree & \\
\hline
35564	&	57852	&	HIP35564	&	HR2814	&	8.691	&	22.89	&	2.14	&	10	&	0.00E+00	&	16.96	&		&		&	0.88	\\
	&	57853	&		&		&	22.137	&		&		&		&		&		&	9.05	&	26.72	&		\\
36414	&	59704	&	J07293141-3807217	&		&	0.941	&	26.65	&	0.23	&	24	&	0.00E+00	&	3.13	&		&		&	1.25	\\
37563	&	62850	&	J07423606-5917507	&		&	0.960	&	17.48	&	0.14	&	16	&	1.17E-01	&	1.01	&		&		&	1.41	\\
37635	&	62848	&	J07432149-5209508	&		&	17.534	&	18.33	&	0.30	&	15	&	0.00E+00	&	2.91	&		&		&	45.54	\\
37918	&	63581	&	J07461480-5948506	&		&	17.199	&	17.73	&	0.27	&	18	&	0.00E+00	&	3.20	&	0.065	&	282.1	&	25.39	\\
	&		&		&		&		&		&		&		&		&		&	23.15	&	44.63	&		\\
37923	&	63608	&	J07461694-5948341	&		&	1.176	&	17.43	&	0.14	&	17	&	1.46E-02	&	0.91	&		&		&	0.18	\\
40188	&	69051	&	J08123271-5408473	&		&	0.894	&	28.76	&	0.28	&	18	&	6.31E-05	&	2.74	&	2.59	&	299.85	&	0.9	\\
40774	&		&	J08191905+0120198	&	V0397 Hya	&	1.053	&		&		&	10	&		&	0.05	&		&		&	0.88	\\
42074	&	72760	&	J08343166-0043339	&		&	0.992	&	35.00	&	0.12	&	25	&	0.985	&	0.44	&		&		&	78.51	\\
58240	&	103742	&	J11564232-3216053	&		&	1.058	&	6.35	&	0.13	&	29	&	6.59E-02	&	1.13	&	18.70	&	81.71	&	0.89	\\
58241	&	103743	&	J11564379-3216026	&		&	1.027	&	6.98	&	0.13	&	30	&	3.01E-01	&	1.38	&		&		&	2.1	\\
60831	&	108574	&	J12280445+4447394	&		&	0.905	&	-2.31	&	0.13	&	27	&	1.98E-02	&	1.02	&		&		&	3.12	\\
	&		&		&		&		&		&		&		&		&		&	9.71	&	156.98	&		\\
60832	&	108575	&	J12280480+4447305	&		&	1.067	&	-2.05	&	0.17	&	4	&	5.33E-01	&	0.48	&		&		&	5.94	\\
60994	&	108799	&	J12300479-1323350	&	HR4758	&	0.927	&	1.30	&	0.14	&	21	&	1.57E-04	&	1.00	&	2.12	&	346.77	&	84.19	\\
68030	&	121560	&	J13554999+1403234	&		&	0.926	&	-10.27	&	0.12	&	20	&	6.88E-01	&	0.75	&		&		&	1.89	\\
\hline
\end{tabular}
\normalsize
\label{tab:binary_carinanear}
\end{table*}

\begin{table*}
\caption{Binary data for stars in the Columba moving group}
\scriptsize
\begin{tabular}{ccccccccccccc}
\hline
HIP	&	HD	&	2MASS	&	Others/WDS	&	RUWE	&	RV	&	RV err	&	Nb	&	pvalue	&	RV ampl	&	sep	&	PA	&	S/N PMA	\\
&&&&& \kms & \kms &&& \kms & arcsec & degree & \\
\hline
3210	&	3888	&	J00405159-5312357	&		&	1.081	&	9.08	&	0.18	&	32	&	2.41E-10	&	2.62	&	0.05	&		&	1.55	\\
6165	&	8077	&	J01190560-5351020	&		&	0.951	&	12.16	&	0.43	&	13	&	3.35E-09	&	4.19	&		&		&	1.22	\\
7805	&	10472	&	J01402406-6059566	&		&	0.945	&	14.29	&	0.25	&	24	&	3.08E-06	&	3.58	&		&		&	2.21	\\
	&	TYC8047-232-1	&	J01521461-5219332	&	CD-52 381	&	0.869	&	13.35	&	1.25	&	27	&	3.39E-10	&	11.47	&	3.27	&	358.84	&		\\
11029	&	14691	&	J02220152-1046390	&	HR692	&	0.893	&	12.79	&	0.25	&	18	&	6.11E-09	&	2.75	&		&		&	2.87	\\
12413	&	16754	&	J02394796-4253300	&	HR789	&	5.363	&		&		&		&		&		&	0.11	&		&	45.58	\\
\hline
\end{tabular}
\normalsize
\label{tab:binary_columba}
\end{table*}

\begin{table*}
\caption{Binary data for stars in the Tuc-Hor moving group}
\scriptsize
\begin{tabular}{ccccccccccccc}
\hline
HIP	&	HD	&	2MASS	&	Others/WDS	&	RUWE	&	RV	&	RV err	&	Nb	&	pvalue	&	RV ampl	&	sep	&	PA	&	S/N PMA	\\
&&&&& \kms & \kms &&& \kms & arcsec & degree & \\
\hline
490	&	105	&	J00055255-4145109	&		&	0.912	&	2.05	&	0.16	&	19	&	5.44E-04	&	1.54	&		&		&	2.85	\\
1113	&	987	&	J00135300-7441178	&		&	0.834	&	9.09	&	0.19	&	19	&	1.11E-02	&	2.41	&		&		&	0.70	\\
1481	&	1466	&	J00182612-6328389	&		&	0.994	&	6.53	&	0.16	&	27	&	4.86E-06	&	1.57	&		&		&	3.46	\\
2484	&	2884	&	J00313267-6257297	&	bet1 Tuc	&	0.839	&	10.77	&	0.45	&	17	&	3.28E-01	&	5.44	&	2.4	&		&	1.63	\\
2487	&	2885	&	J00313347-6257560	&	bet2 Tuc	&		&		&		&		&		&		&	27.02	&	168.28	&		\\
2578	&	3003	&	J00324391-6301533	&	bet3 Tuc	&	1.025	&	3.60	&	0.25	&	20	&	8.65E-01	&	3.08	&	0.1	&	172	&	0.47	\\
\hline
\end{tabular}
\normalsize
\label{tab:binary_tuc-hor}
\end{table*}

\begin{table*}
\caption{Binary data for stars in the Volans/Crius 221 moving group}
\scriptsize
\begin{tabular}{ccccccccccccc}
\hline
HIP	&	HD	&	2MASS	&	Others/WDS	&	RUWE	&	RV	&	RV err	&	Nb	&	pvalue	&	RV ampl	&	sep	&	PA	&	S/N PMA	\\
&&&&& \kms & \kms &&& \kms & arcsec & degree & \\
\hline
	&		&	J00301560-8613552	&	TYC 9494-799-1	&	0.929	&	14.21	&	0.21	&	15	&	0.00314	&	1.79	&		&		&		\\
	&	67928	&	J08075272-5135016	&	TYC 8148-343-1	&	2.251	&		&		&		&		&		&	0.70	&	183.00	&		\\
	&	69406	&	J08134777-5837139	&	CPD-58 1080B	&	1.661	&	27.87	&	0.33	&	22	&	0.00002	&	5.08	&		&		&		\\
	&		&			&		&		&		&		&		&		&		&	6.95	&	172.40	&		\\
	&		&	J08363202-6024116	&	TYC 8926-337-1	&	0.996	&	25.96	&	0.30	&	25	&	0.79582	&	5.16	&		&		&		\\
45096	&	299896	&	J09111005-5501535	&	TYC 8587-323-1	&	0.801	&	26.85	&	0.25	&	8	&	0.06292	&	1.57	&		&		&	0.92	\\
\hline
\end{tabular}
\normalsize
\label{tab:binary_Volans}
\end{table*}

\begin{table*}
\caption{Mass and separation of companions in the AB Dor moving group}
\scriptsize
\begin{tabular}{cccccccccl}
\hline
HIP	&	HD	&	2MASS	&	Others/WDS	&	$M_A$	&	$M_B$	&	a	&	q	&	Method	&	Source	\\
	&		&		&		&	\MSun	&	\MSun	&	au	&		&		&		\\
\hline
 &	1405	&	J00182085+3057224	&	PW And	&	0.877	&		&		&		&		&		\\
3589	&	4277	&	J00455088+5458402	&		&	1.106	&	0.848	&	131.996	&	0.767	&	VIS	&	\citet{Cvetkovic2011}	\\
5191	&	6569	&	J01062614-1417468	&		&	0.911	&	0.204	&	1025.517	&	0.224	&	VIS	&	Gaia	\\
6276	&		&	J01203226-1128035	&	BD-12 243	&	0.952	&		&		&		&		&		\\
	&	236717	&	J01242709+5751062	&		&	0.956	&		&		&	&		&		\\
	&		&	J02051802+5217027	&	TYC 3293-1912-1	&	0.862	&		&		&	&		&		\\
\hline
\end{tabular}
\normalsize
\label{tab:mass_abdor}
\end{table*}

\begin{table*}
\caption{Mass and separation of companions in the Argus moving group}
\scriptsize
\begin{tabular}{cccccccccl}
\hline
HIP	&	HD	&	2MASS	&	Others/WDS	&	$M_A$	&	$M_B$	&	a	&	q	&	Method	&	Source	\\
	&		&		&		&	\MSun	&	\MSun	&	au	&		&		&		\\
\hline
7345	&	9672	&	J01343776-1540348	&	49 Cet	&	1.910	&		&		&		&		&		\\
	&		&	J05345923-2954041	&	CD-29 2360	&	0.853	&	0.144	&	0.313	&	0.169	&	DYN	&	RUWE-Pma-RV	\\
	&		&	J07015339-4227561	&	CD-42 2906	&	0.935	&	0.315	&	549.947	&	0.337	&	VIS	&	Gaia	\\
	&		&	J07282203-4908377	&	CD-48 2972	&	1.002	&		&		&		&		&		\\
36312	&	59659	&	J07283006-4908589	&		&	1.160	&	1.002	&	7150.207	&	0.864	&	VIS	&	Gaia	\\
	&		&	J07311991-5447027	&	CPD-54 1295	&	0.981	&	0.180	&	369.863	&	0.184	&	VIS	&	Gaia	\\
\hline
\end{tabular}
\normalsize
\label{tab:mass_argus}
\end{table*}

\begin{table*}
\caption{Mass and separation of companions in the BPMG}
\scriptsize
\begin{tabular}{cccccccccl}
\hline
HIP	&	HD	&	2MASS	&	Others/WDS	&	$M_A$	&	$M_B$	&	a	&	q	&	Method	&	Source	\\
	&		&		&		&	\MSun	&	\MSun	&	au	&		&		&		\\
\hline
560	&	203	&	J00065008-2306271	&		&	1.342	&	0.0017	&	16.6	&	0.0013	&	DYN	&		\\
	&	TYC1208-468-1	&		&		&	0.884	&	0.738	&	86.3	&	0.835	&	VIS	&	Gaia DR3	\\
10679	&	14082B	&	J02172472+2844305	&		&	1.103	&		&		&		&		&	Gaia DR3	\\
10680	&	14082	&	J02172527+2844423	&		&	1.169	&	1.016	&	544.0	&	0.869	&	VIS	&	Gaia DR3	\\
	&		&	J02272924+3058246	&	AG Tri	&	0.820	&	0.490	&	900.860	&	0.598	&	VIS	&		\\
21547	&	29391	&	J04373613-0228248	&	51 Eri	&	1.486	&	0.005	&	11.1	&	0.0031 	&	VIS	&	\citet{DeRosa2020, Samland2017}	\\
\hline
\end{tabular}
\normalsize
\label{tab:mass_bpmg}
\end{table*}

\begin{table*}
\caption{Mass and separation of companions in the Carina moving group}
\scriptsize
\begin{tabular}{cccccccccl}
\hline
HIP	&	HD	&	2MASS	&	Others/WDS	&	$M_A$	&	$M_B$	&	a	&	q	&	Method	&	Source	\\
	&		&		&		&	\MSun	&	\MSun	&	au	&		&		&		\\
\hline
15201	&	21024	&	J03155768-7723184	&	iot Hyi	&	1.413	&	0.0135	&	17.8	&	0.0096	&	DYN	&	PMa+RUWE+RV	\\
	&		&		&		&	1.422	&	0.116	&	8042.698	&	0.082	&	VIS	&	Gaia	\\
26144	&	37402	&	J05342614-6006151	&		&	1.149	&	0.473	&	0.915	&	0.412	&	VIS	&	SB	\\
	&		&		&		&	1.622	&	0.270	&	20754.354	&	0.167	&	VIS	&	Gaia	\\
	&	42270	&	J05532917-8156532	&		&	0.989	&	0.571	&	122.322	&	0.577	&	VIS	&	Gaia	\\
29314	&	43199	&	J06105289-6129594	&		&	1.446	&		&		&		&		&		\\
30034	&	44627	&	J06191291-5803156	&	AB Pic	&	0.965	&	0.0080	&	3.570	&	0.0071	&	DYN	&	PMa+RUWE+RV	\\
	&		&		&		&	1.140	&	0.022	&	270.247	&	0.019	&	VIS	&	\citet{Vigan2021}	\\
32235	&	49855	&	J06434625-7158356	&		&	0.971	&		&		&		&		&		\\
33737	&	55279	&	J07003046-7941459	&		&	0.928	&		&		&		&		&		\\
40916	&	70703	&	J08210046-5213406	&		&	1.658	&		&		&		&		&		\\
41081	&	71043	&	J08225514-5207254	&	HR 3300	&	1.952	&		&		&		&		&		\\
45585	&	80950	&	J09172755-7444045	&	HR 3721	&	2.072	&	0.00176	&	6.8	&	0.0085	&	DYN	&	PMa+RUWE+RV	\\
46063	&	81544	&	J09233498-6111359	&	V479 Car	&	1.014	&		&		&		&		&	\citet{Elliott2015} is background using Gaia data	\\
	&	TYC 9200-446-1	&	J09303148-7041479	&		&	0.989	&		&		&		&		&		\\
46720	&	83096	&	J09312500-7344496	&		&	1.448	&	1.024	&	151.807	&	0.707	&	VIS	&	Gaia	\\
	&		&		&		&	2.472	&	0.333	&	4294.486	&	0.135	&	VIS	&	Gaia	\\
47391	&	83944	&	J09392101-6119410	&	m Car	&	3.007	&	0.353	&	8.440	&	0.117	&	DYN	&	PMa+RUWE+RV	\\
	&	TYC 8602-718-1	&	J09534760-5453540	&		&	0.911	&	0.237	&	2.53	&	0.260	&	DYN	&	PMa+RUWE+RV	\\
	&	298936	&	J10131476-5230540	&	TWA-21	&	0.920	&		&		&		&		&		\\
	&		&	J10300355-7432320	&	CD-73 584	&	0.932	&		&		&		&		&		\\
\hline
\end{tabular}
\normalsize
\label{tab:mass_carina}
\end{table*}

\begin{table*}
\caption{Mass and separation of companions in the Carina-Near moving group}
\scriptsize
\begin{tabular}{cccccccccl}
\hline
HIP	&	HD	&	2MASS	&	Others/WDS	&	$M_A$	&	$M_B$	&	a	&	q	&	Method	&	Source	\\
	&		&		&		&	\MSun	&	\MSun	&	au	&		&		&		\\
\hline
35564	&		&	J070202142-5218413	&	HR2814	&	1.387	&	0.683	&	1.046	&	0.492	&	DYN	&	RUWE+Pma+RV	\\
	&		&		&		&	2.070	&	1.626	&	316.003	&	0.785	&	VIS	&	Gaia	\\
	&	57853	&		&		&	1.226	&	0.400	&	0.567	&	0.326	&	VIS	&	\citet{Tokovinin2018}	\\
36414	&	59704	&	J07293141-3807217	&		&	1.153	&		&		&		&		&		\\
37563	&	62850	&	J07423606-5917507	&		&	1.060	&		&		&		&		&		\\
37635	&	62848	&	J07432149-5209508	&		&	1.108	&	0.180	&	3.84	&	0.163	&	DYN	&	RUWE+Pma+RV	\\
37918	&	63581	&	J07461480-5948506	&		&	0.972	&	0.280	&	1.931	&	0.253	&	VIS	&	\citet{Bonavita2022} 	\\
	&		&		&		&	1.252	&	0.961	&	794.662	&	0.989	&	VIS	&	Gaia	\\
37923	&	63608	&	J07461694-5948341	&		&	0.958	&		&		&		&		&		\\
40188	&	69051	&	J08123271-5408473	&		&	1.671	&	0.148	&	217.40	& 0.089	& VIS &	Gaia \\
40774	&		&	J08191905+0120198	&	V0397 Hya	&	0.867	&		&		&		&		&		\\
42074	&	72760	&	J08343166-0043339	&		&	0.952	&	0.130	&	31.95	& 0.137	& VIS &	\citet{Hirsch2021} \\
58240	&	103742	&	J11564232-3216053	&		&	1.022	&	1.006	&	664.037	&	0.984	&	VIS	&	Gaia	\\
58241	&	103743	&	J11564379-3216026	&		&	1.005	&		&		&		&		&		\\
60831	&	108574	&	J12280445+4447394	&		&	1.147	&	0.0034	&	9.5	&	0.0030	&	DYN	&	RUWE+Pma+RV	\\
	&		&		&		&	1.156	&	1.041	&	440.072	&	0.908	&	VIS	&	Gaia	\\
60832	&	108575	&	J12280480+4447305	&		&	1.040	&	0.01074	&	9.3	&	0.0071	&	DYN	&	RUWE+Pma+RV	\\
60994	&	108799	&	J12300479-1323350	&	HR 4758	&	1.080	&	0.781	&	52.309	&	0.723	&	VIS	&	Gaia	\\
68030	&	121560	&	J13554999+1403234	&		&	1.129	&		&		&		&		&		\\
\hline
\end{tabular}
\normalsize
\label{tab:mass_carinanear}
\end{table*}

\begin{table*}
\caption{Mass and separation of companions in the Columba moving group}
\scriptsize
\begin{tabular}{cccccccccl}
\hline
HIP	&	HD	&	2MASS	&	Others/WDS	&	$M_A$	&	$M_B$	&	a	&	q	&	Method	&	Source	\\
	&		&		&		&	\MSun	&	\MSun	&	au	&		&		&		\\
\hline
3210	&	3888	&	J00405159-5312357	&		&	1.146	&	0.400	&	2.245	&	0.349	&	VIS	&	\citet{Bonavita2016}	\\
6165	&	8077	&	J01190560-5351020	&		&	1.137	&		&		&		&		&		\\
7805	&	10472	&	J01402406-6059566	&		&	1.457	&		&		&		&		&		\\
	&	TYC8047-232-1	&	J01521461-5219332	&	CD-52 381	&	0.912	&	0.020	&	282.193	&	0.022	&	VIS	&	\citet{Chauvin2010}	\\
11029	&	14691	&	J02220152-1046390	&	HR692	&	1.445	&		&		&		&		&		\\
12413	&	16754	&	J02394796-4253300	&	HR789	&	1.958	&	0.600	&	4.490	&	0.306	&	VIS	&	\citet{Bonavita2016}	\\
\hline
\end{tabular}

$^1$This star is classified as an EB apparently from \citet{Gaposchkin1932} with a period of about seven days and dm=3.1 mag. However HARPS RV is constant with 150 ms (PtV) over 80 days \citep{Trifonov2020}. In addition, it is very difficult to produce the light curve required to produce such a deep eclipse.
\normalsize
\label{tab:mass_columba}
\end{table*}
 
\begin{table*}
\caption{Mass and separation of companions in the Tuc-Hor moving group}
\scriptsize
\begin{tabular}{cccccccccl}
\hline
HIP	&	HD	&	2MASS	&	Others/WDS	&	$M_A$	&	$M_B$	&	a	&	q	&	Method	&	Source	\\
	&		&		&		&	\MSun	&	\MSun	&	au	&		&		&		\\
\hline
490	&	105	&	J00055255-4145109	&		&	1.062	&		&		&		&		&		\\
1113	&	987	&	J00135300-7441178	&		&	0.965	&		&		&		&		&		\\
1481	&	1466	&	J00182612-6328389	&		&	1.109	&	0.0016	&	14.2	&	0.0014	&	DYN	&	\citet{Mesa2022}	\\
2484	&	2884	&	J00313267-6257297	&	bet1 Tuc	&	2.280	&	0.480	&	105.275	&	0.211	&	VIS	&	\citet{Tokovinin2018}	\\
2487	&	2885	&	J00313347-6257560	&	bet2 Tuc	&	2.760	&	2.126	&	1192.927	&	0.770	&	VIS	&	Gaia	\\
2578	&	3003	&	J00324391-6301533	&	bet3 Tuc	&	2.280	&	2.148	&	4.379	&	0.942	&	VIS	&	\citet{Mason2001}	\\
\hline
\end{tabular}
\normalsize
\label{tab:mass_tuchor}
\end{table*}

\begin{table*}
\caption{Mass and separation of companions in the Volans/Crius 221 moving group}
\scriptsize
\begin{tabular}{cccccccccl}
\hline
HIP	&	HD	&	2MASS	&	Others/WDS	&	$M_A$	&	$M_B$	&	a	&	q	&	Method	&	Source	\\
	&		&		&		&	\MSun	&	\MSun	&	au	&		&		&		\\
\hline
	&		&	J08264964-6346369	&	TYC 8933-327-1	&	0.908	&	0.557	&	1.45	&	0.614	&	DYN	&	RUWE/Pma/RV	\\
	&	73129	&	J08324622-6508272	&		&	1.034	&	0.307	&	31730.892	&	0.297	&	VIS	&	Gaia	\\
43783	&	76728	&	J08550282-6038406	&	HR3571	&	5.061	&	0.911	&	3.742	&	0.180	&	DYN	&	RUWE/Pma/RV	\\
	&		&		&		&	5.972	&	0.586	&	1472.161	&	0.098	&	VIS	&	\citet{Mason2001}	\\
45594	&	80563	&	J09173415-6323145	&		&	1.369	&		&		&		&		&		\\
46460	&	82406	&	J09283051-6642067	&	HR3783	&	2.176	&	0.851	&	0.82	&	0.391	&	DYN	&	RUWE/Pma/RV	\\
\hline
\end{tabular}
\normalsize
\label{tab:mass_Volans}
\end{table*}

\end{appendix}

\end{document}